\documentclass{aa}  
\usepackage{dblfloatfix}
\usepackage{placeins}
\usepackage{graphicx}
\usepackage{txfonts}
\usepackage{comment}
\usepackage{soul}
\usepackage{amsmath}
\usepackage[colorlinks=true,linkcolor=blue,filecolor=blue,urlcolor=blue,citecolor=blue,breaklinks=true]{hyperref}
\usepackage[flushleft]{threeparttable}
\usepackage{lastpage}
\usepackage{float}
\usepackage{multirow}
\usepackage{makecell}
\usepackage{xcolor}
\usepackage{cleveref}
\usepackage{appendix}
\usepackage{multicol}

\begin{document} 

   \title{Simulation-based inference of galaxy properties from JWST pixels}
   \titlerunning{Simulation-based inference of galaxy properties from JWST pixels}

   \author{Patricia Iglesias-Navarro \inst{1,2}   \and Marc Huertas-Company \inst{1,2,3,4}  \and Pablo Pérez-González \inst{5} \and Johan H. Knapen \inst{1,2}  \and ChangHoon Hahn \inst{7} \and Anton M. Koekemoer\inst{6} \and Steven L. Finkelstein\inst{8} \and Natalia Villanueva \inst{8} \and Andrés Asensio Ramos\inst{1,2}} 

   \institute{Instituto de Astrofísica de Canarias, C/ Vía Láctea s/n, 38205 La Laguna, Tenerife, Spain \and Departamento de Astrofísica, Universidad de La Laguna, 38200 La Laguna, Tenerife, Spain \and Observatoire de Paris, LERMA, PSL University, 61 avenue de l'Observatoire, F-75014 Paris, France \and  Universit\'e Paris-Cit\'e, 5 Rue Thomas Mann, 75014 Paris, France \and Centro de Astrobiolog\'{\i}a (CAB), CSIC-INTA, Ctra. de Ajalvir km 4, Torrej\'on de Ardoz, E-28850, Madrid, Spain \and Space Telescope Science Institute, 3700 San Martin Drive, Baltimore, MD 21218, USA \and Department of Astronomy and Steward Observatory, University of Arizona, 933 North Cherry Avenue, Tucson, AZ 85721, USA \and Department of Astronomy, The University of Texas at Austin, Austin, TX, USA}

      \date{}

   \abstract 
   {
The spectral energy distributions (SEDs) of galaxies offer detailed insights into their stellar populations, capturing key physical properties such as stellar mass, star formation history (SFH), metallicity, and dust attenuation. However, inferring these properties from SEDs is a highly degenerate inverse problem, particularly when using integrated observations across a limited range of photometric bands. We present an efficient Bayesian SED-fitting framework tailored to multiwavelength pixel photometry from the JWST Advanced Deep Extragalactic Survey (JADES). Our method employs simulation-based inference to enable rapid posterior sampling across galaxy pixels, leveraging the unprecedented spatial resolution, wavelength coverage, and depth provided by the survey. It is trained on synthetic photometry generated from MILES stellar population models, incorporating both parametric and non-parametric SFHs, realistic noise, and JADES-like filter sensitivity thresholds. We validated this amortised inference approach on mock datasets, achieving robust and well-calibrated posterior distributions, with an $R^2$ score of 0.99 for stellar mass. Applying our pipeline to real observations, we derived spatially resolved maps of stellar population properties down to $\mathrm{S/N}_{\rm{pixel}}=5$ (averaged over F277W, F356W, and F444W) for $1083$ JADES galaxies and $\sim$2 million pixels with spectroscopic redshifts. These maps enable the identification of dusty or starburst regions, offering insights into mass growth and structural assembly. We assessed the outshining phenomenon by comparing pixel-based and integrated stellar mass estimates, finding a limited impact only in low-mass galaxies ($<10^8\,M_{\odot}$), but with systematic differences of $\sim$0.20 dex linked to SFH priors. With an average posterior sampling speed of $10^{-4}$ seconds per pixel and a total inference time of $\sim$1 CPU-day for the full dataset, our model offers a scalable solution for extracting high-fidelity stellar population properties from HST+JWST datasets, paving the way for statistical studies on sub-galactic scales.}


   \maketitle

\section{Introduction}

Spatially resolved spectral energy distribution (SED) fitting boasts a long history of use as a powerful tool for understanding internal galaxy structure and evolution. Early works by \citet{abraham1999} and \citet{zibetti2009} demonstrated how spatial variations in stellar populations and mass-to-light ratios can impact global galaxy properties. With high-resolution multi-wavelength imaging from the Hubble Space Telescope (HST), pixel-by-pixel SED fitting techniques were further developed (e.g. \citet{Wuyts2012}, \citet{LanyonFoster2012}) revealing radial gradients in star formation and dust attenuation at intermediate redshifts ($z \sim 1$–2). These studies laid the groundwork for understanding internal galaxy diversity beyond integrated light measurements.
A key challenge in performing unresolved SED fittings is the outshining effect, where bright, young star-forming regions dominate the observed light and mask older, more massive, but less luminous stellar populations, leading to systematic underestimations of total stellar mass \citep{Sorba2015,Sorba2018,Abdurro2017,Smith2018}. This effect is particularly significant in galaxies with high specific star formation rates (sSFRs) and bursty star formation histories (SFHs).\\

Recent advances with HST and especially the James Webb Space Telescope (JWST) have transformed this field by enabling high-resolution, multi-band SEDs for hundreds of thousands of galaxies, extending pixel-level analyses to high redshifts ($z \sim 2$–10) \citep{Perez-Gonzalez2023,Watson2024,Lines2024,Gibson2024,gimenez2024,harvey2025}. These data have revealed that outshining can cause up to ~1 dex discrepancies between integrated and resolved stellar masses in low-mass, high-sSFR galaxies \citep{gimenez2023,Gimnez-Arteaga2024,fujimoto24}, while more massive galaxies show smaller offsets likely due to a smoother SFHs \citep{Lines2024,Perez-Gonzalez2023}. Theoretical studies support these observations, with hydrodynamic simulations demonstrating how bursty SFHs cause mass estimate biases \citep{Narayanan24}, showing that resolved fitting recovers stellar masses well when deep multiband photometry is available \citep{Cochrane2025,Mosleh2025}. Model assumptions, priors, and spatial resolution further influence the interpretation of resolved SED fits \citep{harvey2025}.
Together, these observational and theoretical developments, powered by HST and JWST data, open up a new window onto galaxy evolution by enabling detailed, spatially resolved stellar population analyses over a wide range of cosmic timescales.\\

This vast amount of high-resolution data continues to grow, while the methods we use to analyse it have largely remained the same. Traditional SED analysis techniques are computationally intensive and often prohibitively slow. Standard optimisation approaches \citep[e.g.][]{sawicki98,heavens2000,tojeiro07,dye08, Iyer2017} can take several minutes per galaxy, while Bayesian methods \citep[e.g.][]{pacifici12,Carnall_2018,Boquien2019, Johnson2021} that are critical for accounting for uncertainties and degeneracies in stellar properties can demand up to 10-100 CPU hours per galaxy \citep{Tacchella2021}. This bottleneck in computational efficiency stands in the way of a scalable analysis of galaxy stellar populations.

A solution to this challenge lies in combining Bayesian inference with machine learning tools to achieve rapid and accurate posterior distributions for galaxy physical properties. In particular, simulation-based inference \citep[SBI;][]{Cranmer_2020} has emerged as a powerful framework for learning complex, non-parametric probability distributions from synthetic data \citep[e.g.][]{Hahn2022,2022Khullar,Li2023,iglesias2024,kwon2024}. This method enables an amortised inference: after an initial training phase on synthetic galaxy samples, the model can estimate posterior distributions for new observations in a fraction of the traditional time, bypassing the need for retraining. This approach captures degeneracies among parameters and their uncertainties effectively, even in the context of the sparse photometric data typical of deep-field surveys.

In this paper, we extend this novel and flexible approach to estimate the stellar population properties of resolved galaxies in the Hubble Ultra Deep Field \citep[HUDF;][]{beckwith2006}. Using filters in the Advanced Camera for Surveys on the HST \citep[ACS;][]{ryon2023} and the Near Infrared Camera on the JWST \citep[NIRCam;][]{rieke2023a}, our method reconstructs distributions of stellar mass, SFR, metallicity, and dust content per pixel for thousands of galaxies. Special attention is given to the uncertainties in flux measurements, depth variations, and PSF matching between the HST and JWST filters, optimising the training to reflect realistic observational conditions. Our model is applicable to galaxies across a range of morphologies and redshifts up to \( z \sim 10\), offering a comprehensive framework for studying stellar population properties across cosmic time.

We obtain high-resolution, two-dimensional maps of stellar population properties of over $1000$ galaxies in the JWST Advanced Deep Extragalactic Survey \citep[JADES;][]{eisenstein2023}. The pixel-by-pixel precision achieved by this approach enables us to identify and analyse regions of dusty starbursts, uncovering clues about their mass growth and structural assembly processes and gaining broader insights into the star formation in galaxies during the earliest cosmic epochs.

The outline of the paper is as follows. In Sect.~\ref{data}, we detail the observational data. In Sect.~\ref{methods}, we describe the Bayesian inference framework, using stellar population synthesis (Sect.~\ref{forward}) and a neural density estimator (Sect.~\ref{normflows}). In Sect.~\ref{test}, we test the model with mock observations and then with individual pixels from the JADES survey Sect.~ \ref{JADES_galaxies}. We describe our pixel-by-pixel analysis of the full dataset in Sect.~\ref{pixel-by-pixel}. Finally, in Sects.~\ref{sec:limitations} and~\ref{discussion}, we discuss our model's inferred galaxy properties and potential  future applications of this technique. We assumed a flat cosmology with $\Omega_{M} = 0.3$, $\Omega_{\Lambda} = 0.7$, and a Hubble constant $H_0=70$~km s$^{-1}$Mpc$^{-1}$. We used AB magnitudes \citep{oke1983} and a Chabrier initial mass function \citep{chabrier2003}.

\section{Observational data}
\label{data}

We selected a sample of $1083$ galaxies in the HUDF \citep{beckwith2006} within the GOODS-South field \citep[GOODS-S;][]{Giavalisco2004}, where the JADES \citep{eisenstein2023}, JEMS \citep{williams2023}, and  FRESCO \citep{Oesch2023} surveys overlap. The MUSE \citep{bacon2017,bacon2023} survey also covers this region, which guarantees spectroscopic redshifts for a large fraction of the galaxies. We used no particular criterion and opted to select all the galaxies from the JADES segmentation map \citep{Rieke2023,DEUGENIO2024} in a region with all the available $19$ filters included in Table \ref{filters}, as well as the spectroscopic redshifts. Later, in the inference step, we  used a signal-to-noise ratio (S/N) criterion, as described in Sect.~\ref{JADES_galaxies}. By relying only on spectroscopic redshifts we reduce one major uncertainty of the SED fitting, namely, the uncertainty on the photometric redshift.

\begin{table}[h]
\centering
\caption{Filters  (19), cameras, and surveys used, along with the 5$\sigma$ depth limits per pixel.}
\footnotesize
    \begin{tabular}{cccc}
        \hline
        Filter & Instrument & Survey & 5$\sigma$ [nJy] \\
        \hline
        F435W & ACS & HUDF & 0.394 \\
        F606W & ACS & HUDF & 0.611 \\
        F775W & ACS & HUDF & 0.376 \\
        F814W & ACS & HUDF & 1.388 \\
        F850LP & ACS & HUDF & 0.724 \\
        F090W & NIRCam & JADES & 0.601 \\
        F115W & NIRCam & JADES & 0.496 \\
        F150W & NIRCam & JADES & 0.489 \\
        F182M & NIRCam & JEMS + FRESCO & 1.361 \\
        F200W & NIRCam & JADES & 0.509 \\
        F210M & NIRCam & JEMS + FRESCO & 0.987 \\
        F277W & NIRCam & JADES & 0.367 \\
        F335M & NIRCam & JEMS & 0.616 \\
        F356W & NIRCam & JADES & 0.416 \\
        F410M & NIRCam & JEMS & 0.547 \\
        F430M & NIRCam & JEMS & 1.327 \\
        F444W & NIRCam & JADES + FRESCO & 0.472 \\
        F460M & NIRCam & JEMS & 1.813 \\
        F480M & NIRCam & JEMS & 1.351 \\
        \hline

    \end{tabular}
\label{filters}
\end{table}

We used photometry from the HST ACS filters and JWST NIRCam wide-band and medium-band filters (see Table~\ref{filters}). This provides a wide wavelength coverage from 0.4\;$\mu$m to 4.8\;$\mu$m, from rest-frame ultraviolet (UV) and optical to near-infrared (NIR)\ light. The HST mosaics include data from GOODS \citep{Giavalisco2004} that overlap with imaging data from the HUDF \citep{beckwith2006,ellis2013,illingworth2013,koekemoer2013} and CANDELS \citep{grogin2011,koekemoer2011}. These were combined with other archival HST imaging data,  all been reprocessed using up-to-date calibrations to produce mosaics at a scale of 0$\farcs$03 per pixel, astrometrically aligned to Gaia-DR3\footnote{\url{https://www.cosmos.esa.int/web/gaia/dr3}}. For the JADES mosaics, we used the DR2 of JADES of the GOODS field, which are described in more detail by the JADES team \citep{Rieke2023,eisenstein2023b,eisenstein2023}.

We subtracted the background signal using the \texttt{Background2D} function from \texttt{Photutils} \citep{bradley2024}. We estimated the background by first dividing the mosaics into grids of $200 \times 200$ pixels $(0.1'\times 0.1')$. We then computed, for each grid, the background level using the $\sigma$-clipping method, averaging the flux of pixels whose signal was not above $3\sigma$ of the total flux distribution in each grid.

We convolved all the mosaics with the point spread function (PSF) of the reddest wide-band filter: F444W, except for the F460M and F480M mosaics, which were left unconvolved. We built the PSFs using \texttt{Photutils} from a list of unresolved objects in the HUDF \citep{Pirzkal2005}. We generated the kernels with the function \texttt{Create\_matching\_kernel}, again from the \texttt{Photutils} package, using a \texttt{CosineBellWindow} filter with \( \alpha = 0.5 \). We normalised the kernels so that they would not affect the average flux values.
 
\section{Simulation-based inference}
\label{methods}

Our approach is aimed at inferring the posterior distribution of galaxy properties, denoted as \( \theta \), based on photometric observations, \( \{X_i\} \). Traditional Bayesian inference methods depend on the likelihood function  \( P(X|\theta) \), which describes the probability of observing the data given the parameters, to compute the posterior distributions through the Bayes' theorem. However, such calculations become challenging in high-dimensional systems with complex degeneracies between those parameters. To address this issue, we employ SBI, which utilises forward models to simulate data from physical parameters and produce pairs of parameters and data points \( (\theta, X') \) via a forward model. By comparing the synthetic data points,  \( X '\), with the actual observations,  \( X \), SBI enables the estimation of the posterior distribution of the parameters.

This comparison is performed with a backward or inverse model. Techniques such as approximate Bayesian computation refine this process by simulating and selecting parameter values that are closely aligned with observed data \citep[see][for a review]{marin2011}. Advanced methods employ neural networks to backward model complex posterior distributions, for example normalising flows (NFs) \citep{durkan2019}. Although SBI avoids the direct calculation of the likelihood function and provides flexibility in model assumptions, it requires high-fidelity simulated data. As in any inference, the accuracy of the properties depends on the quality of the forward models, and specific modelling considerations, such as prior selection, wavelength coverage, and resolution, can have a significant impact on the inferred properties of stellar populations \citep{pacifi2023}. It is therefore essential to develop a reliable forward model in order to ensure reliable parameter inference.

\subsection{Forward model}
\label{forward}

First, we generated a dataset of simulated composite stellar populations (CSP; interpretable as galaxies or galaxy pixels) with known ACS and NIRCam photometry, associated errors, and properties, \( \theta \). We relied on the stellar population synthesis (SPS) approach: starting from a library of empirical stellar spectra and assuming an initial mass function (IMF) and a set of isochrones, we retrieved the SEDs of simple stellar populations (SSPs) across varying ages and metallicities. We combined these using SFHs, assuming a constant metallicity, and we applied a dust attenuation law. For a more detailed review, we refer to \cite{Conroy_2013} and \cite{iyer2025spectralenergydistributionsgalaxies}.

We used the pipeline provided by \texttt{FSPS} \citep{Conroy2009, Conroy_2010} and \texttt{Dense Basis} \citep{Iyer2017,Iyer19}, with MIST isochrones \citep{Choi_2016}, the MILES stellar spectral library \citep{vazdekis2010}, and the DL07 dust emission library \citep{draine2007}, as well as a Chabrier IMF \citep{chabrier2003}. We considered a Calzetti dust attenuation law \citep{Calzetti2000}, coupling the birth-cloud attenuation to that from older stars, and we generated both nebular continuum and emission using \texttt{CLOUDY} \citep{Ferland2013, Ferland2017}, with the gas-phase metallicity set equal to stellar metallicity and held constant throughout the galaxy’s lifetime. We used two sets of simulations with different SFHs. First, we considered a parametric tau-delayed model,
\begin{equation}
\operatorname{SFR}(t)\propto\left(t-t_i\right) e^{-\left(t-t_i\right) / \tau} \quad \text { for } t \geq t_i,
\end{equation}
where $t_i$ is the cosmic time at which the star formation begins, $\tau$ is the characterisic timescale for the exponential decline, and  $t$ is the current cosmic time. For $t<t_i,$ we have\;$\operatorname{SFR}(t)=0$, since no star formation occurs before $t_i$.

We also constructed non-parametric SFHs with the \texttt{Dense Basis} module \texttt{GP-SFH} \citep{Iyer2017,Iyer19}, allowing for complex behaviours such as rejuvenation events, bursts, or sudden quenching without relying on a fixed functional form. Our selected prior assumes that the  sSFR  (i.e. the SFR normalised by stellar mass) across three equally spaced time bins, follows a Dirichlet distribution with a concentration parameter of \( \alpha=1 \). This  SFH prior was studied in detail by \cite{Leja2019}. Here, we also included the AGB circumstellar dust model from \cite{Villaume2015} and accounted for mass loss in our total calculations.

The priors for the parameters that control these components for both simulations are specified in Table~\ref{priors_combined}. Rather than matching the observational distributions, we opted to keep the priors as uniform as possible, as inference with unbalanced priors is more prone to systematic effects and biases \citep{hahn2023}. The exception is the redshift distribution, which remains uniform but whose ranges were chosen according to the dataset described in Sect.~\ref{data}. We set a lower limit of \( \log_{10} (M_{*}/\rm{M}_{\odot}) = 4.0 \) for the prior for the stellar mass. At these very low stellar masses per pixel, SSP models become unreliable due to statistical fluctuations in the stellar population, particularly from stochastic sampling of the initial mass function (IMF, \citealt{cervino2003}). This limitation is especially relevant in the outskirts of low-redshift galaxies, where the surface mass density can be extremely low. In these regions, inferred stellar population properties such as age and metallicity, are highly uncertain. Therefore, instead of fitting each pixel independently, more robust strategies such as binning adjacent pixels or analysing radial profiles should be used to improve the signal and reduce stochastic noise.

\begin{table}
\caption{Priors for the parameters of the simulation using the $\tau$-delayed and Dirichlet models.}
\label{priors_combined}
\centering
\footnotesize
\begin{tabular}{p{2.5cm}p{2.0cm}p{2.6cm}}
\hline
Parameter & $\tau$-delayed & Dirichlet \\ 
\hline
$\log_{10} (M_{*}^{\rm{formed}}/\rm{M}_{\odot})^{\dagger}$ & U$(4.0,12.0)$ & U$(4.0,12.0)$ \\
$\tau$~[Gyr] & LU$(10^{-2},100)$ & -- \\
$t_i$~[Gyr] & U$(0,\mathrm{age}_{\mathrm{Universe}}(z))$ & -- \\
$\log_{10}$(sSFR) & -- & U$(-11.5,-7.5)$ \\
$t_{25\%,50\%,75\%}$$^{\dagger\dagger}$ & -- & Dir$(\alpha=1.0,N=3)$ \\
$[M/\mathrm{H}]$ & U$(-2.3,0.4)$ & U$(-2.3,0.4)$ \\
$A_{V}$ & U$(0.0,4.0)$ & U$(0.0,4.0)$ \\
$z$ & U$(0.0,7.5)$ & U$(0.0,7.5)$ \\
\hline
\multicolumn{3}{c}{{Fixed parameters}} \\
\hline
$\gamma_{\rm{dust}}$ & \multicolumn{2}{c}{$0.01$} \\
$U_{\rm{min}}$ & \multicolumn{2}{c}{$1.0$} \\
$Q_{\rm{PAH}}$ & \multicolumn{2}{c}{$2.0$} \\
log($U$) & \multicolumn{2}{c}{$-2.0$} \\
\hline
\end{tabular}
\tablefoot{For the Dirichlet SFHs, we selected the prior on $\log_{10}$(sSFR/(yr$^{-1}$)), but we inferred $\log_{10}$(SFR/($\rm{M}_{\odot}$yr$^{-1}$)) instead. Fixed parameters include dust emission, nebular continuum, and emission information. \textbf{U}: Uniform, \textbf{LU}: Log-uniform,  and \textbf{Dir}: Dirichlet.}
\begin{tablenotes}
\footnotesize
\item{ $\dagger$ Formed stellar mass (integral of SFH). We also inferred the surviving stellar mass.}
\item{ $\dagger\dagger$ Cosmic times at which 25\%, 50\%, and 75\% of total stellar mass was formed, normalised to the age of the Universe.}
\end{tablenotes}
\end{table}

The output of the forward model were the fluxes in the 19 filters included in Table~\ref{filters}, including ACS filters and NIRCam wide-band and medium-band filters.  To simulate realistic observational conditions, we modelled the flux uncertainties as a function of flux in each filter. We used the JADES segmentation map to identify all pixels classified as galaxies and, for each filter, we binned the pixel fluxes into five flux bins. Within each bin, we collected the flux uncertainties of the real observations and fit their distribution with a Gaussian. These fitted Gaussians, capturing the empirical uncertainty-flux relation, were then used to add noise to the simulated fluxes: for each simulated SED, the noise was sampled from the appropriate Gaussian based on its flux in each filter. This approach captures a broad range of uncertainties consistent with the observational data. Following the procedure described in \cite{Hahn2022}, the network inputs included the fluxes with noise sampled from the distributions, as well as the standard deviations of the Gaussian distributions, to condition the inference by producing noise-aware posteriors.

We transformed both fluxes and standard deviations to AB magnitudes, avoiding numerical instabilities in the training process caused by the wide dynamic range of possible flux values. We also took into account the  $1 \sigma$ depth limit in each filter $m_{i}^{b}$. We used the values obtained in \cite{Hainline2024} for the JADES mosaics. Although we repeated the calculations of the $1 \sigma$ depth limits in the region were the galaxies were selected, we found negligible differences. To amortise this process,  we used a mean value for the depth limit that was not dependent on the specific regions where the galaxies were located. If the simulated magnitude in a filter, $m_{i}$, for a CSP is fainter than that depth limit ($m_i>m_{\sigma_i}^b$), we transformed it to $m_i^{\prime}=100$ and and the noise in the input was $m_{\sigma_i}^b$. While $m_i^{\prime}=100$ is not physically meaningful (since it should be infinite), it is a convenient choice that avoids convergence problems in the calculations and allows for a clear identification of dropouts (i.e. cases where the signal is not detected in a given filter). These dropouts typically occur at bluer wavelengths than the Lyman break for high-z galaxies, although non-detections can also occur in redder filters, especially in low-S/N photometry. Identifying these dropouts could also ensure a correct redshift estimate, although we have not addressed it in this work. We obtained posterior distributions from the photometry to constrain the stellar population parameters included in Table \ref{priors_combined} for CSPs, using the neural density estimator, as explained below.

\subsection{Amortised neural inference}
\label{normflows}

We used a neural density estimator known as 'NFs' to estimate the posterior distributions for the stellar population properties. This method transforms variables described by a simple base distribution, such as a multivariate Gaussian, through a series of parametrised invertible transformations. The parameters of these transformations are trained to minimise the Kullback–Leibler divergence between the target distribution and the model’s estimate. For a more detailed explanation, we refer to \cite{rezende2015} and \cite{Hahn2022}.

We used a specific implementation of NFs known as MAF \citep[masked autoregressive flow;][]{papamakarios2017}, implemented with the \texttt{SBI} module \citep{tejero2020}. Following the example of \cite{Hahn2022}, we selected 15 MADE blocks (masked autoencoder for distribution estimation), each containing two hidden layers with 500 units. We split the data into training and validation sets in a 90/10 ratio and used the Adam optimiser \citep{kingma2017} with a learning rate of \( 5 \cdot 10^{-4} \). To prevent overfitting, we monitored the model's performance on the validation set and halted the training if the validation likelihood did not improve after $20$ epochs.

We performed two trainings for the two different simulations, with parametric~(a) and non-parametric~(b) SFHs. The output of the network are the posterior distributions for the following properties: the formed stellar mass,  surviving stellar mass including remnants,  average SFR for the last $100$ Myr,  metallicity $[M/H],$ and  dust attenuation index, $A_{V}$. For the SFHs, in (a) we inferred $\tau$ and $t_i$, both in Gyr, while in (b) we inferred the normalised time $(t_{x}/t_{\rm{universe(z)}})$ at which $25\%$, $50\%,$ and $75\%$ of the total stellar mass were formed.

We discuss the possibility of also retrieving the redshift pixel-by-pixel, as done in \citep{Perez-Gonzalez2023} in Sect.\ref{discussion}. However, in this work we introduce the aforementioned integrated spectroscopic redshifts as an input in the network together with the photometry. The training was performed with $1.000.000$ simulated CSPs. We generated another $10.000$ for testing purposes.    

\subsection{Fitting of JADES galaxies}

We use the trained model to fit the galaxies introduced in Sect.~\ref{data}. We computed the noise of the galaxies in each pixel and filter as $\sqrt{\sigma^2 + \sigma_{b}^2}$, where $\sigma$ is the flux uncertainty from the mosaics and  $\sigma_b$ the standard deviation of the background signal. We refer to the NIRCam IDs of the galaxies throughout the paper. All spectroscopic redshifts are extracted from catalogues \citep{dahlen2010,guo12,Oesch2023,bunker24}. We input  the photometry with uncertainties (in AB magnitudes) into the network, again repeating  the cut at a $1 \sigma$ depth limit for each filter as done for the simulations, as well as the redshifts. Thus, we obtained the posterior distributions for each pixel for (a) $\log_{10} (M_{*}/\rm{M}_{\odot})$, $\log_{10} (M_{*}^{\rm{formed}}/\rm{M}_{\odot})$, $\log_{10}$(SFR/($\rm{M}_{\odot}$ yr$^{-1}$)), $\tau$,  $t_i$, [$M$/H] and $A_{V}$; and (b) $\log_{10} (M_{*}/\rm{M}_{\odot})$, $\log_{10} (M_{*}^{\rm{formed}}/\rm{M}_{\odot})$, $\log_{10}$(SFR/($\rm{M}_{\odot}$ yr$^{-1}$)), $t_{25\%}$,  $t_{50\%}$,  $t_{75\%}$, [$M$/H] and $A_{V}$.

The results are organised as follows. In Sect.~\ref{test}, we analyse the performance of the model in the simulated test set. In Sect.~\ref{JADES_galaxies}, we describe our fit of the individual pixels from galaxies of the JADES survey. We  inspect the posterior distributions at different redshifts (Sect.~\ref{grid_posteriors}), comparing them with other SED fitting codes. Then, in Sect.~\ref{pixel-by-pixel}, we present the maps for stellar population properties of selected galaxies. Finally, in Sect.~\ref{outshining}, we describe our statistical study and compare the integrated stellar masses with their pixel-based analogues to investigate the outshining phenomenon.

\section{Recovering properties of the simulated stellar populations}
\label{test}

We trained the neural density estimator over $\sim 1$~hour and $30$~min. We obtained the posterior probability distributions for each stellar population parameter for the full test set, from the photometry with the associated uncertainties. It takes $\sim 0.04$\,s to obtain $500$ samples of the distributions for all the properties for one simulation. All of the time computations were performed on a single core of an Apple M3 Max CPU, part of an ARM-based 14-core system with 36 GB of unified memory. However, this model natively supports GPU execution, which can further reduce inference time.

To assess the accuracy of the model, we computed the medians of the posteriors inferred for the  parameters and compared them with the true simulated values for the full test sample with S/N averaged in the filters F277W, F356W, and F444W (hereafter, $\overline{\rm{S/N}}$) higher than 1. This ensures we did not show any simulations that are completely above of the observational capabilities of ACS and NIRCam, with upper limits for most of the filters. In Table~\ref{R2}, we include the $R^{2}$ values\footnote{Also known as coefficient of determination $\displaystyle R^2=1-\frac{\sum_{i=1}^n\left(\theta_i-\widehat{\theta}_i\right)^2}{\sum_{i=1}^n\left(\theta_i-\overline{\theta}_i\right)^2}$, not to be confused with the square of the Pearson's coefficient.} for the training using parametric (a) and non-parametric (b) SFHs. The performance is relatively good for the stellar mass and the dust attenuation index, but worse for the SFR, metallicity, and ages as a result of the degeneracies that we know exist being reproduced when inferring these properties from photometry. As expected,  the $R^2$ values for the SFR and [$M$/H] are higher for parametric SFHs. However, while $t_i$  can be well constrained, the accuracy of the medians of the posterior distribution for $\tau$ are extremely low. This can be explained not only because it is difficult to retrieve from photometry, but also because it is very degenerated with $t_i$ \citep{Conroy_2013}. Instead, we find that by combining $\tau$ and $t_i$ samples to obtain mass-weighted ages, we are able to constrain better the age of the galaxy, with $R^2$ scores of $0.94$.

\begin{figure*}[h]
    \centering
    \includegraphics[width=0.3\linewidth]{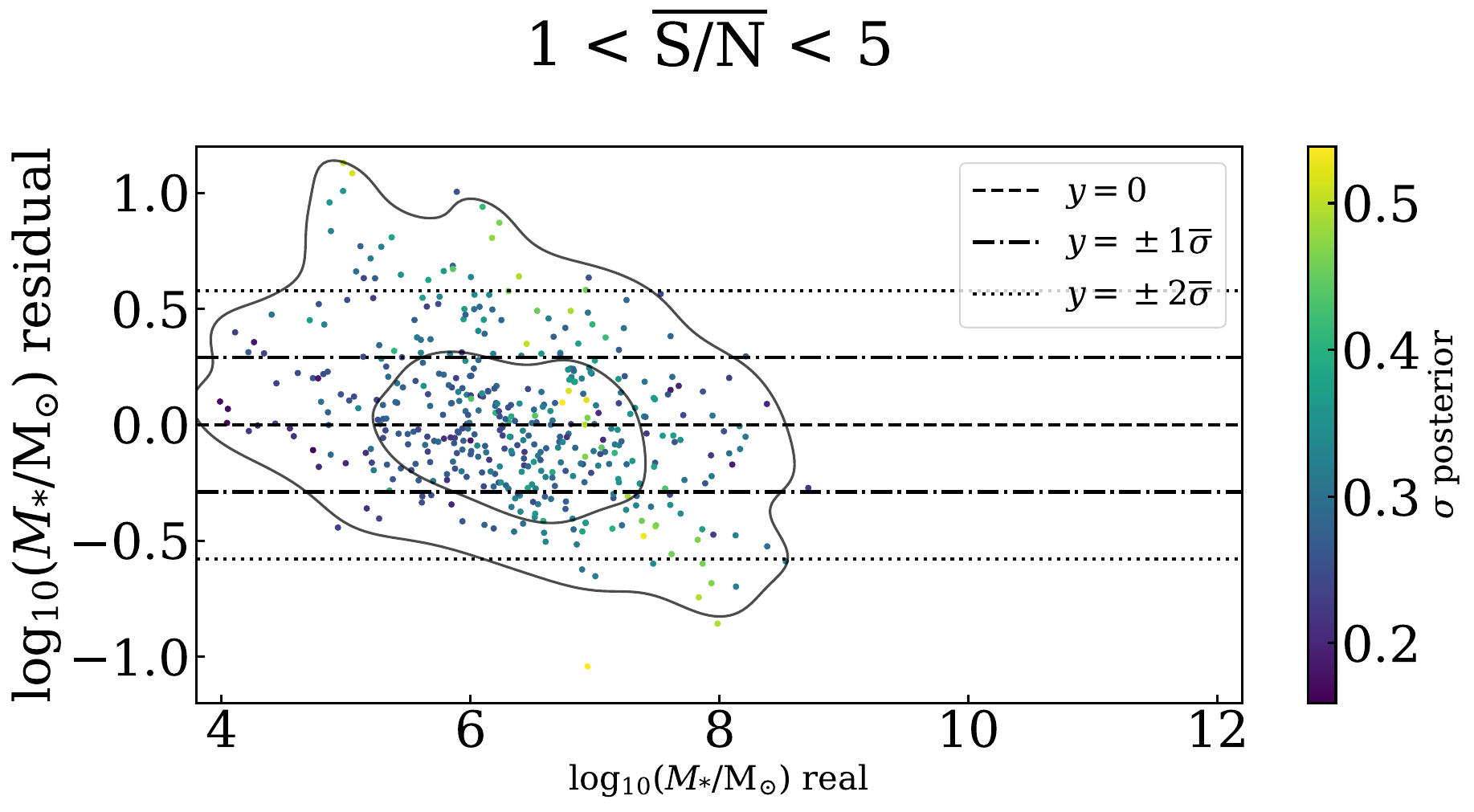}
    \includegraphics[width=0.3\linewidth]{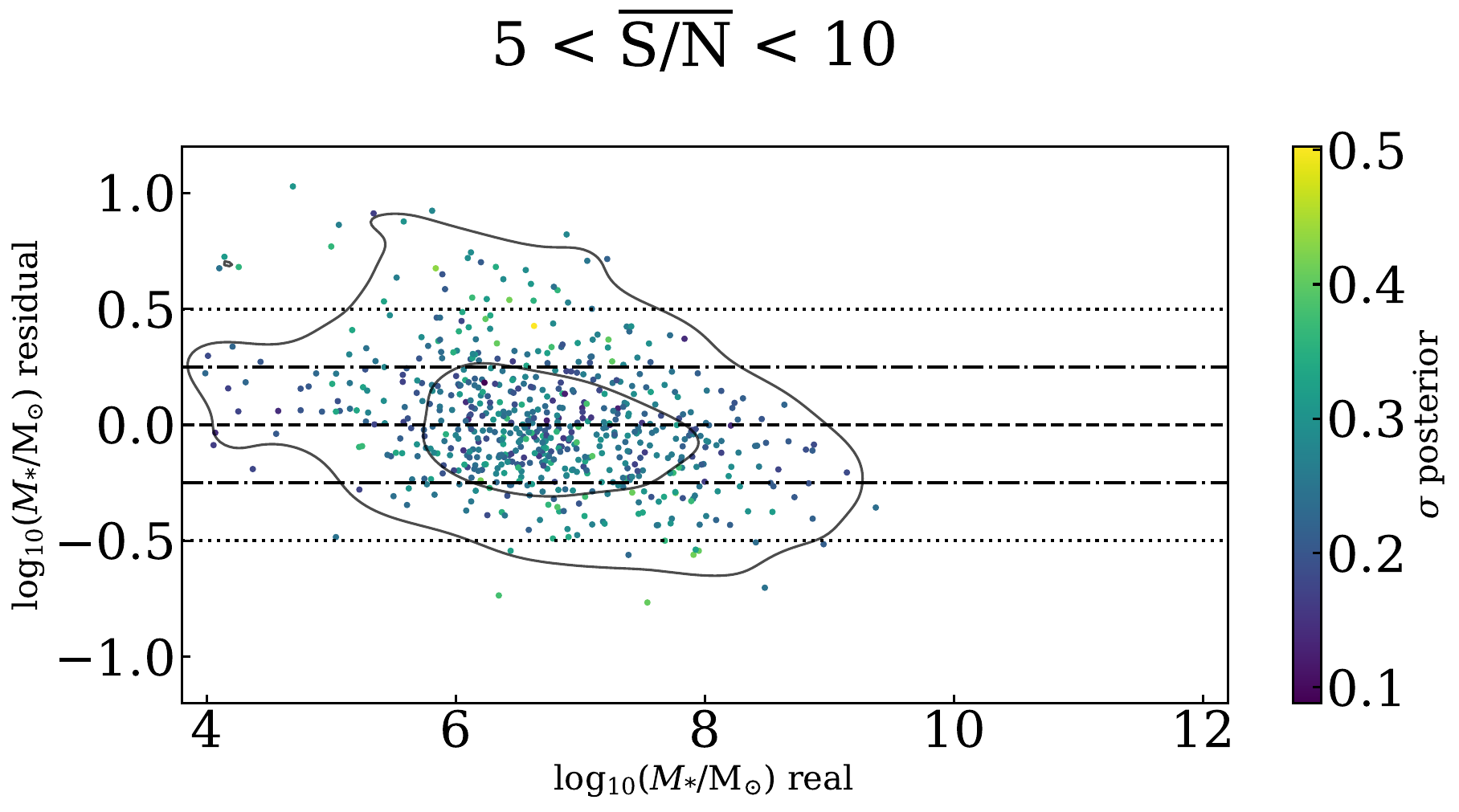}
    \includegraphics[width=0.3\linewidth]{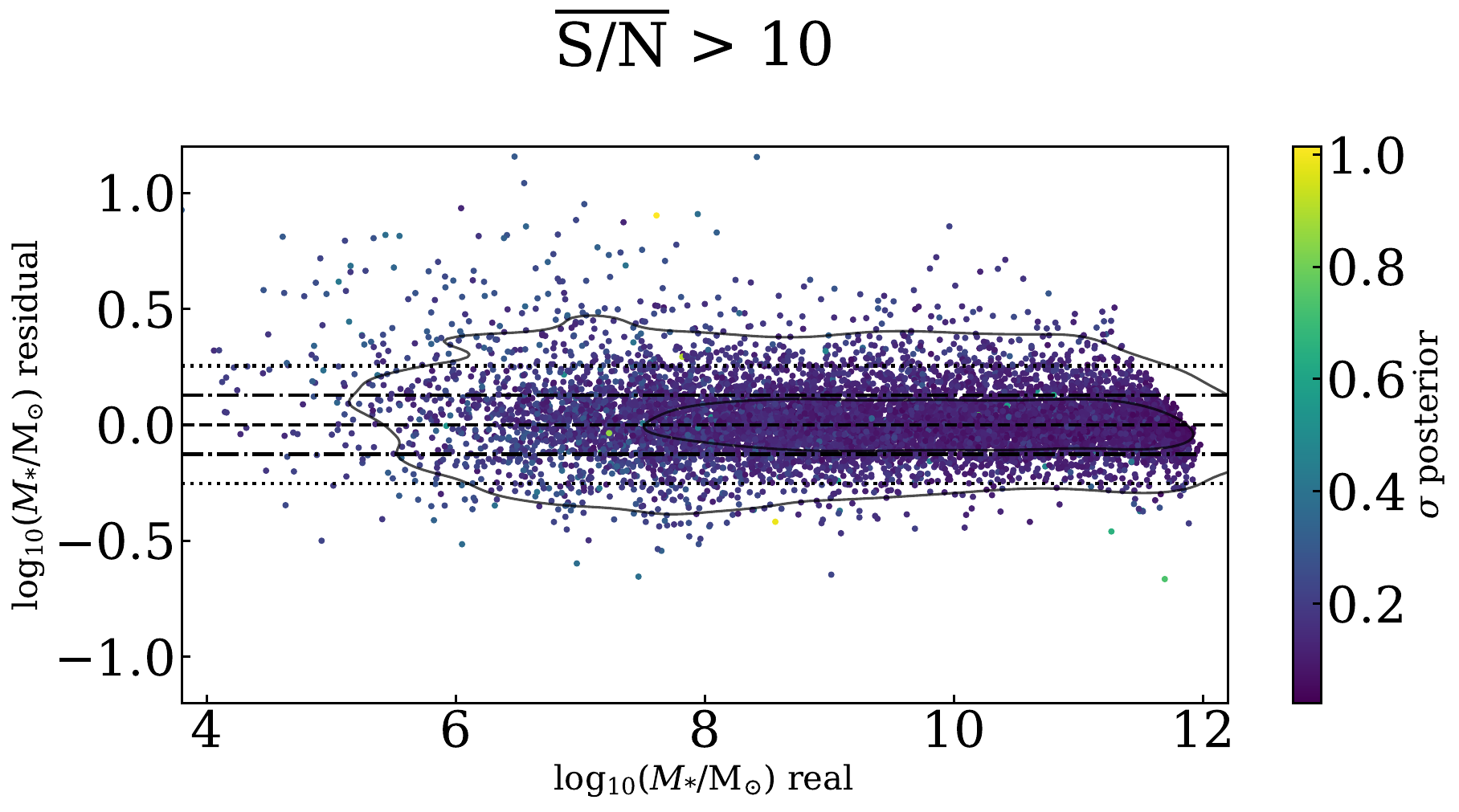}
    \includegraphics[width=0.3\linewidth]{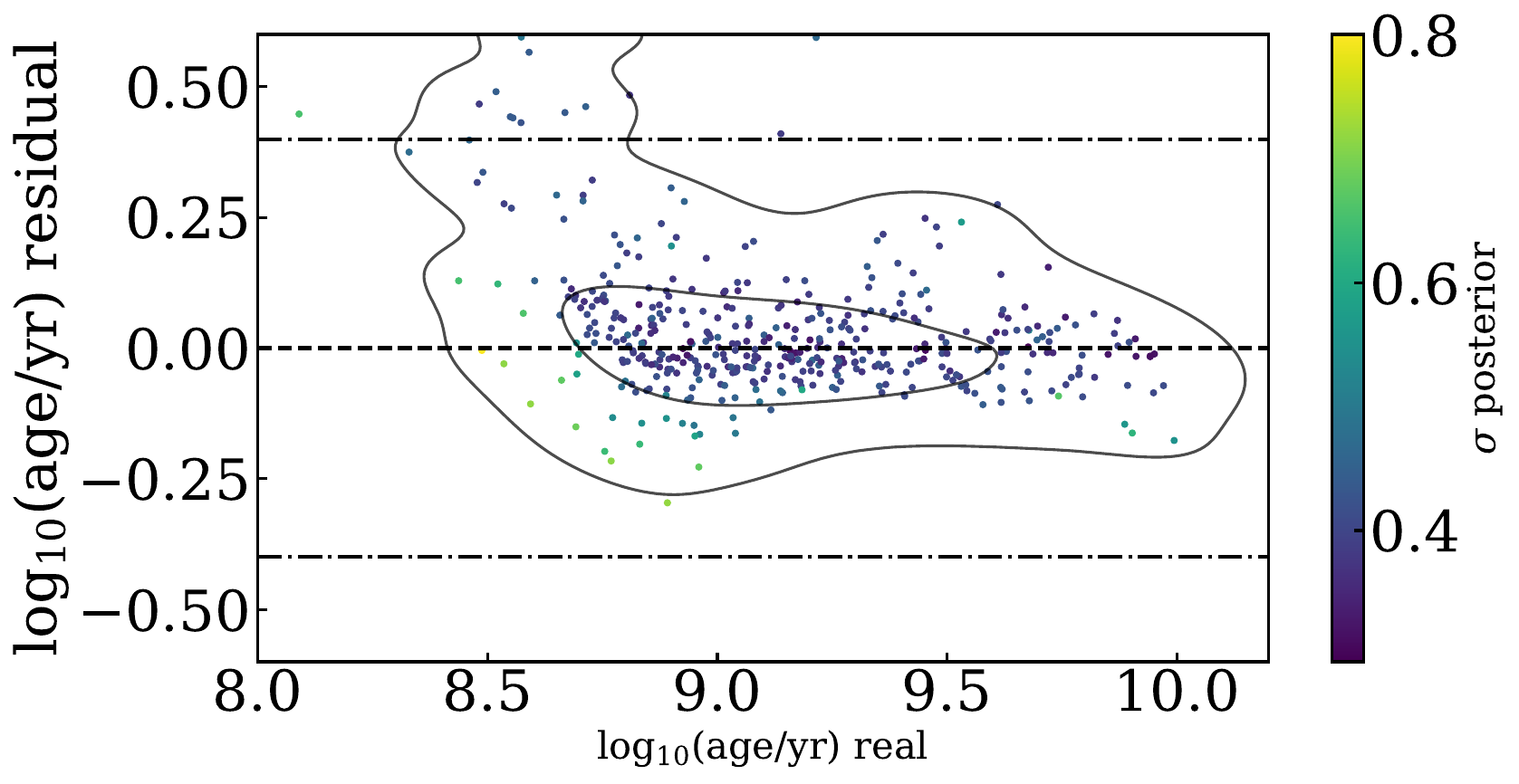}
    \includegraphics[width=0.3\linewidth]{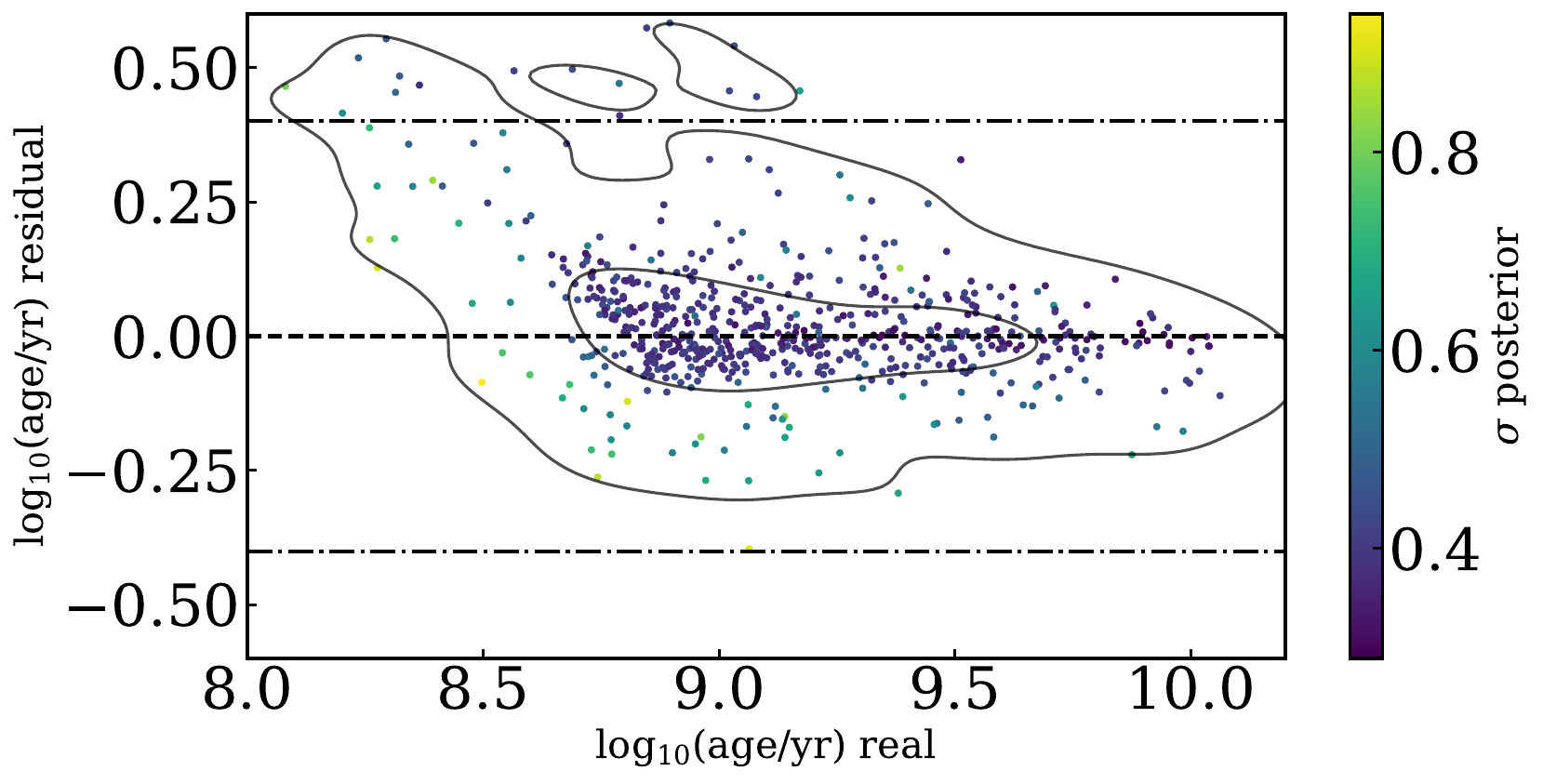}
    \includegraphics[width=0.3\linewidth]{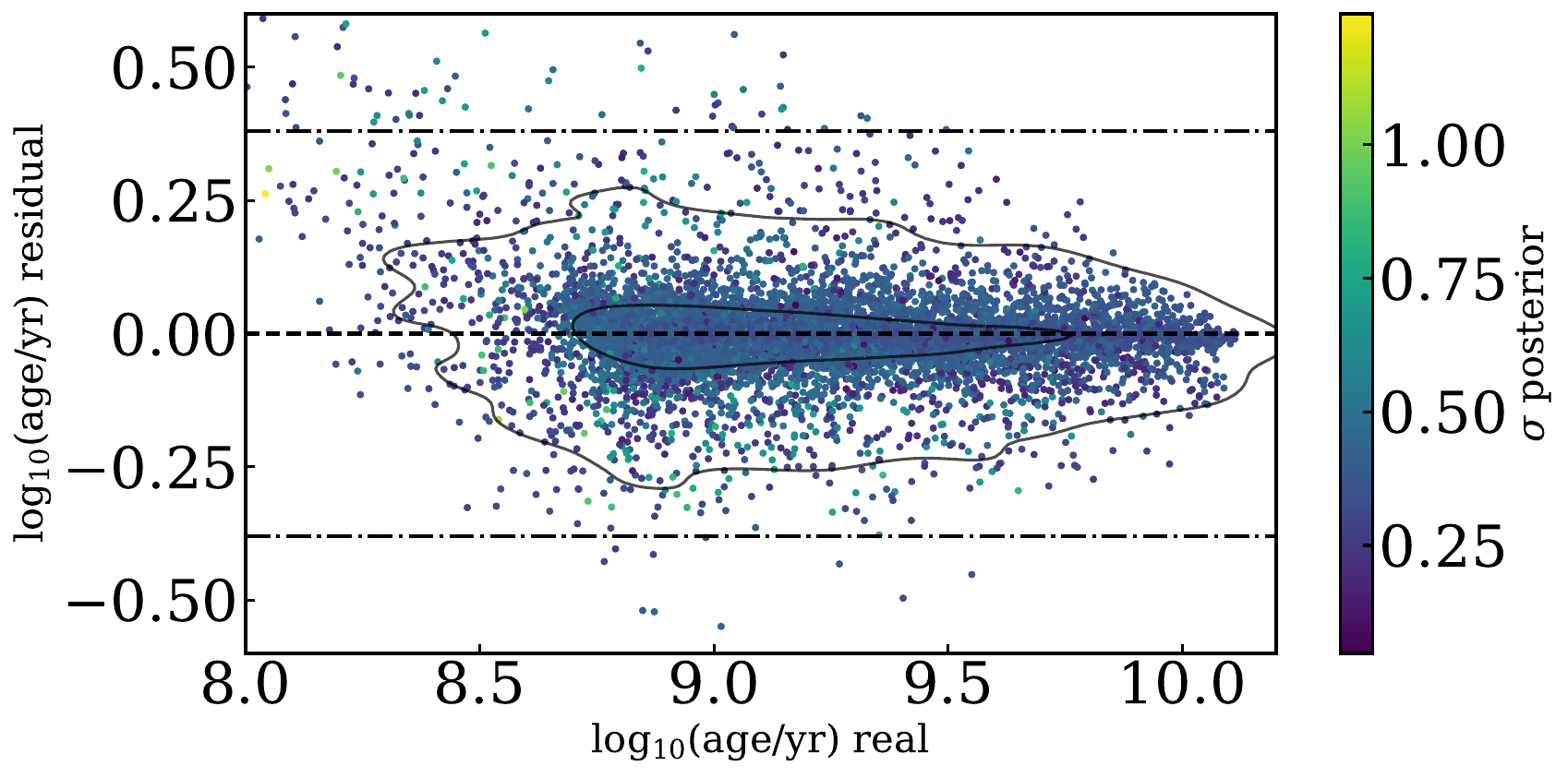}
    \includegraphics[width=0.3\linewidth]{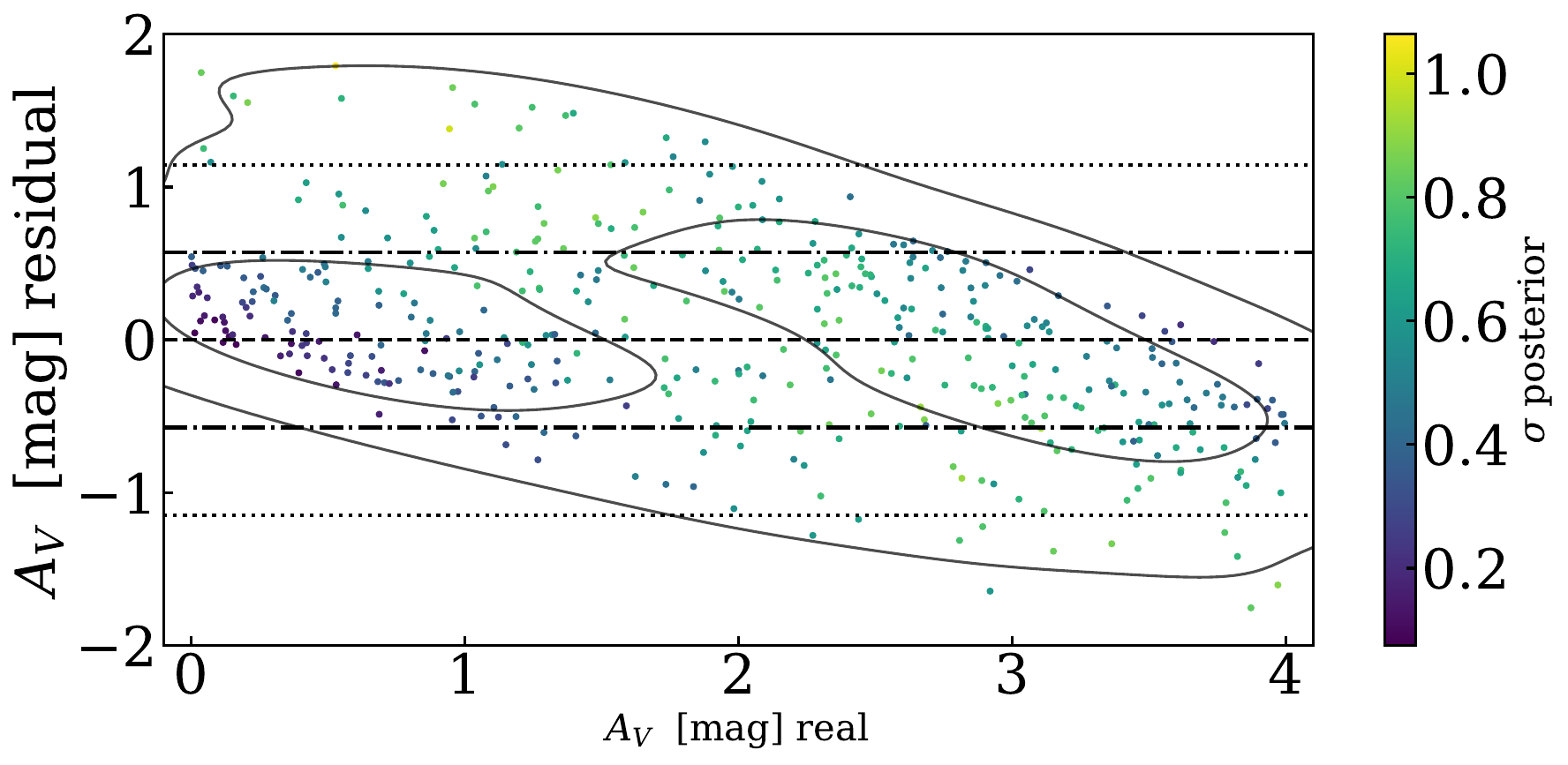}
    \includegraphics[width=0.3\linewidth]{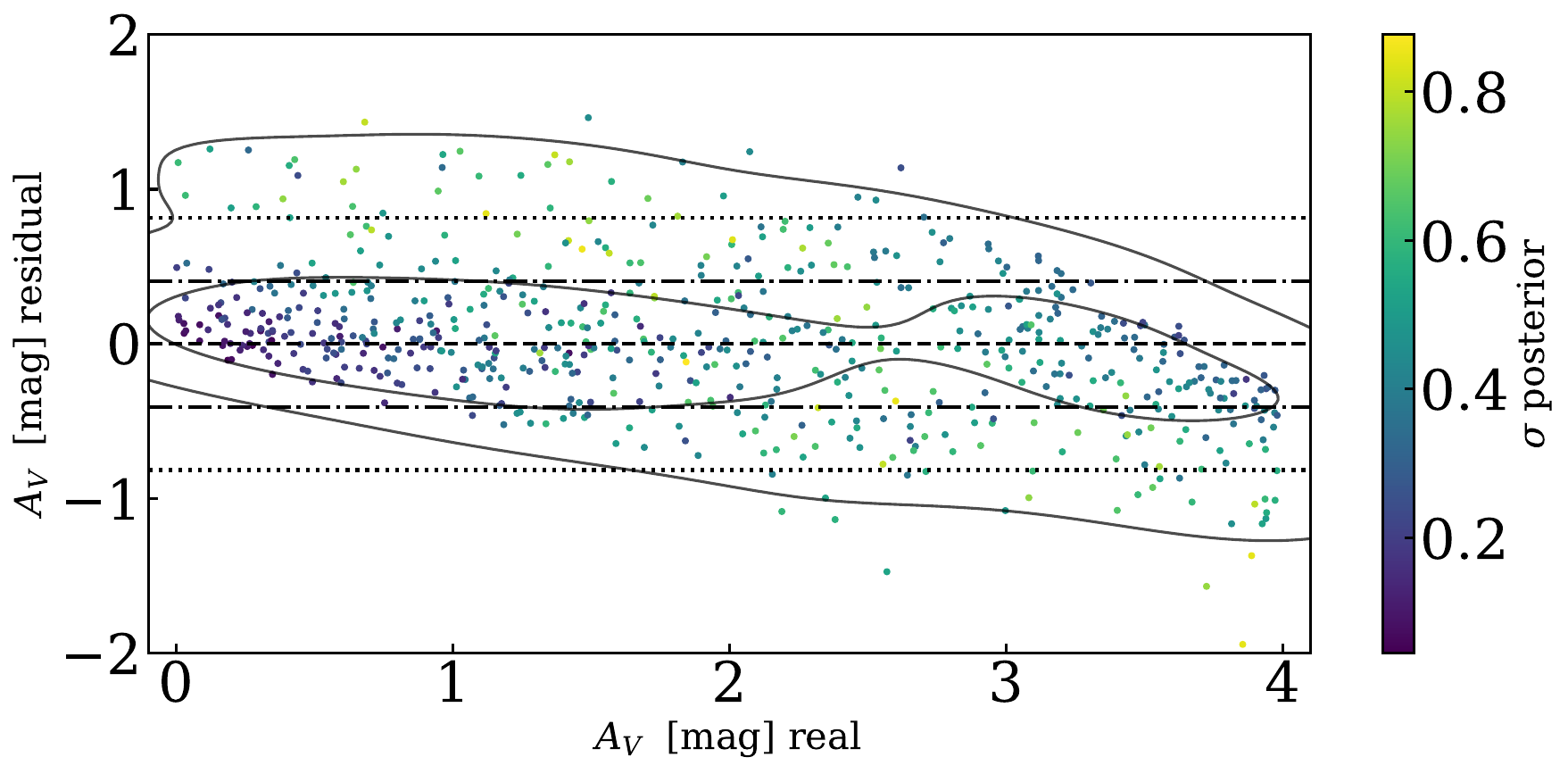}
    \includegraphics[width=0.3\linewidth]
    {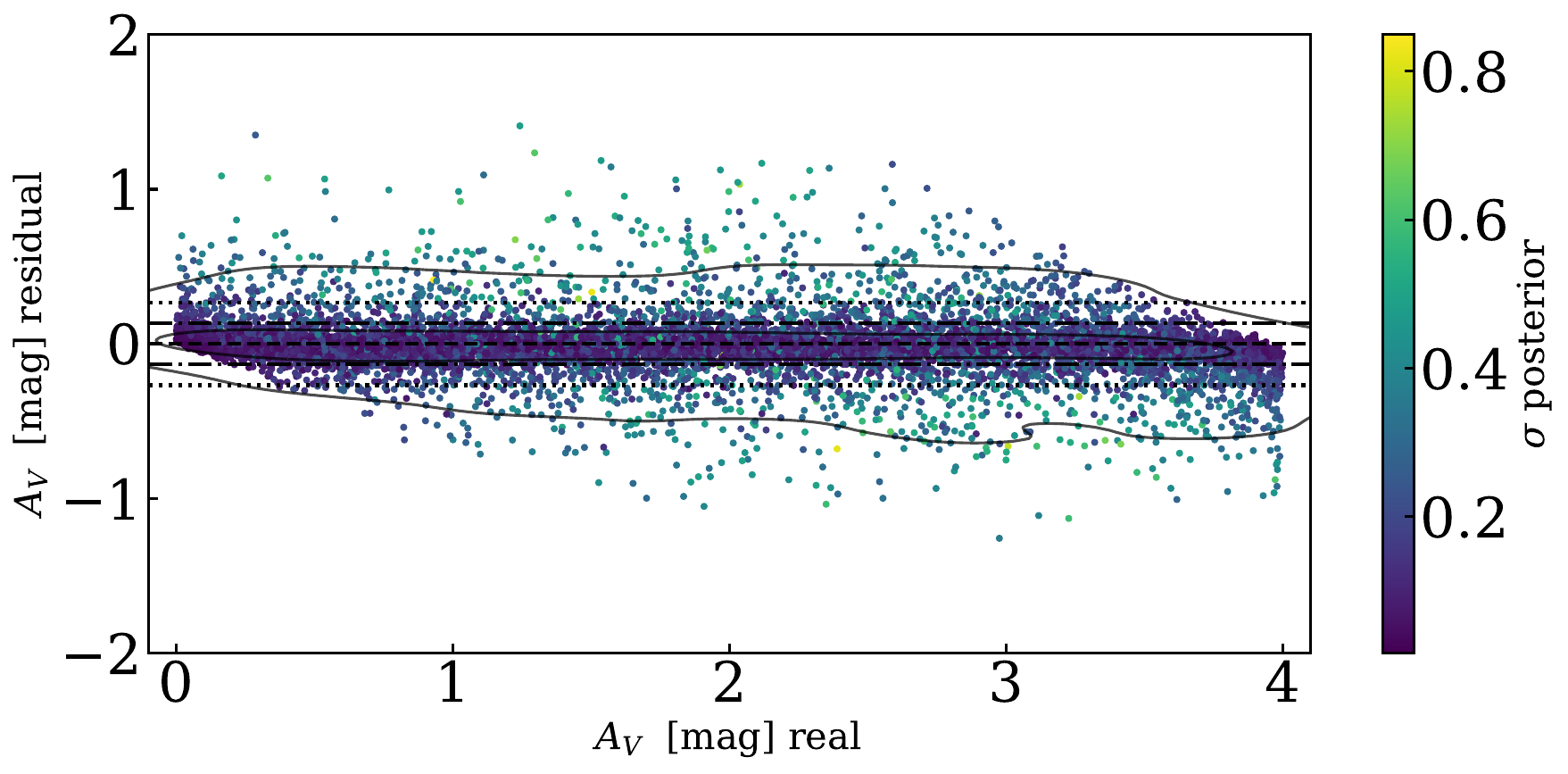}
    \caption{Residuals of properties compared to the medians of the posterior distributions obtained with the $\tau$-delayed model, for $\log_{10} (M_{*}/\rm{M}_{\odot})$, the $\log_{10}(\rm{age}/{yr})$, referring to the mass-weighted age, and $A_V$ [mag]. We split the simulated test sample in bins of mean S/N in the filters F277W, F356W, and F444W, 1-5 (left), 5-10 (middle) and $>10$ (right). We colour-coded each simulation with the standard deviation of the posterior distribution for the three properties and included dashed and dotted lines corresponding to the average one and two standard deviations of the posterior distributions respectively. We also plotted kernel density distribution contours with black solid lines for clarity.}
    \label{true_vs_pred}
\end{figure*}

We show in Fig.~\ref{true_vs_pred} the residuals (true property:\ median posterior distribution) for the test sample as a function of $\log_{10} (M_{*}/\rm{M}_{\odot})$, the mass-weighted age, and $A_V$, for the $\tau$-delayed prior. We colour-coded the samples with the standard deviation of the posterior distributions for each of the property, showing that simulations with higher residuals have also higher uncertainty in the posterior distributions. We split the data into bins of $\overline{\rm{S/N}}$, including $1<\overline{\rm{S/N}}<5$, $5<\overline{\rm{S/N}}<10$, and $\overline{\rm{S/N}}>10$. Although there are many more simulations in the last bin, we can see how the performance improves considerably for higher $\overline{\rm{S/N}}$, especially for the mass and dust attenuation. For these three properties, we find a good agreement between the inferred and simulated properties without significant systematics, except for a mild apparent trend in $A_V$. The latter reflects the influence of the prior shape under low S/N conditions, with the posteriors clustering around the midpoint of the uniform prior range.

\begin{figure*}[ht]
    \centering
    \includegraphics[width=0.3\linewidth]{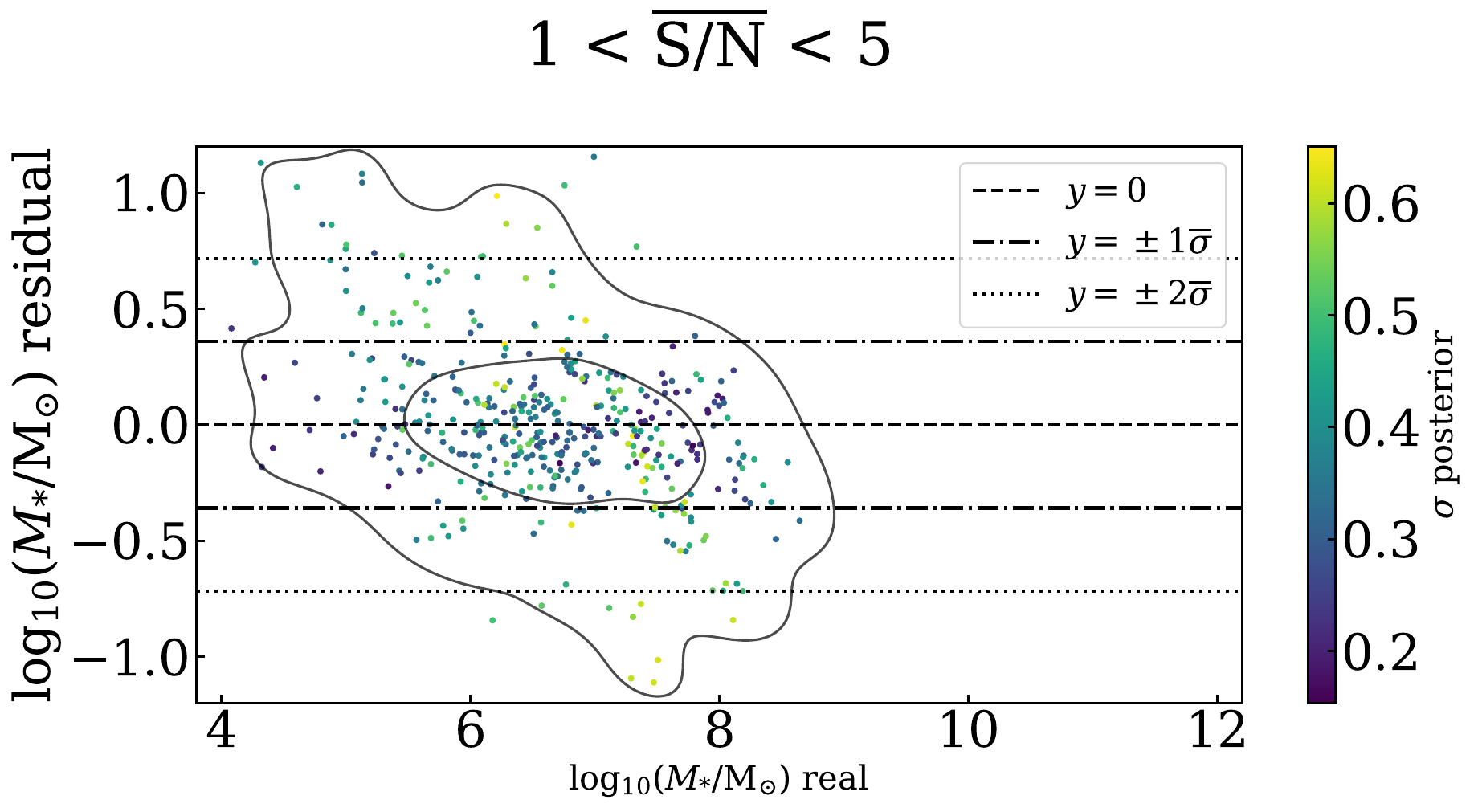}
    \includegraphics[width=0.3\linewidth]{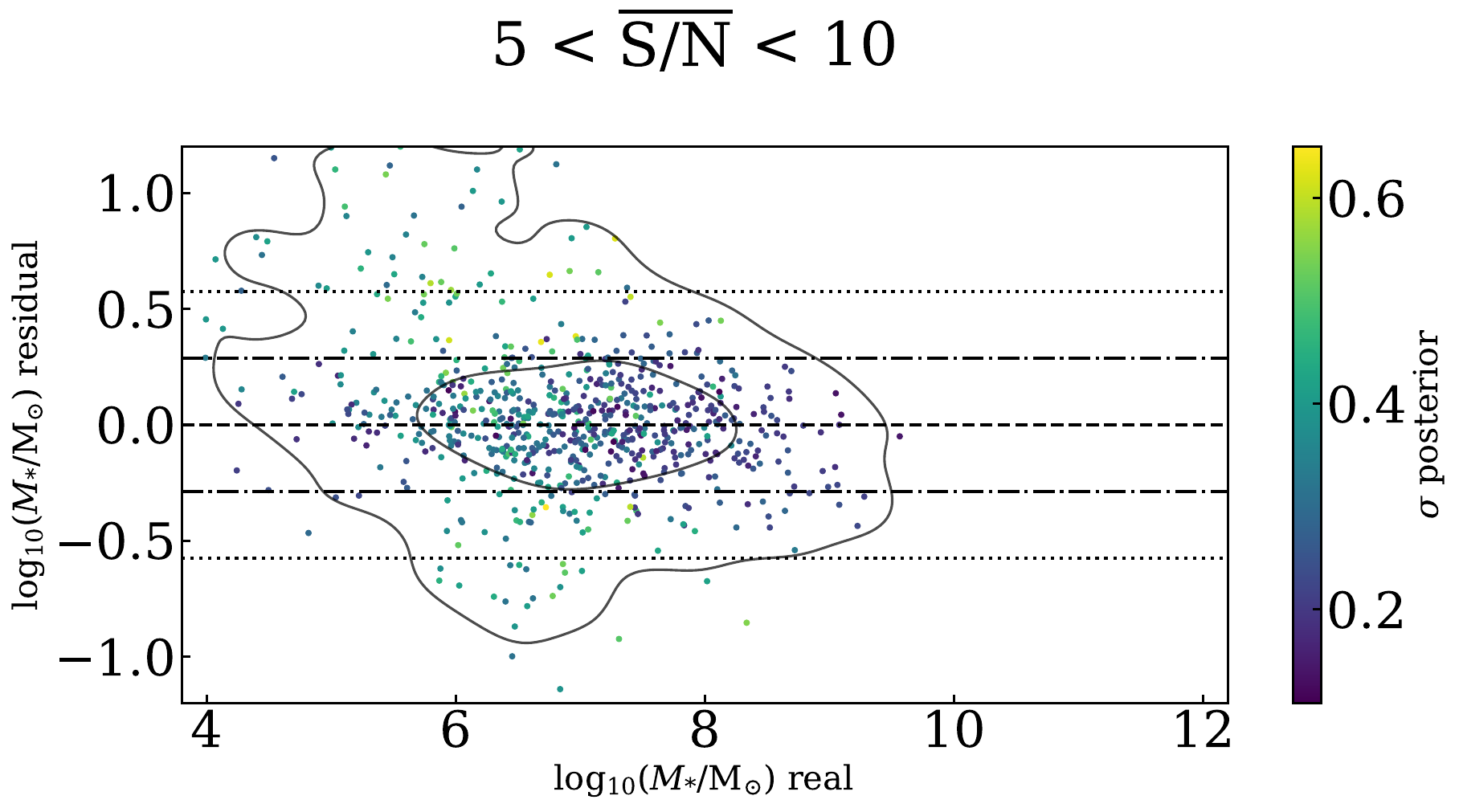}
    \includegraphics[width=0.3\linewidth]{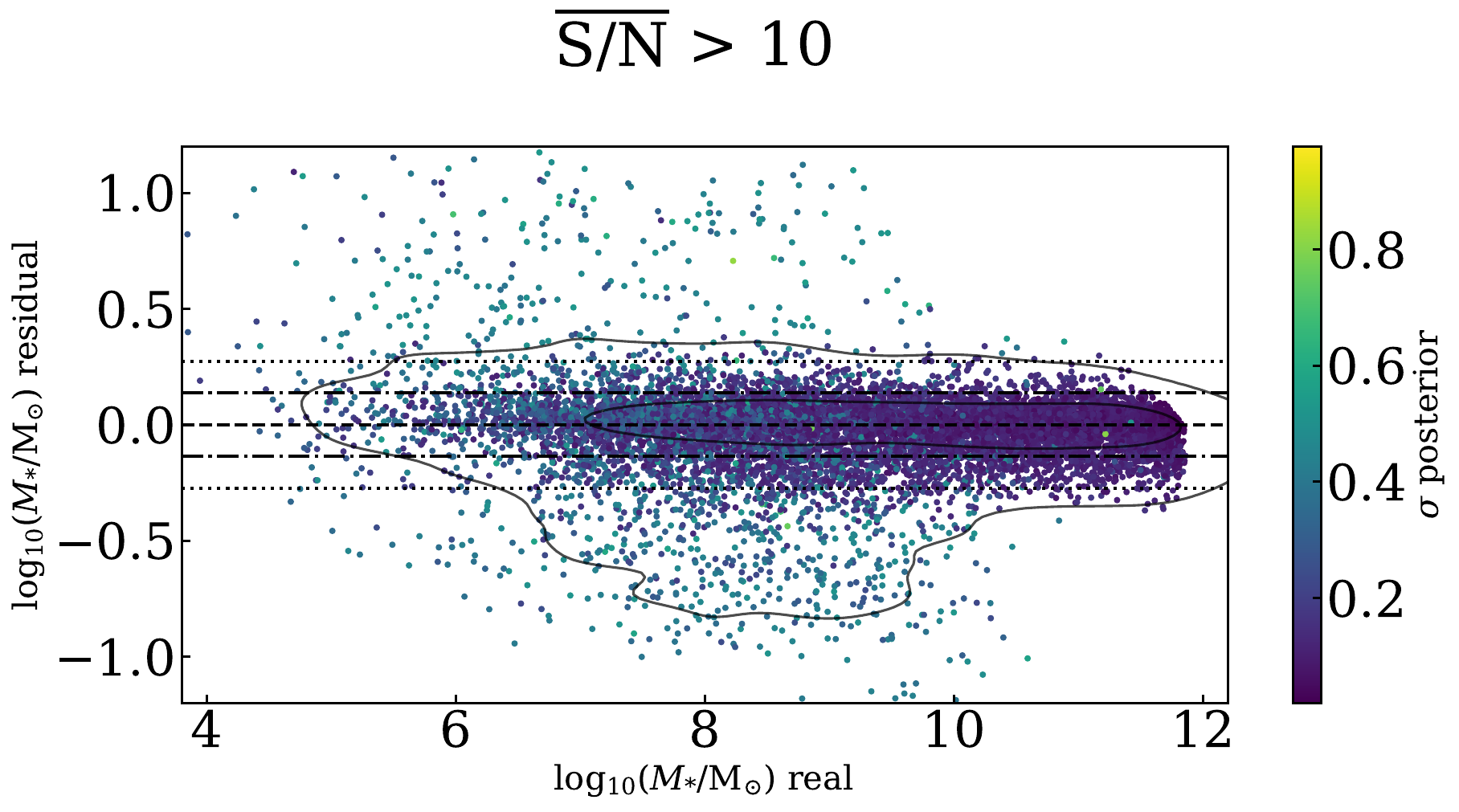}
    \includegraphics[width=0.3\linewidth]{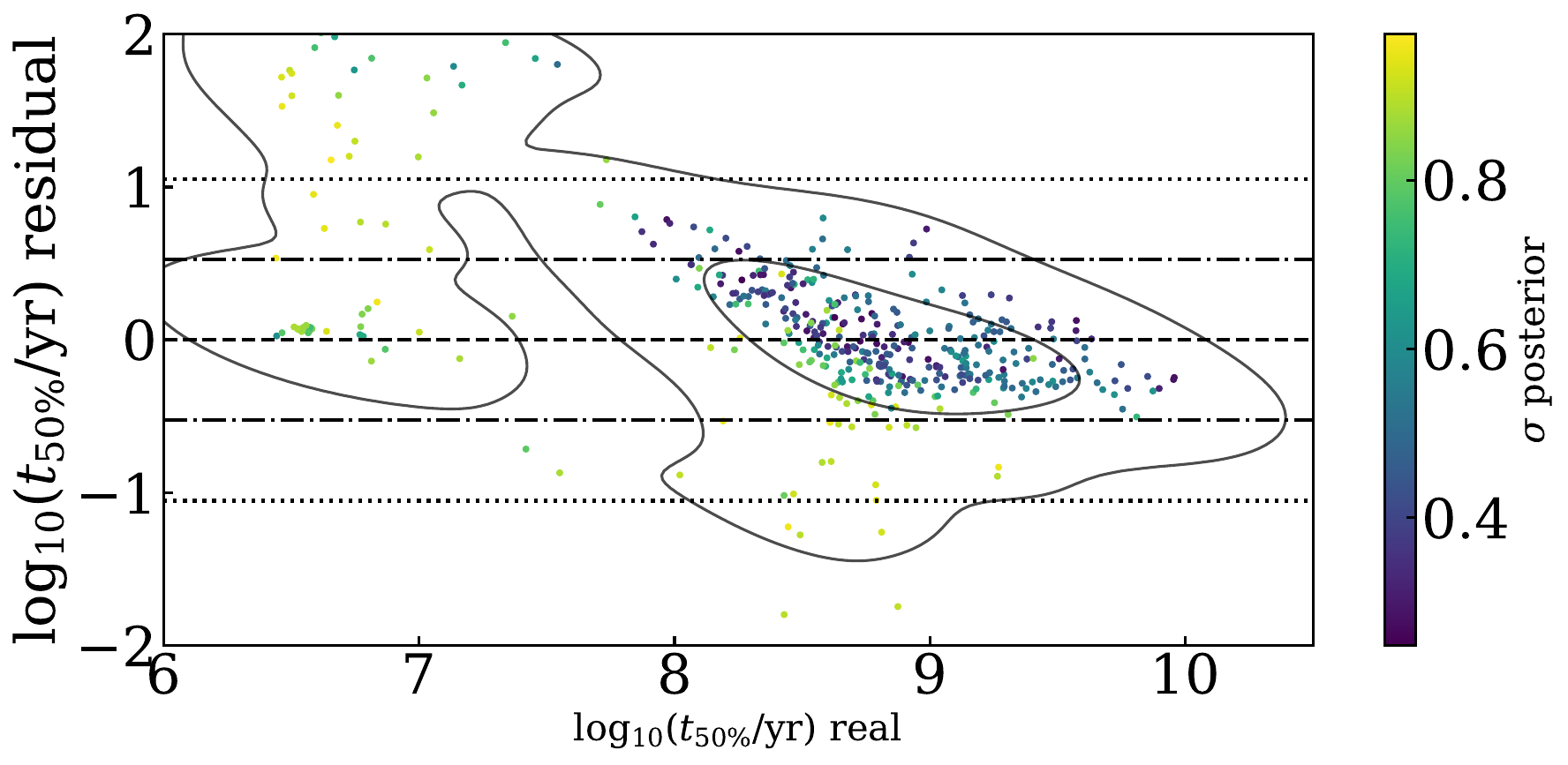}
    \includegraphics[width=0.3\linewidth]{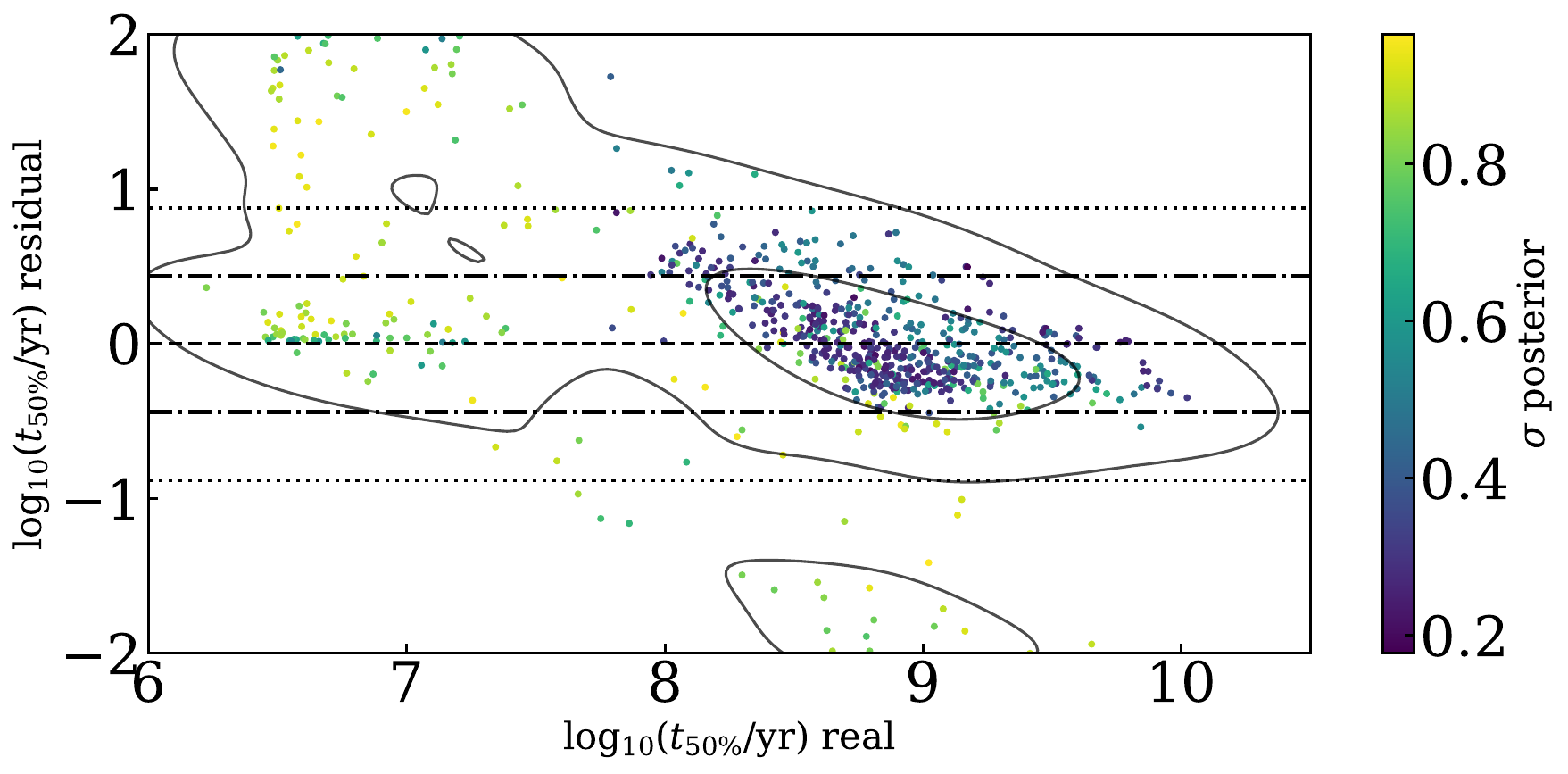}
    \includegraphics[width=0.3\linewidth]{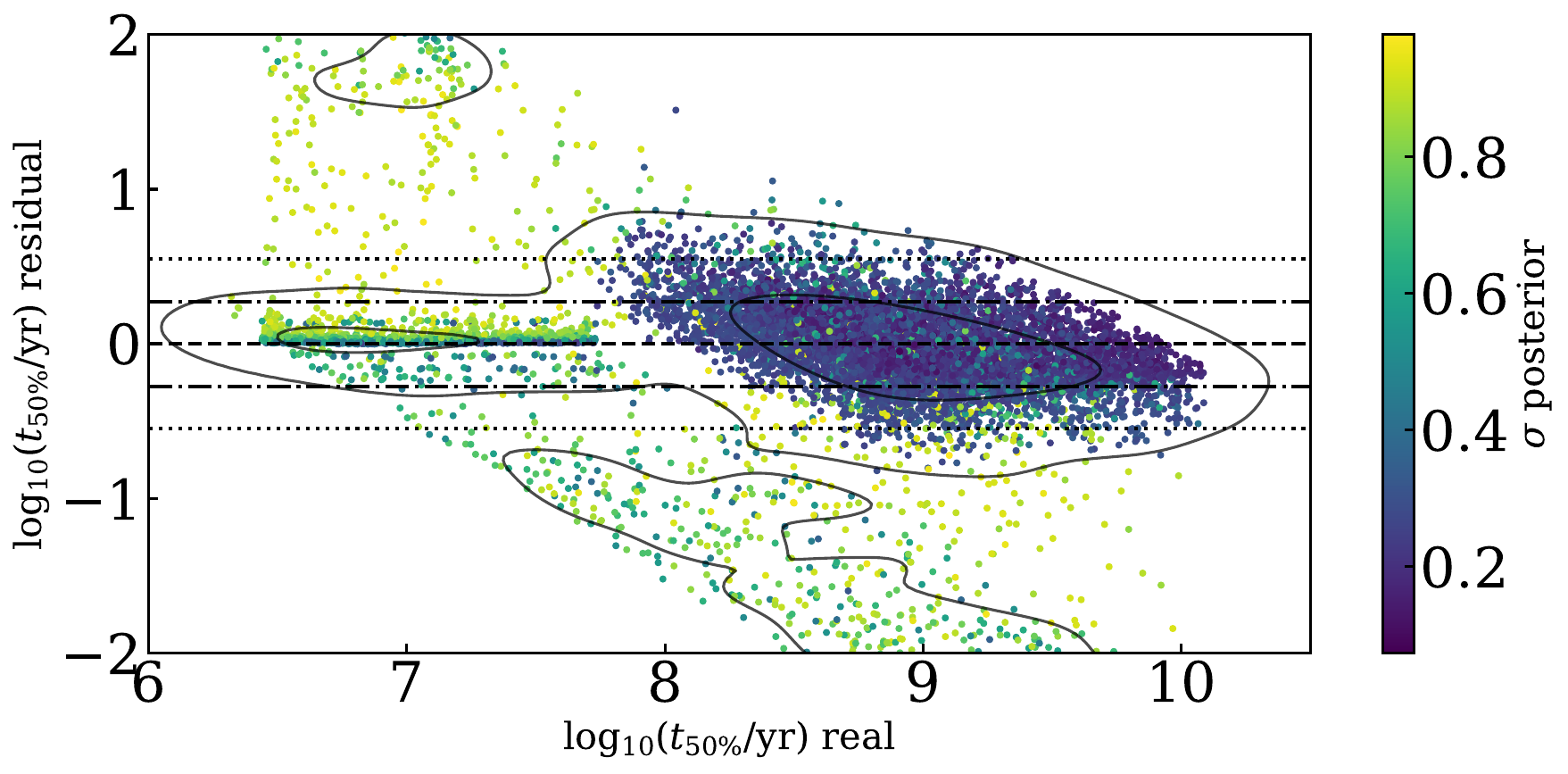}
    \includegraphics[width=0.3\linewidth]{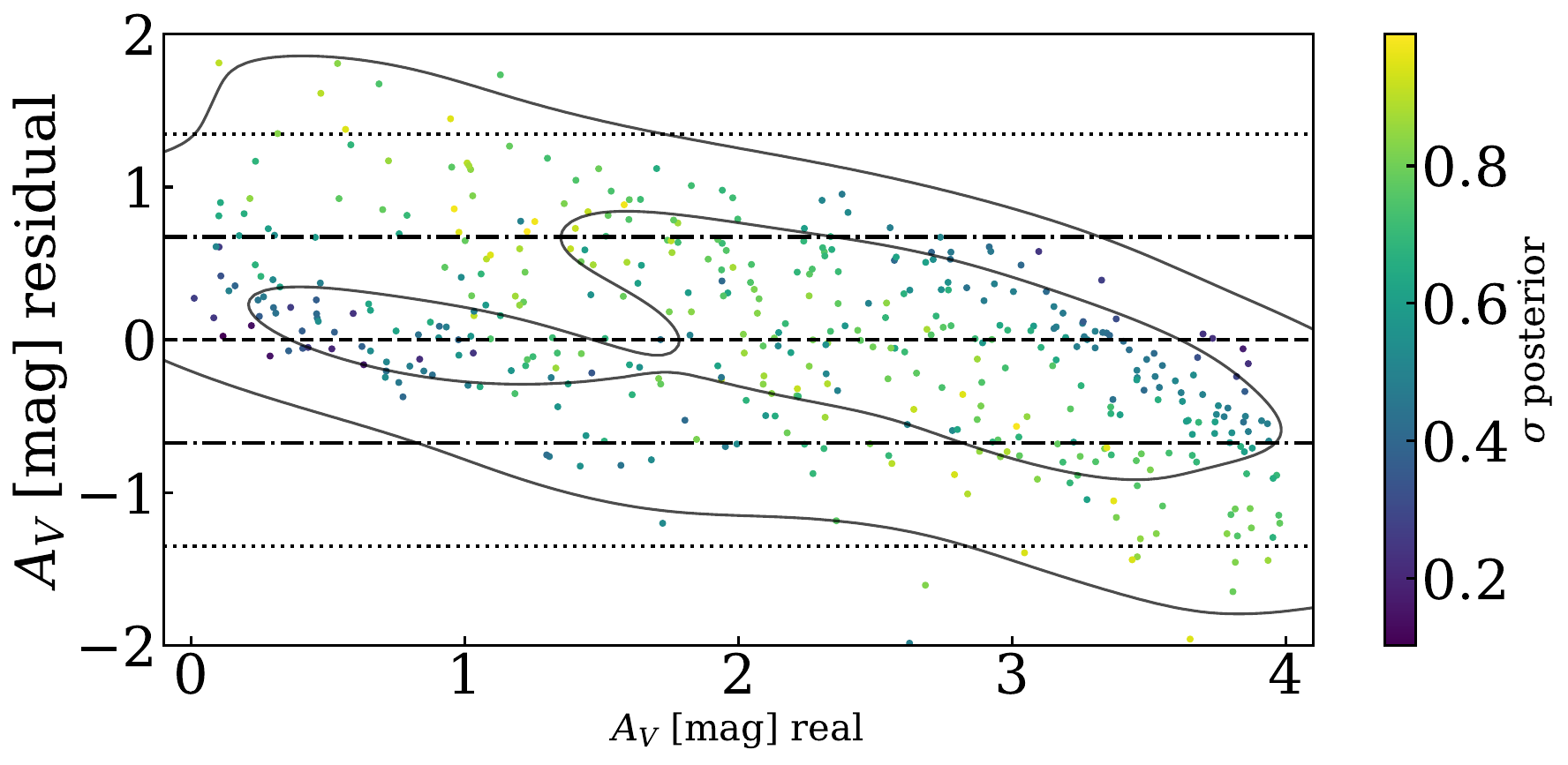}
    \includegraphics[width=0.3\linewidth]{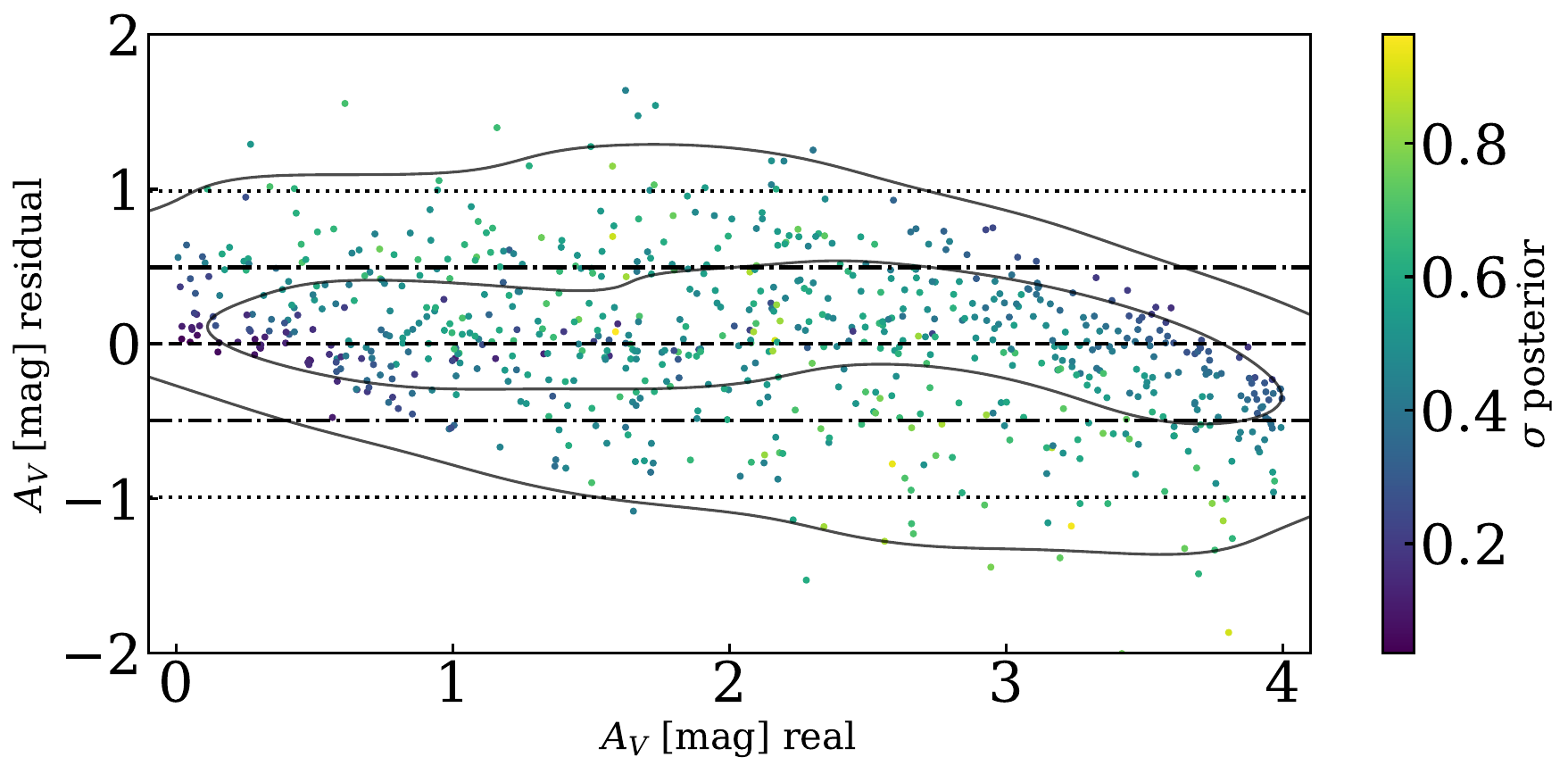}
    \includegraphics[width=0.3\linewidth]
    {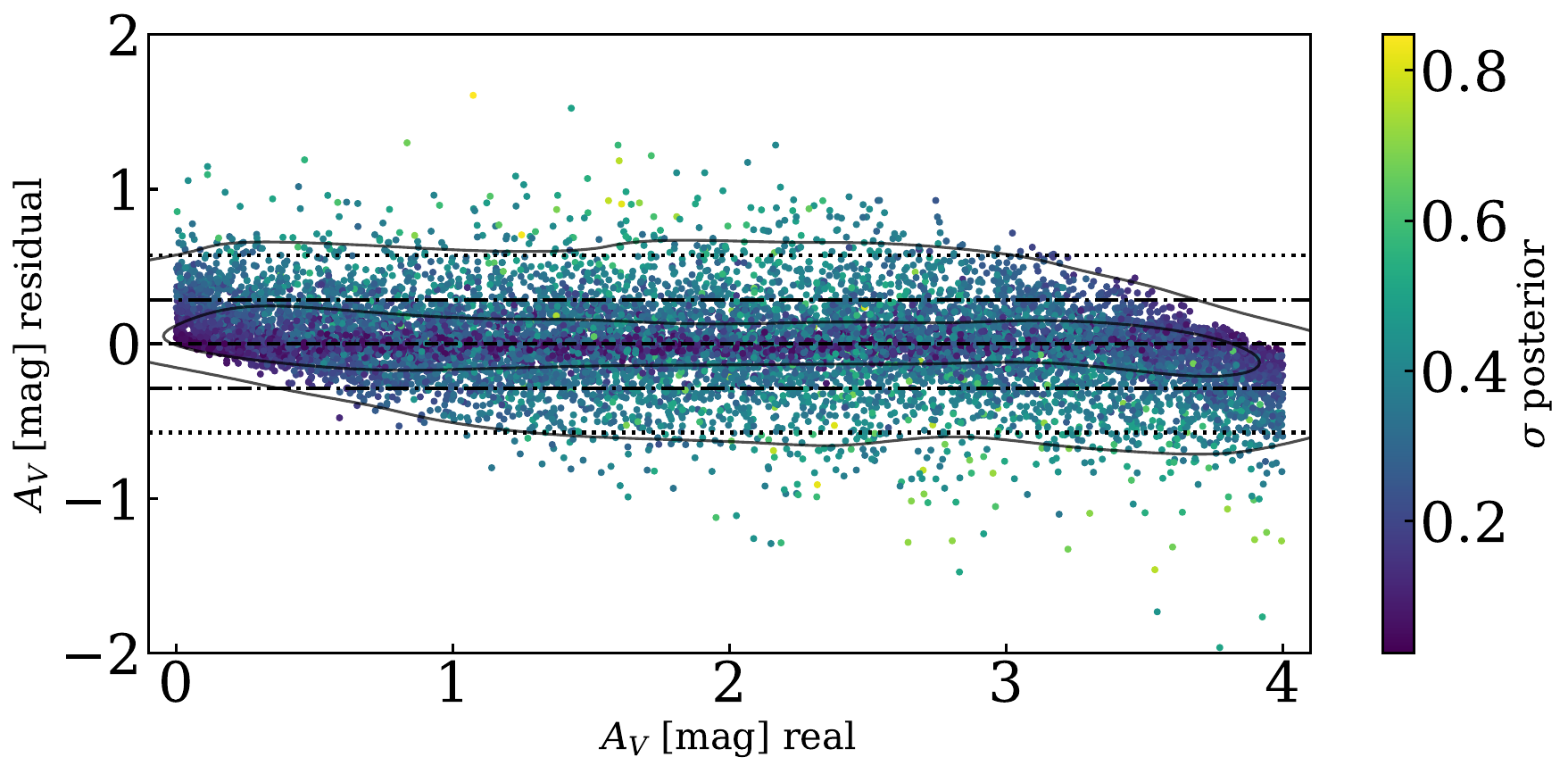}
    \caption{Same as Fig~\ref{true_vs_pred} for the Dirichlet model. We show $\log_{10} (M_{*}/\rm{M}_{\odot})$, log$_{10}(t_{50\%})$, referring to the lookback time in Gyr at which 50\% of the total stellar mass was formed, and $A_V$ [mag].}
    \label{plot_residual_dirichlet}
\end{figure*}

We repeat the analysis in Fig.~\ref{plot_residual_dirichlet} for the model trained with the Dirichlet prior, finding similar results, but with slightly higher scatter for $\log_{10} (M_{*}/\rm{M}_{\odot})$ and  $A_V$. Our simulations also include very young populations with $\log _{10}\left(t_{50 \%} / \mathrm{yr}\right)<8$. These cases correspond to SFHs with very recent bursts, in the last $\sim 100$~Myr of lookback time; furthermore, the brightness from young and hot stars prevents us from making a proper estimation of their previous SFHs, leading to higher residuals and standard deviations in the posteriors for $\log _{10}\left(t_{50 \%} / \mathrm{yr}\right)<8$. 

The galaxies exhibiting larger scatter typically show degeneracies between stellar mass, age, and dust attenuation: older, dustier stellar populations can resemble younger, but less massive, less extinguished ones in their observed SEDs. Conversely, very young, dust-free populations can mimic older and dustier systems with higher stellar masses. This degeneracy often leads to bimodal posterior distributions, where the data cannot distinguish between two distinct combinations of physical parameters. Instead of collapsing this ambiguity into a single best-fit estimate (or the median, as done in Fig.~\ref{plot_residual_dirichlet}) that is not representative, our model returns the full posterior. This captures the complete range of plausible solutions, including bimodal ones, allowing us to directly characterise the associated uncertainties

We analyse in Appendix~\ref{app_corner} the posterior distributions obtained for the same $\tau$-delayed simulation using the models trained on different priors. We find the expected correlations between parameters and the widths of the posteriors for the different properties (compared to the widths of the prior distributions) match the $R^2$ scores discussed previously. We also performed a simulation-based calibration (SBC) test in Appendix~\ref{sbc} to verify that the NFs are well trained and, thus, the posteriors are correctly calibrated.

\begin{table}
    \centering
    \caption{$R^2$ values for the medians of the posteriors using the $\tau$-delayed prior and the Dirichlet prior, obtained for the full test set with $\overline{\rm{S/N}}>1$, with respect to the simulated values.}
    \footnotesize
\begin{tabular}{cc}
    \hline
    Properties & $R^2$ value ($\overline{\rm{S/N}}>1$) \\ \hline
    \multicolumn{2}{l}{\textbf{(a) $\tau$-delayed prior}} \\ \hline
    $\log_{10}(M_{*}/\rm{M}_{\odot})$         & 0.99 \\ 
    $\log_{10}$(SFR/($\rm{M}_{\odot}$ yr$^{-1}$)) & 0.67 \\ 
    $\tau$~[Gyr]                          & 0.07 \\ 
    $t_i$~[Gyr]                           & 0.77 \\ 
    $[M/$H$]$                             & 0.60 \\ 
    $A_V$~[mag]                           & 0.95 \\ \hline
    \multicolumn{2}{l}{\textbf{(b) Dirichlet prior}} \\ \hline
    $\log_{10}(M_{*}/\rm{M}_{\odot})$         & 0.98 \\ 
    $\log_{10}$(SFR/($\rm{M}_{\odot}$ yr$^{-1}$)) & 0.54 \\ 
    $t_{25\%}$                            & 0.49 \\ 
    $t_{50\%}$                            & 0.40 \\ 
    $t_{75\%}$                            & 0.40 \\ 
    $[M/$H$]$                             & 0.39 \\ 
    $A_V$~[mag]                           & 0.92 \\ \hline
\end{tabular}
\label{R2}
\end{table}

\section{Inferring properties of JWST galaxies}
\label{JADES_galaxies}
\subsection{Sample selection}

We selected $1083$ galaxies of the HUDF measured in the 19 filters listed in Table~\ref{filters} and with spectroscopic redshifts. We first examined, across stellar masses and redshifts, the fraction of pixels within two effective radii with a sufficient S/N in order to be fitted with our model. We extracted these properties from a catalogue that includes stellar masses from surviving stars into a Kron aperture and redshifts from \cite{dahlen2010,guo12,Oesch2023}, and \cite{bunker24}, assuming SFHs described by delayed exponentials, with timescales between 1~Myr and 10~Gyr, free metallicity, ages between~1 Myr and the age of the Universe at the redshift of each source, the Calzetti attenuation law, and a Chabrier IMF.

\begin{figure}[h]
    \centering
    \includegraphics[width=\linewidth]{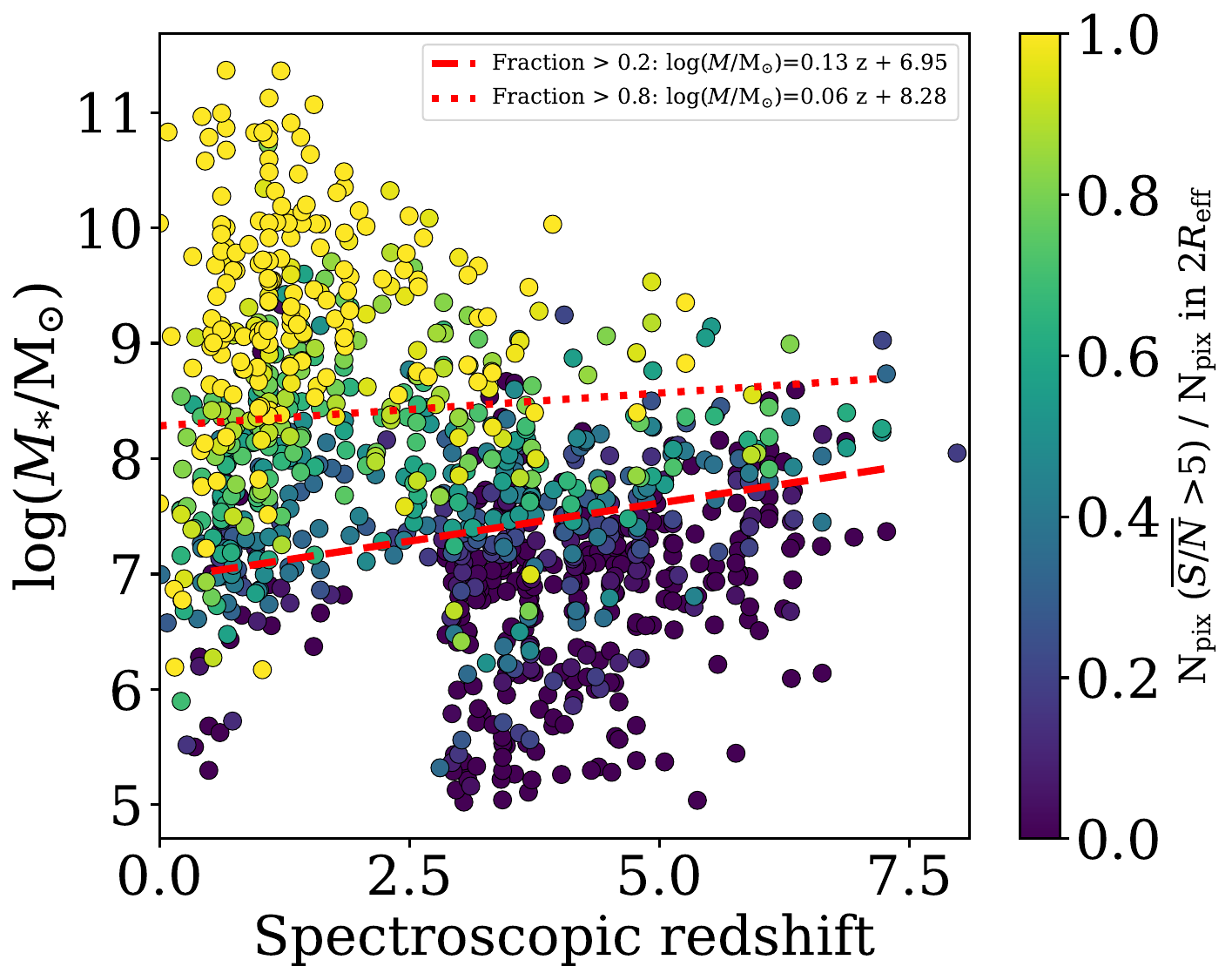}
    \caption{Distribution of $1.083$ galaxies from the JADES survey in the plane of integrated stellar mass versus spectroscopic redshift, with colours indicating the fraction of pixels within two effective radii (in F444W) that have a mean S/N of 5 or greater in the filters F277W, F356W, and F444W. The lines represent the parameter-space limits used to set thresholds at pixel fractions of 0.20 (red dashed line) and 0.80 (red dotted line) averaged across the full sample. These thresholds indicate, for the ranges of integrated masses and redshifts, the fraction of pixels within the galaxies that can be effectively fitted with our model.}
    \label{z_m_reff}
\end{figure}

Again, instead of using the mean or maximum S/N across all available filters, an approach susceptible to dropout effects due to the galaxy redshift, we focus on the mean S/N in the three reddest wide-band filters: F277W, F356W, and F444W. Specifically, we analysed the fraction of pixels with a mean S/N in these three bands equal to or exceeding 5. This threshold, initially selected based on model performance on simulated data, will be further validated with observational data in the following section. Fig.~\ref{z_m_reff} shows that up to redshifts of \( z \approx 5-6 \) most galaxies with total stellar masses above \( 10^8 \) M\(_\odot\) contain a substantial fraction of pixels above the S/N threshold, allowing them to be analysed effectively with our model.

We fit all the pixels of the $1083$ galaxies with the trained model and obtained posterior distributions with $500$ samples for each of them.  We carried out a statistical analysis of these posteriors according to the S/N of the pixels and the redshift in Appendix~\ref{sigma_sn}. By comparing the posterior and prior widths, we found a strong broadening of the posteriors for most of the properties at $\overline{\rm{S/N}} < 5$, consistent with the previous results in the synthetic observations. This showcases the regime where our inference is prior-dominated, namely, where the observational data provide limited constraining power and the resulting posterior distribution closely follows the form of the adopted prior. Moreover, the constraining power for each of the posteriors, measured as the ratio $\sigma_{\rm{posterior}} / \sigma_{\rm{prior}}$, agrees with the $R^2$ scores given in Table~\ref{R2}.

\subsection{Analysis of sample galaxies and comparison with other codes}
\label{grid_posteriors}

We selected six galaxies from the sample to show the performance of the model across different morphologies, orientations, and redshifts.  In Table~\ref{galaxies_table}, we include their JADES ID, right ascension, declination, spectroscopic redshift, and number of pixels in the segmentation map of the JADES survey with $\overline{\rm{S/N}}>5$.

\begin{table}
    \centering
    %\small % Reduce font size
    \setlength{\tabcolsep}{2pt} % Reduce column spacing
    \renewcommand{\arraystretch}{0.9} % Reduce row spacing
    \caption{JADES IDs, RA, Dec, spectroscopic redshifts, and number of pixels with $\overline{\rm{S/N}}>5$ in the filters F277W, F356W, and F444W for the six example galaxies analysed with the model.}
    \label{galaxies_table}
    \begin{tabular}{ccccc}
        \hline
        ID  & RA [$^\circ$]  & dec [$^\circ$]  & $z_{\rm{spec}}$  & $N_{\rm{pix}}(\overline{\rm{S/N}} > 5)$ \\ \hline
        254985  & 53.169961  & -27.771119  & 0.63  & 25208 \\ 
        205449  & 53.166186  & -27.787576  & 1.08  & 11069  \\ 
        211273  & 53.165590  & -27.769882  & 1.55  & 8233  \\ 
        117960  & 53.144489  & -27.790426  & 2.44  & 388   \\ 
        118081  & 53.178183  & -27.790264  & 3.30  & 147    \\ 
        206146  & 53.146669  & -27.786211  & 5.52  & 201   \\ \hline
    \end{tabular}
\end{table}

\begin{figure*}[h!]
\centering
    \includegraphics[width=0.99\linewidth]{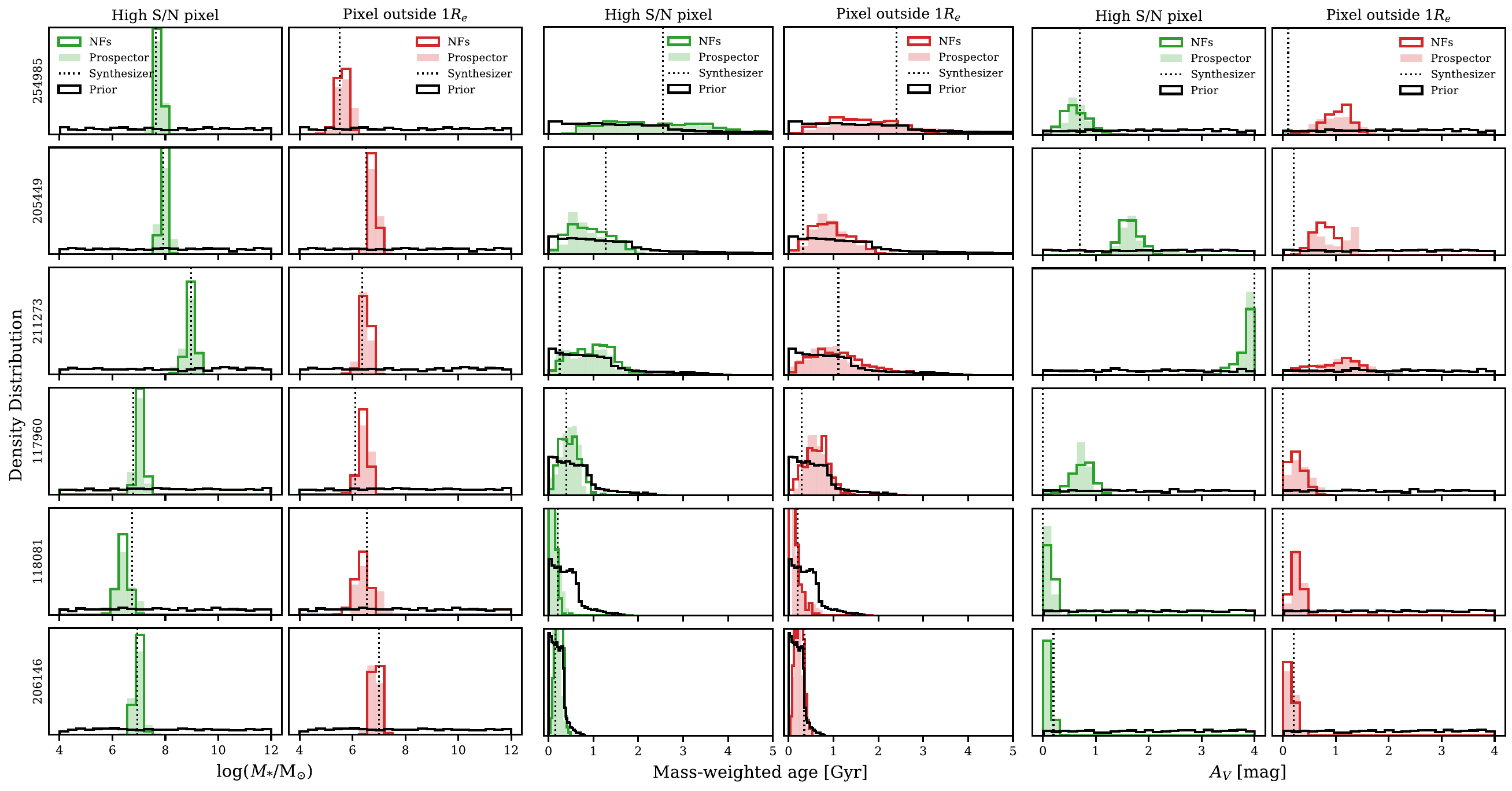}
    \caption{Probability distributions for $\log_{10}(M_{*}/\rm{M}_{\odot})$, the mass-weighted age [Gyr], and $A_{V}$ [mag] for the pixel with highest mean S/N in the filters F277W, F356W, and F444W (in green), and a random pixel out of $1R_{\rm{eff}}$ (in red) for the six galaxies. Step histograms (solid lines) represent the prior distributions (in black) and our posterior distributions (in green and red). The posterior distributions from \texttt{Prospector} are shown with lower opacity (in green and red). The dotted lines correspond to the minimum $\chi^2$ parameters obtained with the optimisation code \texttt{Synthesizer}.}
    \label{posteriors_six}
\end{figure*}

In Fig.~\ref{posteriors_six}, we show the posterior distributions we obtain for two pixels of each galaxy: one with the highest $\overline{\rm{S/N}}$ and other selected randomly beyond $1R_{\rm{eff}}$, together with the priors, for the model using the $\tau$-delayed prior. We show the surviving stellar mass and the dust attenuation,  combining the samples of the posteriors of $\tau$ and $t_i$ to derive samples for the mass-weighted age which, as seen in the simulated test sample, is more robust. We omitted from this analysis the metallicity, for which we obtained a  low $R^2$ score and and wide posterior distributions in the simulations.

We compared our posterior distributions with those obtained with \texttt{Prospector} \citep{prospector}, which are also shown in Fig.~\ref{posteriors_six}. We ran \texttt{Prospector} with the same assumptions: MILES SSPs, Chabrier IMF and MIST isochrones, $\tau$-delayed SFHs and Calzetti dust attenuation, and the same dust emission, nebular continuum, and emission libraries. We also used the priors given in Table~\ref{priors_combined}. We optimised the inference with \texttt{Dynesty} \citep{dynesty} to reduce the inference time. Despite this optimisation, it takes $\sim 12$~min to fit the posterior distributions with ~$2500$ samples. We repeated the inference for the NFs with the same number of samples and it takes ~$0.20$\;s, which is $\sim3.5\times10^3$ times faster. This demonstrates the potential of the amortised inference for the pixel-based fitting of high-resolution multi-wavelength observations for large samples of galaxies.

We repeated the fit of all the pixels for the six galaxies included in Table~\ref{galaxies_table} with the optimisation code \texttt{Synthesizer} \citep{perez2003,perez2008}. This optimisation model employs the BC03 \cite{bc03} stellar population synthesis framework, utilising the Padova 1994 isochrones \citep{padova94} with high-resolution spectra. Both the spectral library and the isochrones differ from those used by our model.  However, it assumes the same Chabrier IMF and also incorporates a Calzetti attenuation law  and a delayed-$\tau$ model for the SFHs, with the same ranges of parameters than the included in our model, optimising the parameters to minimise the chi-square ($\chi^2$) for SED fitting. More information about the nebular emission and continuum models used by \texttt{Synthesizer} can be found in \cite{perez2003,perez2008}.  We also give the parameter estimations of \texttt{Synthesizer}  in Fig.~\ref{posteriors_six} for comparison.

Our posterior distributions are in perfect agreement with those from \texttt{Prospector}, but differ from \texttt{Synthesizer} estimations. The latter is a possible consequence of using a different SPS model. Although we do not have access to the full posterior distributions in \texttt{Synthesizer}, we further analysed how well the models fit the SEDs in the Appendix~\ref{ppc}. There, we performed a posterior predictive check \citep[PPC;][]{TALTS2018}, repeating the simulation for every sample of the posteriors for the same two pixels of each of the six galaxies using the parametric model.  We compared the results of the PPC with the actual observed photometry, as well as the SED fits from \texttt{Prospector} and \texttt{Synthesizer}, finding an overall good agreement.

\section{Pixel-by-pixel SED fitting of JWST galaxies}
\label{pixel-by-pixel}

\subsection{Maps of stellar population properties}
\begin{figure*}[h!]
    \centering
    \includegraphics[width=0.77\linewidth]{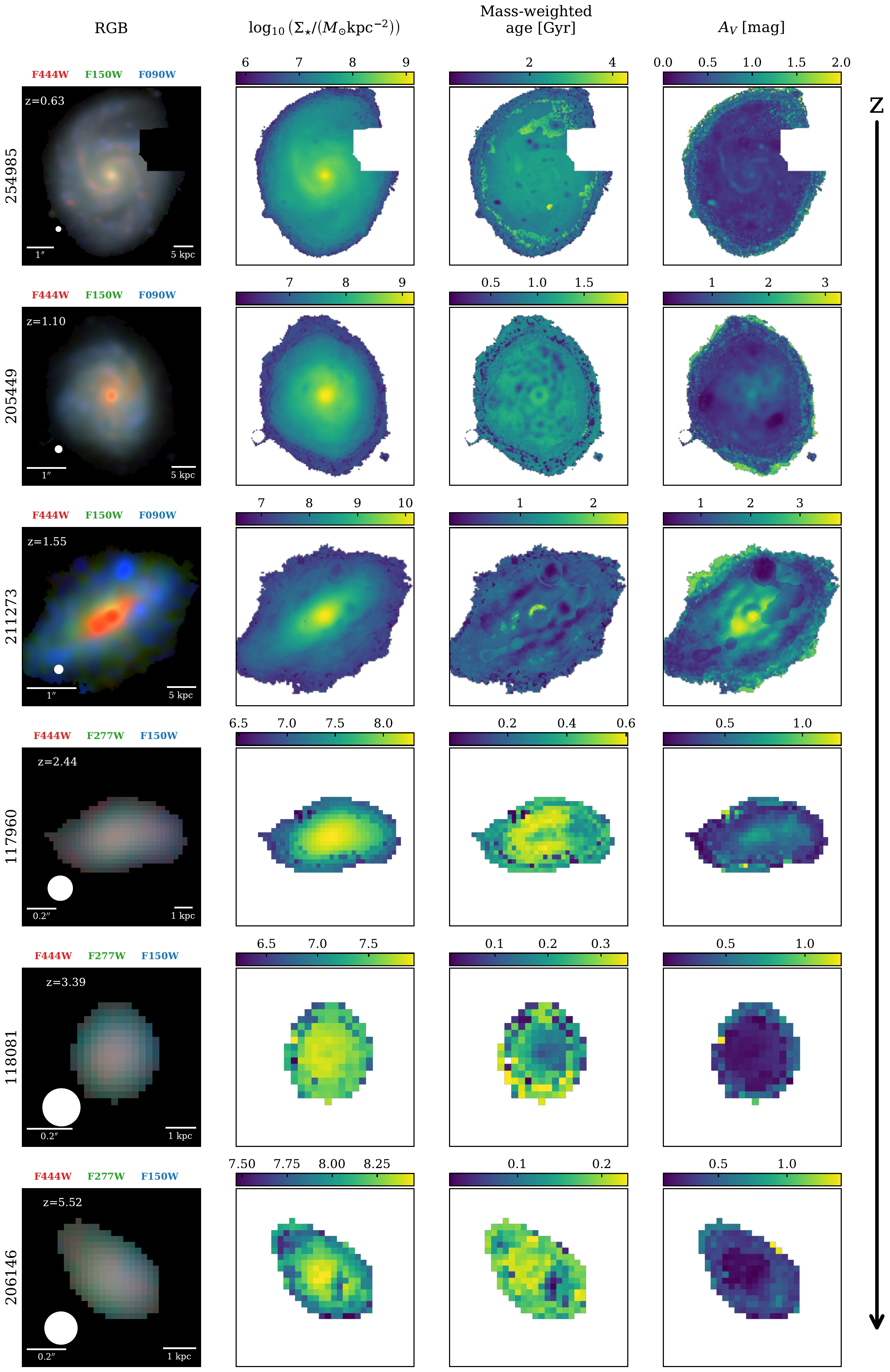}
    \caption{Galaxy property maps. We include only the pixels with a S/N (averaged between F277W, F356W, and F444W) higher than 5. We use a $\tau$-delayed prior for the SFHs. From left to right: RGB images (built using the method described in \cite{lupton04} with the filters specified above each image to enhance the different structures), and the inferred properties per pixel $\log _{10}\left(\Sigma_{\star} /\left(M_{\odot} \mathrm{kpc}^{-2}\right)\right)$, the mass-weighted age [Gyr], and $A_V$ [mag].  In the lower-left corner of the first column, we show circles with diameters equal to the FWHM of the F444W PSF. Scale bars in the first column indicate angular (arcseconds) and physical (kiloparsecs) sizes.}
    \label{maps_tau}
\end{figure*}

\begin{figure*}[h!]
\centering
    \includegraphics[width=0.77\linewidth]{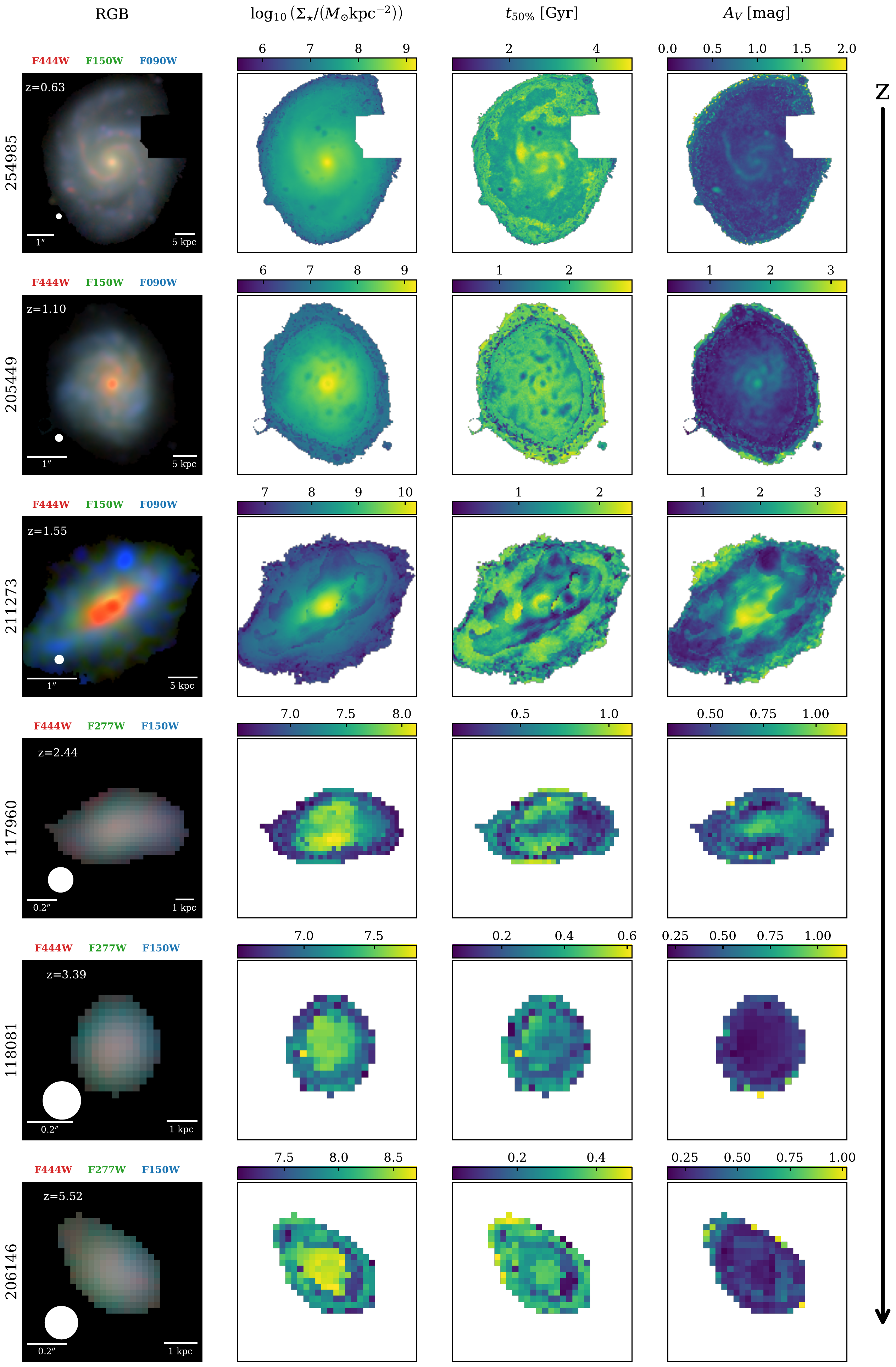}
    \caption{Same as Fig.~\ref{maps_tau}, but for the Dirichlet prior for the SFHs. We infer the properties $\log _{10}\left(\Sigma_{\star} /\left(M_{\odot} \mathrm{kpc}^{-2}\right)\right)$, $t_{50\%}$ [Gyr] in lookback time, and $A_V$.}
    \label{maps_nonparametric}

\end{figure*}

 After comparing our model with \texttt{Prospector} and \texttt{Synthesizer} for a few pixels, we proceeded to use it to create maps of the stellar population properties of the six sample galaxies. We also built RGB images using a colour scheme designed to enhance the structures enabling a better comparison with the population maps. We show the RGB images and the medians of the posteriors retrieved for the three physical properties for the six galaxies in Fig. \ref{maps_tau}, using the parametric SFHs. For the galaxies at lower redshifts (254985, 205449 and 211273), the PSF full width at half maximum (FWHM) is considerably smaller than the visual size of the galaxies. This, when combined with the large number of pixels available, enables clear gradients to be resolved in all the properties. The stellar mass density, $\log _{10}\left(\Sigma_{\star} /\left(M_{\odot} \mathrm{kpc}^{-2}\right)\right)$, is concentrated predominantly in the central regions of galaxies. The ages of the stellar populations are not as symmetric and are, instead, aligned with the morphology of the galaxies, with young populations along the spiral arms for the first two galaxies, which are face-on, and in the disc plane for the more inclined third galaxy. The dust attenuation shows strong variations and is correlated with the age, with young dusty clumps, together with centres and inner spiral arms highly attenuated with values up to $~4$~mag for 211273.
 
 In contrast, for galaxies at higher redshifts (117960, 118081, and 206146), our analysis primarily captures the effective radii, which is of the order of the PSF FWHM for F444W ($0.16^{\prime \prime}$, $5.36$~pix) because of the S/N threshold. Consequently, the maps reveal smooth trends, with the finer structural gradients remaining unresolved. The mass density maps show fewer gradients. The galaxies 117960 and 118081 present signs of inside-outside star formation, with cores younger than the outskirts. The dust attenuation is less strong for these galaxies, with smaller variations and slight negative correlations with the age, while for younger regions, it is more attenuated. Additionally, we show in Appendix~\ref{uncertainty_maps} the maps of uncertainties for the properties of the six galaxies that we obtained from the posterior distributions.
 
Continuing with the study of the six galaxies, we show in Fig.~\ref{maps_nonparametric} the maps of the stellar population properties obtained with the model using the Dirichlet prior for the SFHs. The structures we find in these maps are very similar to those obtained with the $\tau$ delayed prior. However, we find more structure in the stellar mass maps for the high-redshift galaxies. The other differences are mainly in terms of the age, which is expected from the differences in the SFH shapes; in this case, this is computed as $t_{50\%}$, which is the lookback time in Gyr at which 50\% of the total stellar mass was formed (itself a direct parameter of the model).

We did not observe significant variations within the PSF size of the F444W filter and all the variations are consistent with photometric noise levels, demonstrating the robustness of the inference. Although this behaviour was not explicitly enforced in our model (as is sometimes done in some binned modelling approaches), it naturally arises from our methodology. These maps offer multiple pathways for interpreting pixel-by-pixel properties within the global context of each galaxy. For example, rather than binning the photometry, we could directly bin the inferred properties, leveraging the strengths of Bayesian inference and hierarchical modelling to narrow the posterior distributions without losing the high resolution of our photometry. Alternatively, we could analyse gradients or property profiles to explore how these quantities vary across the galaxy. As a proof of concept, we include the radial profiles of the properties shown in the maps in Appendix~\ref{profiles}.

\subsection{Comparison between resolved and integrated stellar masses}
\label{outshining}
\begin{figure}
    \centering
    \includegraphics[width=0.75\linewidth]{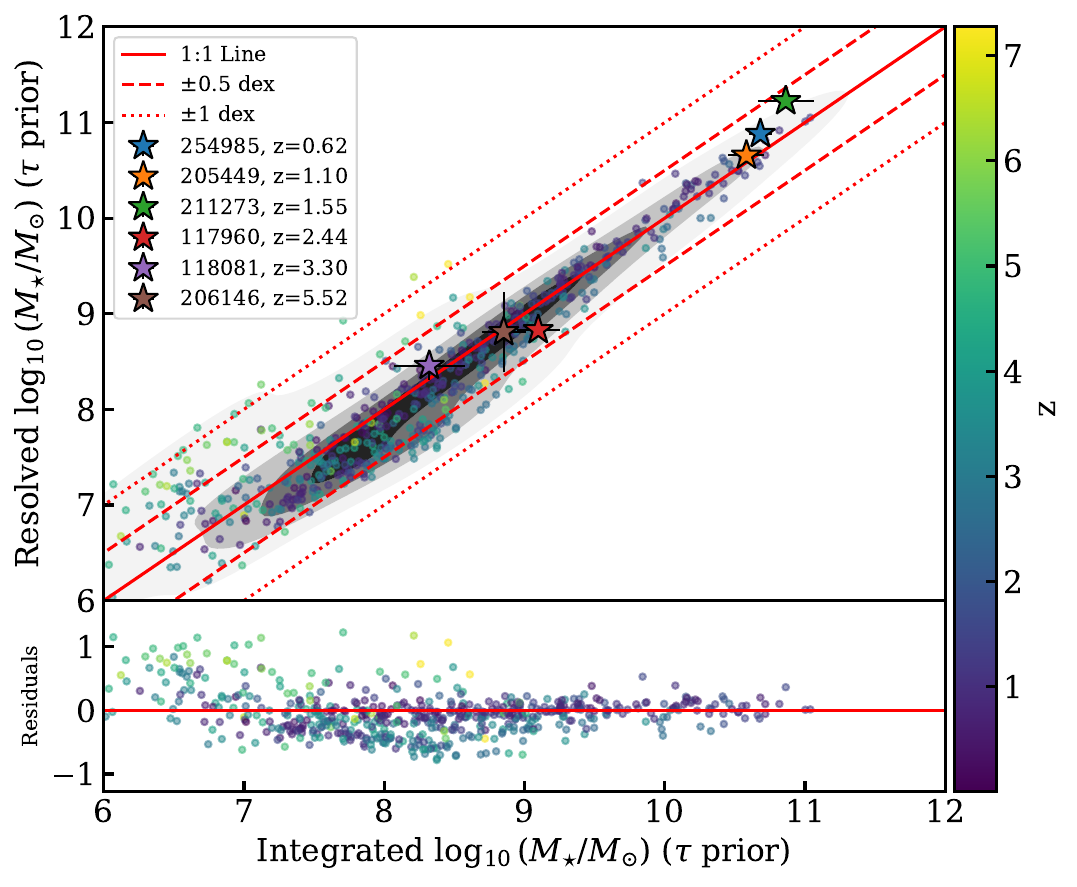}
    \includegraphics[width=0.75\linewidth]{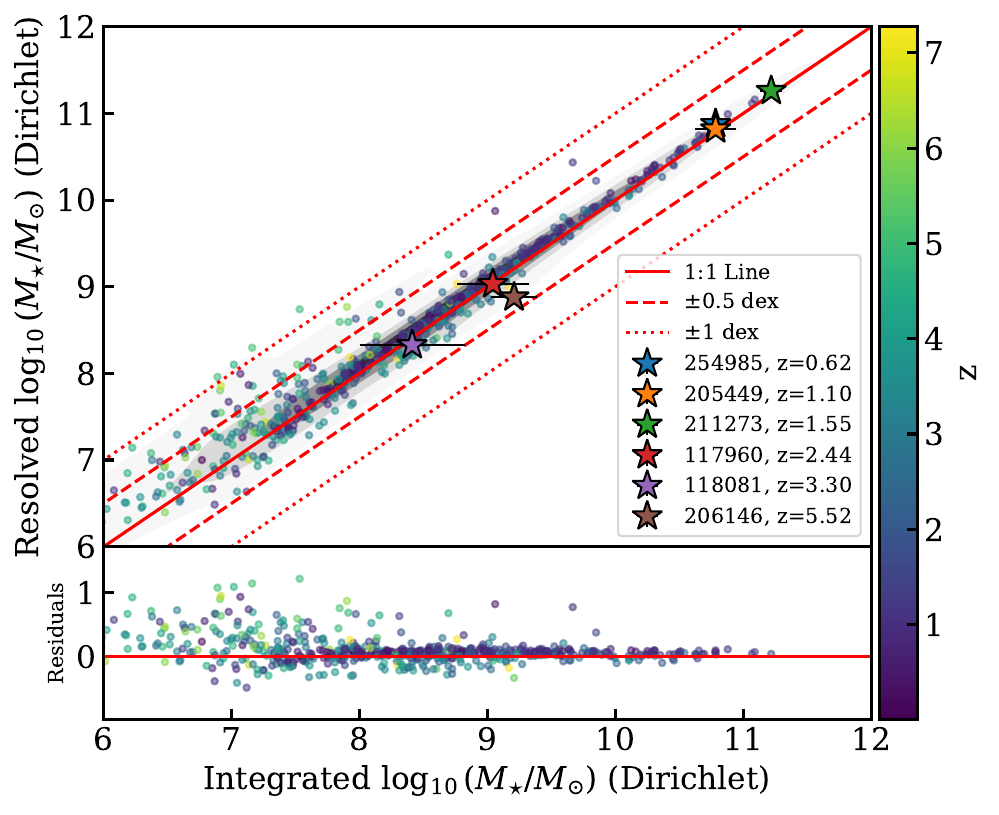} 
    \includegraphics[width=0.75\linewidth]{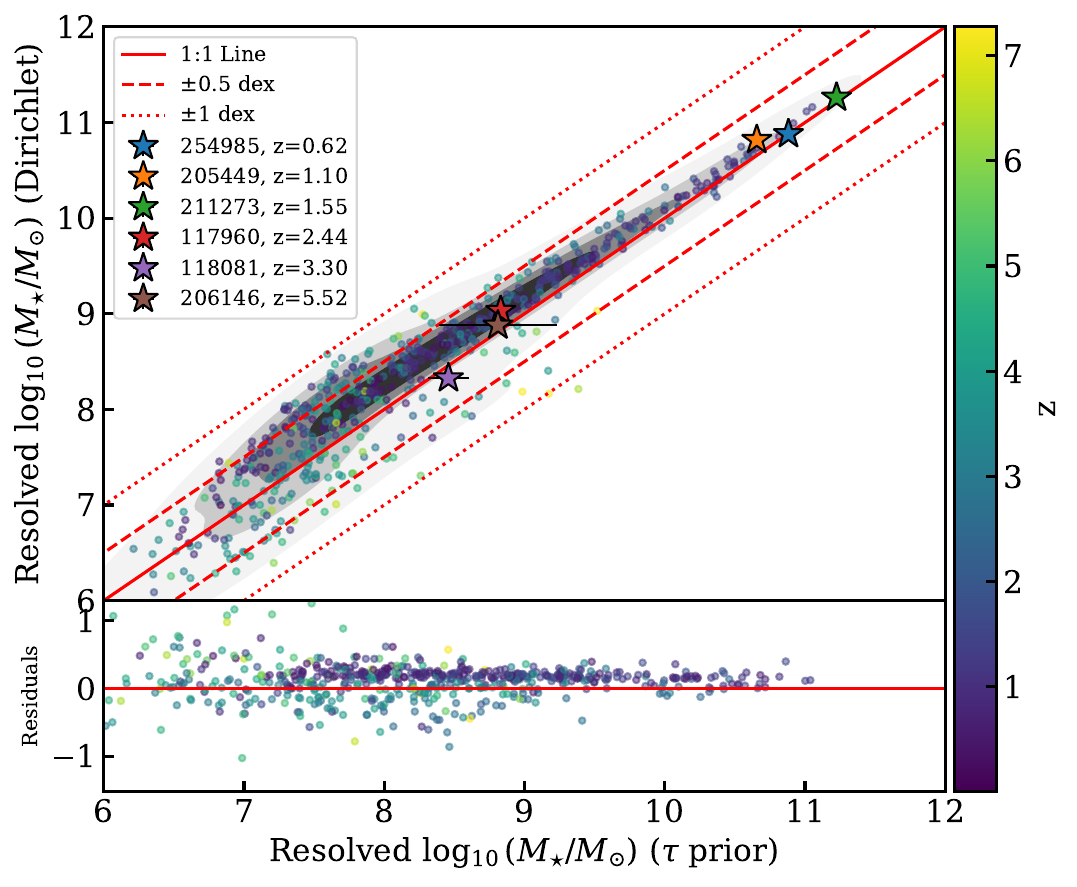}   
    \includegraphics[width=0.75\linewidth]{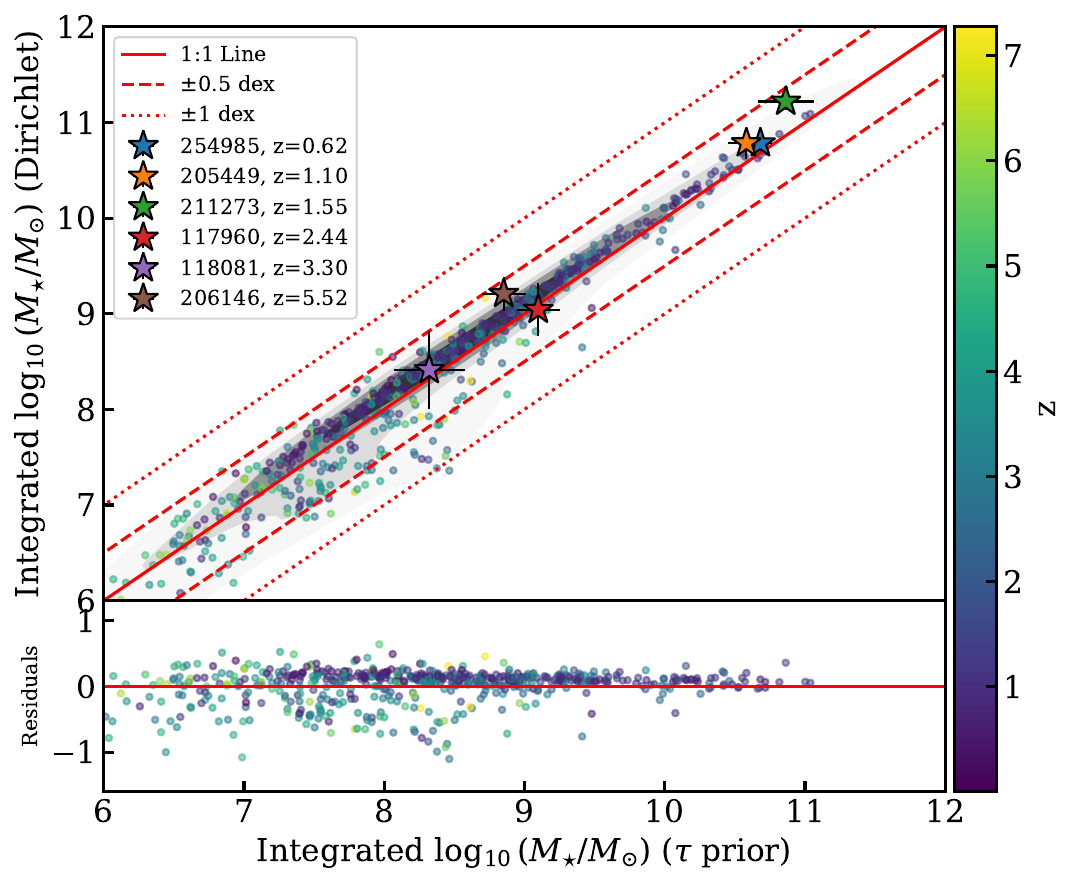}      
    \caption{ Pixel-based  vs the integrated estimations for the stellar masses  for the $\tau$-delayed prior for the SFHs, for the Dirichlet prior, the pixel-based stellar masses for the two different priors, and the integrated stellar masses for the two priors for $1083$ galaxies (from top to bottom). Each galaxy is a different dot colour-coded with the redshift. We also show the six previously studied galaxies as stars, as well as black kernel density distribution contours, and in red the 1:1 ratio (solid line), $\pm 0.5$~dex (dashed lines), and $\pm1$~dex (dotted lines) offsets. For clarity, only the error bars of the six galaxies are shown for both the integrated and resolved stellar masses.}
    \label{resolved_vs_integrated}
\end{figure}

We went on to study all the pixels of the $1083$ galaxies with $\overline{\rm{S/N}}$>5, $\sim2$\;million pixels. We focussed on the surviving stellar mass recovery, leaving the analysis of other properties for future exploration. Our approach treats individual pixels as statistically independent for simplicity and computational feasibility. While we recognise that the PSF inherently introduces spatial correlations between neighbouring pixels, we plan to thoroughly investigate this in upcoming works. For the purposes of this work, we chose to compute the total stellar mass by summing the posterior samples across pixels.

Specifically, we drew one sample from the posterior distribution of each pixel and summed these to obtain one sample of the posterior for the galaxy’s total stellar mass. Repeating this process for all posterior samples yields a posterior distribution for the total mass. We then used the medians of this total-mass posteriors to compare them with the integrated fit. For the latter, we summed the photometry of each of the above pixels and computed the integrated stellar masses for all the galaxies. We did this for the two models with different priors for the SFHs to compare the differences between these two ways of computing total stellar masses.

Several works attribute systematic differences between those stellar masses due to outshining of young stellar populations in the integrated analysis  \citep{Sorba2015, Sorba2018,jain2023, gimenez2024,fujimoto24,Narayanan24}, as a consequence of bright O-type and B-type stars that obscure the old populations, which dominate the total stellar mass. This outshining phenomenon is thought to be stronger in high-redshift low-mass galaxies, due to the stochasticity in their SFHs. Other recent works claim that this problem is mainly solved due to the large wavelength coverage of JWST \citep{Song2023, Lines2024} and medium bands of NIRCam \citep{adams2025,Cochrane2025,harvey2025}. Our approach allowed us to perform such a study in a Bayesian framework for a statistically significant set of observed galaxies, comprising at least one order of magnitude more galaxies than previous works.

We show in Fig.~\ref{resolved_vs_integrated} the resolved versus integrated stellar masses for all galaxies colour-coded with redshift, as well as the six galaxies studied in detail before with symbols of stars, for both the $\tau$-delayed and the Dirichlet prior for the SFHs. We did not find any strong outshining effects for any of the priors for stellar masses from $10^8 M_{\odot}$ to $10^{12} M_{\odot}$. However, the scatter is larger for the less flexible $\tau$-delayed SFHs. For this prior, we found ten outliers with their offsets (computed as $\log_{10} \left( M_*^{\text{resolved}} \right) - \log_{10} \left( M_*^{\text{integrated}} \right)$) higher than $1$~dex (one outlier with $z<3$ and nine with $z>3$). The median offset for this prior is $-0.068$~dex for $z<3$ and $0.063$~dex for $z>3$, while the median standard deviation for the posterior distributions of integrated stellar mass are  $0.14$~dex and $0.20$~dex, respectively. The uncertainties associated to the resolved masses, derived from the uncertainty propagation of the standard deviation of the posterior distributions of each of the pixels, are instead $0.044$~dex ($z<3$) and $0.14$~dex ($z>3$). For the Dirichlet prior, we found five outliers, all of them with $z>3$. The median offsets are $0.091$~dex for $z<3$ and $0.016$ for $z>3$, while the median standard deviation for the posterior distributions of integrated stellar mass are  $0.20$~dex and $0.31$~dex, respectively, and the uncertainties of the resolved analysis are $0.014$~dex and $0.089$~dex (from the uncertainty propagation of the standard deviations of the posteriors). The largest positive offsets and all the outliers were found for low-mass galaxies (below $10^{8.5} M_{\odot}$) at all redshifts, reporting offsets of 0.40~dex and 0.25~dex for the parametric and non-parametric models, respectively; this can be compared with the offsets for larger stellar masses, namely, of  0.11~dex and 0.09~dex. This is consistent with the findings of \cite{Lines2024} and \cite{harvey2025}.

In Fig.~\ref{resolved_vs_integrated}, we also compare the resolved and integrated stellar masses estimated for the same galaxies using the two different SFH priors. We measured a significant offset between the resolved stellar masses estimated with a $\tau$-delayed and a Dirichlet prior. The median offsets are of $0.204$~dex at $z<3$ and of $0.028$~dex at $z>3$. The differences are less pronounced  when comparing stellar masses obtained from integrated photometry. We reported median offsets of $0.044$~dex at $z<3$ and $-0.072$~dex at $z>3$.

To further investigated this, we performed a cross-test in the simulations, which we describe in Appendix~\ref{offset_simulations}. The results in the simulations show that due to the flexibility of the model trained on the Dirichlet simulation, it can correctly estimate the masses from a simulation performed with a $\tau$-delayed prior, marginalising over all parameters that the model cannot constrain from photometry (e.g. possible SFHs) and accurately reflecting this in the uncertainties. In contrast, the model using the $\tau$-delayed prior systematically underestimates the stellar masses when applied to the Dirichlet simulation, on average, by $0.147$~dex; this is roughly consistent with the reported offsets in observations. By inspecting the inferred and true SFHs, we conclude that the $\tau$-delayed prior leads to model misspecification: it only recovers the last burst of star formation and tends to ignore significant fractions of stellar mass formed in early epochs, without accounting for them in the uncertainties of the recovered properties.

\section{Limitations} 
\label{sec:limitations}

The amortised nature of our methodology, while offering a significant computational advantage over traditional Bayesian approaches such as MCMC, imposes constraints on its applicability. The model's training is the only computationally intensive step, enabling inference speeds orders of magnitude faster. However, it requires retraining the model for different filter sets or observational setups, as the simulation must precisely match the observed data, including uncertainties and background noise. Expanding the simulation to encompass multiple surveys would significantly increase the computational burden of training. Nonetheless, the substantial sample sizes available within surveys such as JADES DR2 (approximately $94,000$ objects, with $\sim 11,000$ meeting our filter requirements) justify this approach, marking a notable advancement.

A primary limitation in our galaxy sample selection was the reliance on spectroscopic redshifts, which are sparsely available. While this choice minimises redshift uncertainty in SED fitting, it restricts the number of galaxies analysed. To address this, we could incorporate photometric redshifts, either as direct inputs without retraining the model or retraining it with their associated uncertainties as input of the network. However, this will introduce a larger uncertainty into other property estimations. Furthermore, our assumption of equal redshift across all pixels within a galaxy introduces another source of potential error. While we generally trust the segmentation maps, galaxy overlaps or ambiguous pixel assignments can occur. Inferring redshifts per pixel or clump, as demonstrated in \cite{Perez-Gonzalez2023}, offers a potential solution. Although this approach may be susceptible to issues of low S/N, we intend to investigate it in future works, seeking to detect and resolve redshift discrepancies indicative of contamination or to refine photometric redshift estimates through Bayesian combination of posterior distributions.

The choice of SFH prior, whether restrictive (e.g. $\tau$-delayed) or flexible (e.g. Dirichlet), influences the retrieved properties and their uncertainties. While this study explored these extremes, real galaxy SFHs may exhibit intermediate complexity. More flexible priors, while potentially more realistic, often lead to wider posterior distributions and increased uncertainty. In the recent results of \cite{harvey2025}, the continuity and power-law priors were the ones that lead to less outshining effects. We aim to explore these priors in future works. Furthermore, the choice of SPS model can significantly influence results, as seen in comparisons with \texttt{Synthesizer}. These limitations inherent in SED models reflect our incomplete understanding of stellar evolution, a factor that equally affects conventional MCMC sampling approaches and extends beyond the scope of this work.

An important consideration to take into account is whether fitting individual pixels without modelling the PSF introduces biases. Our framework does not require explicit PSF characterisation in the forward model because each pixel is treated as an independent stellar population, rather than as part of a resolved spatial structure. However, oversampling the PSF can introduce statistical dependencies between neighbouring pixels. This does not bias individual posterior estimates but could affect summed quantities, such as total stellar mass or SFRs, due to correlated errors or partial double-counting of flux, especially in compact sources. To quantify this effect, we would need cosmological hydrodynamical simulations that properly model galaxy morphology, physical properties, and observational effects, allowing us to compare pixel-by-pixel and integrated estimates against known ground truth. We could also address this issue with appropriate spatial binning or post-processing to account for PSF scale correlations. The SBI framework allows for posterior-based binning after inference, for instance, using measures such as the Kullback-Leibler divergence between posteriors rather than simple spatial proximity, enabling more adaptive grouping strategies that respect both the S/N level and physical continuity. We plan to explore these two approaches in future investigations.

\section{Summary and future work}
\label{discussion}

We demonstrate the efficiency and robustness of simulation-based inference for recovering the properties of simulated and observed stellar populations from multi-wavelength photometric data. A key strength of our approach lies in its computational efficiency. The amortised inference enabled by NFs makes it possible to carry out rapid analyses of vast datasets. We achieved a fitting time of approximately $0.04$ seconds per pixel, which is a speed that is $\sim 3.5 \times 10^3$ times faster than traditional Bayesian codes such as \texttt{Prospector}. This speed is crucial for the pixel-by-pixel analysis of thousands of galaxies, enabling a statistical Bayesian approach to resolved stellar population studies that would be prohibitively time-consuming with traditional methods.

Applying our model to 1083 galaxies with spectroscopic redshifts, we explored the properties per pixel and their dependence on S/N at different redshifts. Stellar masses, SFRs, mass-weighted ages, and dust attenuation can be well constrained down to $\overline{\rm{S/N}}=5$, while the recovery of other properties, such as metallicity, is more challenging. The choice of SFH prior impacts the inferred properties, including the stellar mass, which shows an average offset of 0.15 dex in our cross-validation study. This highlights the importance of considering prior assumptions when assessing the uncertainties. 

We showcased the model's ability to resolve spatial variations in galaxy physical properties via a pixel-by-pixel analysis of six example JADES galaxies across different redshifts and morphologies. We observed clear gradients in terms of stellar mass, age, and dust attenuation, particularly in the case of low-redshift galaxies $(z<2),$ where the PSF FWHM for the F444W filter (to which all the photometry mosaics were convolved) is considerably smaller than the galaxy size.

As a proof of concept for future statistical analysis, we made a  pixel-by-pixel comparison and integrated the estimations of stellar masses for the $1083$ galaxies in our sample and their $\sim2$ million pixels with $\overline{\rm{S/N}}>5$. We found no strong evidence of outshining effects, with only a few outliers observed in high-redshift low-mass galaxies. This result is consistent with recent studies that have leveraged the high-wavelength coverage of HST+JWST, including NIRCam's medium bands \citep[e.g.][]{Song2023, Lines2024, Cochrane2025, harvey2025}. However, it deviates from other findings that suggested significant outshining \citep{Sorba2015, Sorba2018,jain2023, gimenez2024,Narayanan24}. We attribute the primary systematic uncertainty in stellar mass estimation to the choice of SFH prior, rather than the resolved or integrated analysis.

A key direction for future works is the implementation of spatial binning strategies to mitigate the effects of PSF oversampling. While traditional binning approaches such as Voronoi tessellation \citep{cappellari03} ensure a constant S/N, they do not fully exploit the spatial resolution of JWST imaging. We plan to incorporate the \texttt{piXedfit} binning methodology \citep{Abdurro2017, Abdurro2021}, which segments galaxies based on $\chi^2$ comparisons of neighbouring SEDs. This method retains detailed spatial information, while respecting PSF limitations and maintaining sufficient S/N within each bin, making it well suited to our framework.

In parallel, we aim to extend the pipeline to other JWST surveys, such as The Cosmic Evolution Early Release Science Survey \citep[CEERS;][]{finkelstein2025}, and to explore datasets with different observational characteristics. Additional developments will include testing more flexible SFH priors and implementing pixel-level redshift inference. These efforts will enhance the robustness and interpretability of resolved stellar population studies in large galaxy samples.

   \label{LastPage} 
   
\begin{acknowledgements}
Co-funded by the European Union (MSCA Doctoral Network EDUCADO, GA 101119830 and Widening Participation, ExGal-Twin, GA 101158446). PIN thanks the LSST-DA Data Science Fellowship Program, which is funded by LSST-DA, the Brinson Foundation, the WoodNext Foundation, and the Research Corporation for Science Advancement Foundation; her participation in the program has benefited this work. MHC and PIN acknowledge financial support from the State Research Agency of the Spanish Ministry of Science and Innovation (AEI-MCINN) under the grants ``Galaxy Evolution with Artificial Intelligence'' with reference PGC2018-100852-A-I00 and ``BASALT'' with reference PID2021-126838NB-I00. JHK acknowledges support from AEI-MCINN under the grant ``The structure and evolution of galaxies and their outer regions'' and the European Regional Development Fund (ERDF) with reference PID2022-136505NB-I00/10.13039/501100011033. PGP-G acknowledges support from grant PID2022-139567NB-I00 funded by Spanish Ministerio de Ciencia e Innovaci\'on MCIN/AEI/10.13039/501100011033, FEDER {\it Una manera de hacer Europa}.
\end{acknowledgements}

%\begin{dataavailability}
The entire pipeline, including the scripts for the simulation and the Bayesian inference framework, is publicly available at \texttt{\href{https://github.com/patriglesias/SBIPIX.git}{https://github.com/patriglesias/SBIPIX.git}}. Catalogues of results and additional files will also be provided upon reasonable request.
%\end{dataavailability}

\bibliographystyle{aa}
   \bibliography{references}

\begin{thebibliography}{105}
\expandafter\ifx\csname natexlab\endcsname\relax\def\natexlab#1{#1}\fi

\bibitem[{{Abdurro'uf} \& {Akiyama}(2017)}]{Abdurro2017}
{Abdurro'uf} \& {Akiyama}, M. 2017, \mnras, 469, 2806

\bibitem[{{Abdurro'uf} {et~al.}(2021){Abdurro'uf}, {Lin}, {Wu}, \& {Akiyama}}]{Abdurro2021}
{Abdurro'uf}, {Lin}, Y.-T., {Wu}, P.-F., \& {Akiyama}, M. 2021, \apjs, 254, 15

\bibitem[{{Abraham} {et~al.}(1999){Abraham}, {Ellis}, {Fabian}, {Tanvir}, \& {Glazebrook}}]{abraham1999}
{Abraham}, R.~G., {Ellis}, R.~S., {Fabian}, A.~C., {Tanvir}, N.~R., \& {Glazebrook}, K. 1999, \mnras, 303, 641

\bibitem[{{Adams} {et~al.}(2025){Adams}, {Austin}, {Harvey}, {Conselice}, {Trussler}, {Li}, {Westcott}, {Ferreira}, {Rusakov}, \& {Goolsby}}]{adams2025}
{Adams}, N.~J., {Austin}, D., {Harvey}, T., {et~al.} 2025, \mnras, 542, 1705

\bibitem[{{Bacon} {et~al.}(2023){Bacon}, {Brinchmann}, {Conseil}, {Maseda}, {Nanayakkara}, {Wendt}, {Bacher}, {Mary}, {Weilbacher}, {Krajnovi{\'c}}, {Boogaard}, {Bouch{\'e}}, {Contini}, {Epinat}, {Feltre}, {Guo}, {Herenz}, {Kollatschny}, {Kusakabe}, {Leclercq}, {Michel-Dansac}, {Pello}, {Richard}, {Roth}, {Salvignol}, {Schaye}, {Steinmetz}, {Tresse}, {Urrutia}, {Verhamme}, {Vitte}, {Wisotzki}, \& {Zoutendijk}}]{bacon2023}
{Bacon}, R., {Brinchmann}, J., {Conseil}, S., {et~al.} 2023, \aap, 670, A4

\bibitem[{{Bacon} {et~al.}(2017){Bacon}, {Conseil}, {Mary}, {Brinchmann}, {Shepherd}, {Akhlaghi}, {Weilbacher}, {Piqueras}, {Wisotzki}, {Lagattuta}, {Epinat}, {Guerou}, {Inami}, {Cantalupo}, {Courbot}, {Contini}, {Richard}, {Maseda}, {Bouwens}, {Bouch{\'e}}, {Kollatschny}, {Schaye}, {Marino}, {Pello}, {Herenz}, {Guiderdoni}, \& {Carollo}}]{bacon2017}
{Bacon}, R., {Conseil}, S., {Mary}, D., {et~al.} 2017, \aap, 608, A1

\bibitem[{{Beckwith} {et~al.}(2006){Beckwith}, {Stiavelli}, {Koekemoer}, {Caldwell}, {Ferguson}, {Hook}, {Lucas}, {Bergeron}, {Corbin}, {Jogee}, {Panagia}, {Robberto}, {Royle}, {Somerville}, \& {Sosey}}]{beckwith2006}
{Beckwith}, S. V.~W., {Stiavelli}, M., {Koekemoer}, A.~M., {et~al.} 2006, \aj, 132, 1729

\bibitem[{{Bertelli} {et~al.}(1994){Bertelli}, {Bressan}, {Chiosi}, {Fagotto}, \& {Nasi}}]{padova94}
{Bertelli}, G., {Bressan}, A., {Chiosi}, C., {Fagotto}, F., \& {Nasi}, E. 1994, \aaps, 106, 275

\bibitem[{{Boquien} {et~al.}(2019){Boquien}, {Burgarella}, {Roehlly}, {Buat}, {Ciesla}, {Corre}, {Inoue}, \& {Salas}}]{Boquien2019}
{Boquien}, M., {Burgarella}, D., {Roehlly}, Y., {et~al.} 2019, \aap, 622, A103

\bibitem[{Bradley {et~al.}(2024)Bradley, Sip{\H o}cz, Robitaille, Tollerud, Vin{\'{\i}}cius, Deil, Barbary, Wilson, Busko, Donath, G{\"u}nther, Cara, Lim, Me{\ss}linger, Burnett, Conseil, Droettboom, Bostroem, Bray, Bratholm, Jamieson, Ginsburg, Barentsen, Craig, Pascual, Rathi, Perrin, Morris, \& Perren}]{bradley2024}
Bradley, L., Sip{\H o}cz, B., Robitaille, T., {et~al.} 2024, astropy/photutils: 1.13.0

\bibitem[{{Bruzual} \& {Charlot}(2003)}]{bc03}
{Bruzual}, G. \& {Charlot}, S. 2003, \mnras, 344, 1000

\bibitem[{{Bunker} {et~al.}(2024){Bunker}, {Cameron}, {Curtis-Lake}, {Jakobsen}, {Carniani}, {Curti}, {Witstok}, {Maiolino}, {D'Eugenio}, {Looser}, {Willott}, {Bonaventura}, {Hainline}, {{\"U}bler}, {Willmer}, {Saxena}, {Smit}, {Alberts}, {Arribas}, {Baker}, {Baum}, {Bhatawdekar}, {Bowler}, {Boyett}, {Charlot}, {Chen}, {Chevallard}, {Circosta}, {DeCoursey}, {de Graaff}, {Egami}, {Eisenstein}, {Endsley}, {Ferruit}, {Giardino}, {Hausen}, {Helton}, {Hviding}, {Ji}, {Johnson}, {Jones}, {Kumari}, {Laseter}, {L{\"u}tzgendorf}, {Maseda}, {Nelson}, {Parlanti}, {Perna}, {Rauscher}, {Rawle}, {Rix}, {Rieke}, {Robertson}, {Rodr{\'\i}guez Del Pino}, {Sandles}, {Scholtz}, {Sharpe}, {Skarbinski}, {Stark}, {Sun}, {Tacchella}, {Topping}, {Villanueva}, {Wallace}, {Williams}, \& {Woodrum}}]{bunker24}
{Bunker}, A.~J., {Cameron}, A.~J., {Curtis-Lake}, E., {et~al.} 2024, \aap, 690, A288

\bibitem[{{Calzetti} {et~al.}(2000){Calzetti}, {Armus}, {Bohlin}, {Kinney}, {Koornneef}, \& {Storchi-Bergmann}}]{Calzetti2000}
{Calzetti}, D., {Armus}, L., {Bohlin}, R.~C., {et~al.} 2000, \apj, 533, 682

\bibitem[{Caplar \& Tacchella(2019)}]{Caplar_2019}
Caplar, N. \& Tacchella, S. 2019, \mnras, 487, 3845–3869

\bibitem[{{Cappellari} \& {Copin}(2003)}]{cappellari03}
{Cappellari}, M. \& {Copin}, Y. 2003, \mnras, 342, 345

\bibitem[{{Carnall} {et~al.}(2019){Carnall}, {Leja}, {Johnson}, {McLure}, {Dunlop}, \& {Conroy}}]{carnall2019}
{Carnall}, A.~C., {Leja}, J., {Johnson}, B.~D., {et~al.} 2019, \apj, 873, 44

\bibitem[{Carnall {et~al.}(2018)Carnall, McLure, Dunlop, \& Davé}]{Carnall_2018}
Carnall, A.~C., McLure, R.~J., Dunlop, J.~S., \& Davé, R. 2018, \mnras, 480, 4379–4401

\bibitem[{{Cervi{\~n}o} \& {Valls-Gabaud}(2003)}]{cervino2003}
{Cervi{\~n}o}, M. \& {Valls-Gabaud}, D. 2003, \mnras, 338, 481

\bibitem[{{Chabrier}(2003)}]{chabrier2003}
{Chabrier}, G. 2003, \pasp, 115, 763

\bibitem[{Choi {et~al.}(2016)Choi, Dotter, Conroy, Cantiello, Paxton, \& Johnson}]{Choi_2016}
Choi, J., Dotter, A., Conroy, C., {et~al.} 2016, \apj, 823, 102

\bibitem[{Cochrane {et~al.}(2025)Cochrane, Katz, Begley, Hayward, \& Best}]{Cochrane2025}
Cochrane, R.~K., Katz, H., Begley, R., Hayward, C.~C., \& Best, P.~N. 2025, \apjl, 978, L42

\bibitem[{Conroy(2013)}]{Conroy_2013}
Conroy, C. 2013, {ARA\&A}, 51, 393–455

\bibitem[{{Conroy} \& {Gunn}(2010)}]{Conroy_2010}
{Conroy}, C. \& {Gunn}, J.~E. 2010, \apj, 712, 833

\bibitem[{{Conroy} {et~al.}(2009){Conroy}, {Gunn}, \& {White}}]{Conroy2009}
{Conroy}, C., {Gunn}, J.~E., \& {White}, M. 2009, \apj, 699, 486

\bibitem[{Cranmer {et~al.}(2020)Cranmer, Brehmer, \& Louppe}]{Cranmer_2020}
Cranmer, K., Brehmer, J., \& Louppe, G. 2020, Proceedings of the National Academy of Sciences, 117, 30055

\bibitem[{{Dahlen} {et~al.}(2010){Dahlen}, {Mobasher}, {Dickinson}, {Ferguson}, {Giavalisco}, {Grogin}, {Guo}, {Koekemoer}, {Lee}, {Lee}, {Nonino}, {Riess}, \& {Salimbeni}}]{dahlen2010}
{Dahlen}, T., {Mobasher}, B., {Dickinson}, M., {et~al.} 2010, \apj, 724, 425

\bibitem[{{D'Eugenio} {et~al.}(2024){D'Eugenio}, {Cameron}, {Scholtz}, {Carniani}, {Willott}, {Curtis-Lake}, {Bunker}, {Parlanti}, {Maiolino}, {Willmer}, {Jakobsen}, {Robertson}, {Johnson}, {Tacchella}, {Cargile}, {Rawle}, {Arribas}, {Chevallard}, {Curti}, {Egami}, {Eisenstein}, {Kumari}, {Looser}, {Rieke}, {Rodr{\'\i}guez Del Pino}, {Saxena}, {{\"U}bler}, {Venturi}, {Witstok}, {Baker}, {Bhatawdekar}, {Bonaventura}, {Boyett}, {Charlot}, {Danhaive}, {Hainline}, {Hausen}, {Helton}, {Ji}, {Ji}, {Jones}, {Joud{\v{z}}balis}, {Maseda}, {P{\'e}rez-Gonz{\'a}lez}, {Perna}, {Pusk{\'a}s}, {Shivaei}, {Silcock}, {Simmonds}, {Smit}, {Sun}, {Villanueva}, {Williams}, \& {Zhu}}]{DEUGENIO2024}
{D'Eugenio}, F., {Cameron}, A.~J., {Scholtz}, J., {et~al.} 2024, arXiv e-prints, arXiv:2404.06531

\bibitem[{{Draine} \& {Li}(2007)}]{draine2007}
{Draine}, B.~T. \& {Li}, A. 2007, \apj, 657, 810

\bibitem[{{Durkan} {et~al.}(2019){Durkan}, {Bekasov}, {Murray}, \& {Papamakarios}}]{durkan2019}
{Durkan}, C., {Bekasov}, A., {Murray}, I., \& {Papamakarios}, G. 2019, arXiv e-prints, arXiv:1906.04032

\bibitem[{{Dye}(2008)}]{dye08}
{Dye}, S. 2008, \mnras, 389, 1293

\bibitem[{{Eisenstein} {et~al.}(2023{\natexlab{a}}){Eisenstein}, {Johnson}, {Robertson}, {Tacchella}, {Hainline}, {Jakobsen}, {Maiolino}, {Bonaventura}, {Bunker}, {Cameron}, {Cargile}, {Curtis-Lake}, {Hausen}, {Pusk{\'a}s}, {Rieke}, {Sun}, {Willmer}, {Willott}, {Alberts}, {Arribas}, {Baker}, {Baum}, {Bhatawdekar}, {Carniani}, {Charlot}, {Chen}, {Chevallard}, {Curti}, {DeCoursey}, {D'Eugenio}, {de Graaff}, {Egami}, {Helton}, {Ji}, {Jones}, {Kumari}, {L{\"u}tzgendorf}, {Laseter}, {Looser}, {Lyu}, {Maseda}, {Nelson}, {Parlanti}, {Rauscher}, {Rawle}, {Rieke}, {Rix}, {Rujopakarn}, {Sandles}, {Saxena}, {Scholtz}, {Sharpe}, {Shivaei}, {Simmonds}, {Smit}, {Topping}, {{\"U}bler}, {Venturi}, {Williams}, {Witstok}, \& {Woodrum}}]{eisenstein2023b}
{Eisenstein}, D.~J., {Johnson}, B.~D., {Robertson}, B., {et~al.} 2023{\natexlab{a}}, arXiv e-prints, arXiv:2310.12340

\bibitem[{{Eisenstein} {et~al.}(2023{\natexlab{b}}){Eisenstein}, {Willott}, {Alberts}, {Arribas}, {Bonaventura}, {Bunker}, {Cameron}, {Carniani}, {Charlot}, {Curtis-Lake}, {D'Eugenio}, {Endsley}, {Ferruit}, {Giardino}, {Hainline}, {Hausen}, {Jakobsen}, {Johnson}, {Maiolino}, {Rieke}, {Rieke}, {Rix}, {Robertson}, {Stark}, {Tacchella}, {Williams}, {Willmer}, {Baker}, {Baum}, {Bhatawdekar}, {Boyett}, {Chen}, {Chevallard}, {Circosta}, {Curti}, {Danhaive}, {DeCoursey}, {de Graaff}, {Dressler}, {Egami}, {Helton}, {Hviding}, {Ji}, {Jones}, {Kumari}, {L{\"u}tzgendorf}, {Laseter}, {Looser}, {Lyu}, {Maseda}, {Nelson}, {Parlanti}, {Perna}, {Pusk{\'a}s}, {Rawle}, {Rodr{\'\i}guez Del Pino}, {Sandles}, {Saxena}, {Scholtz}, {Sharpe}, {Shivaei}, {Silcock}, {Simmonds}, {Skarbinski}, {Smit}, {Stone}, {Suess}, {Sun}, {Tang}, {Topping}, {{\"U}bler}, {Villanueva}, {Wallace}, {Whitler}, {Witstok}, \& {Woodrum}}]{eisenstein2023}
{Eisenstein}, D.~J., {Willott}, C., {Alberts}, S., {et~al.} 2023{\natexlab{b}}, arXiv e-prints, arXiv:2306.02465

\bibitem[{{Ellis} {et~al.}(2013){Ellis}, {McLure}, {Dunlop}, {Robertson}, {Ono}, {Schenker}, {Koekemoer}, {Bowler}, {Ouchi}, {Rogers}, {Curtis-Lake}, {Schneider}, {Charlot}, {Stark}, {Furlanetto}, \& {Cirasuolo}}]{ellis2013}
{Ellis}, R.~S., {McLure}, R.~J., {Dunlop}, J.~S., {et~al.} 2013, \apjl, 763, L7

\bibitem[{{Ferland} {et~al.}(2017){Ferland}, {Chatzikos}, {Guzm{\'a}n}, {Lykins}, {van Hoof}, {Williams}, {Abel}, {Badnell}, {Keenan}, {Porter}, \& {Stancil}}]{Ferland2017}
{Ferland}, G.~J., {Chatzikos}, M., {Guzm{\'a}n}, F., {et~al.} 2017, \rmxaa, 53, 385

\bibitem[{{Ferland} {et~al.}(2013){Ferland}, {Porter}, {van Hoof}, {Williams}, {Abel}, {Lykins}, {Shaw}, {Henney}, \& {Stancil}}]{Ferland2013}
{Ferland}, G.~J., {Porter}, R.~L., {van Hoof}, P.~A.~M., {et~al.} 2013, \rmxaa, 49, 137

\bibitem[{{Finkelstein} {et~al.}(2025){Finkelstein}, {Bagley}, {Arrabal Haro}, {Dickinson}, {Ferguson}, {Kartaltepe}, {Kocevski}, {Koekemoer}, {Lotz}, {Papovich}, {P{\'e}rez-Gonz{\'a}lez}, {Pirzkal}, {Somerville}, {Trump}, {Yang}, {Yung}, {Fontana}, {Grazian}, {Grogin}, {Kewley}, {Kirkpatrick}, {Larson}, {Pentericci}, {Ravindranath}, {Wilkins}, {Almaini}, {Amor{\'\i}n}, {Barro}, {Bhatawdekar}, {Bisigello}, {Brooks}, {Buat}, {Buitrago}, {Burgarella}, {Calabr{\`o}}, {Castellano}, {Cheng}, {Cleri}, {Cole}, {Cooper}, {Cooper}, {Costantin}, {Cox}, {Croton}, {Daddi}, {Davis}, {Dekel}, {Elbaz}, {Fern{\'a}ndez}, {Fujimoto}, {Gandolfi}, {Gardner}, {Gawiser}, {Giavalisco}, {G{\'o}mez-Guijarro}, {Guo}, {Gupta}, {Hathi}, {Harish}, {Henry}, {Hirschmann}, {Hu}, {Hutchison}, {Iyer}, {Jaskot}, {Jha}, {Jung}, {Kassin}, {Kokorev}, {Kurczynski}, {Leung}, {Llerena}, {Long}, {Lucas}, {Lu}, {McGrath}, {McIntosh}, {Merlin}, {Mobasher}, {Morales}, {Napolitano}, {Pacucci}, {Pandya}, {Rafelski}, {Rodighiero}, {Rose}, {Santini}, {Seill{\'e}}, {Simons}, {Shen}, {Straughn}, {Tacchella}, {Taylor}, {Vanderhoof}, {Vega-Ferrero}, {Weiner}, {Willmer}, {Zhu}, {Bell}, {Wuyts}, {Holwerda}, {Wang}, {Wang}, {Zavala}, \& {CEERS Collaboration}}]{finkelstein2025}
{Finkelstein}, S.~L., {Bagley}, M.~B., {Arrabal Haro}, P., {et~al.} 2025, \apjl, 983, L4

\bibitem[{{Fujimoto} {et~al.}(2025){Fujimoto}, {Ouchi}, {Kohno}, {Valentino}, {Gim{\'e}nez-Arteaga}, {Brammer}, {Furtak}, {Kohandel}, {Oguri}, {Pallottini}, {Richard}, {Zitrin}, {Bauer}, {Boylan-Kolchin}, {Dessauges-Zavadsky}, {Egami}, {Finkelstein}, {Ma}, {Smail}, {Watson}, {Hutchison}, {Rigby}, {Welch}, {Ao}, {Bradley}, {Caminha}, {Caputi}, {Espada}, {Endsley}, {Fudamoto}, {Gonz{\'a}lez-L{\'o}pez}, {Hatsukade}, {Koekemoer}, {Kokorev}, {Laporte}, {Lee}, {Magdis}, {Ono}, {Rizzo}, {Shibuya}, {Shimasaku}, {Sun}, {Toft}, {Umehata}, {Wang}, \& {Yajima}}]{fujimoto24}
{Fujimoto}, S., {Ouchi}, M., {Kohno}, K., {et~al.} 2025, Nat. Astron. [\eprint[arXiv]{2402.18543}]

\bibitem[{{Giavalisco} {et~al.}(2004){Giavalisco}, {Ferguson}, {Koekemoer}, {Dickinson}, {Alexander}, {Bauer}, {Bergeron}, {Biagetti}, {Brandt}, {Casertano}, {Cesarsky}, {Chatzichristou}, {Conselice}, {Cristiani}, {Da Costa}, {Dahlen}, {de Mello}, {Eisenhardt}, {Erben}, {Fall}, {Fassnacht}, {Fosbury}, {Fruchter}, {Gardner}, {Grogin}, {Hook}, {Hornschemeier}, {Idzi}, {Jogee}, {Kretchmer}, {Laidler}, {Lee}, {Livio}, {Lucas}, {Madau}, {Mobasher}, {Moustakas}, {Nonino}, {Padovani}, {Papovich}, {Park}, {Ravindranath}, {Renzini}, {Richardson}, {Riess}, {Rosati}, {Schirmer}, {Schreier}, {Somerville}, {Spinrad}, {Stern}, {Stiavelli}, {Strolger}, {Urry}, {Vandame}, {Williams}, \& {Wolf}}]{Giavalisco2004}
{Giavalisco}, M., {Ferguson}, H.~C., {Koekemoer}, A.~M., {et~al.} 2004, \apjl, 600, L93

\bibitem[{Gibson {et~al.}(2024)Gibson, Nelson, Williams, Price, Whitaker, Suess, de~Graaff, Johnson, Bunker, Baker, Bhatawdekar, Boyett, Charlot, Curtis-Lake, Eisenstein, Hainline, Hausen, Maiolino, Rieke, Rieke, Robertson, Tacchella, \& Willott}]{Gibson2024}
Gibson, J.~L., Nelson, E., Williams, C.~C., {et~al.} 2024, \apj, 974, 48

\bibitem[{{Gim{\'e}nez-Arteaga} {et~al.}(2024){Gim{\'e}nez-Arteaga}, {Fujimoto}, {Valentino}, {Brammer}, {Mason}, {Rizzo}, {Rusakov}, {Colina}, {Prieto-Lyon}, {Oesch}, {Espada}, {Heintz}, {Knudsen}, {Dessauges-Zavadsky}, {Laporte}, {Lee}, {Magdis}, {Ono}, {Ao}, {Ouchi}, {Kohno}, \& {Koekemoer}}]{gimenez2024}
{Gim{\'e}nez-Arteaga}, C., {Fujimoto}, S., {Valentino}, F., {et~al.} 2024, \aap, 686, A63

\bibitem[{{Gim{\'e}nez-Arteaga} {et~al.}(2023){Gim{\'e}nez-Arteaga}, {Oesch}, {Brammer}, {Valentino}, {Mason}, {Weibel}, {Barrufet}, {Fujimoto}, {Heintz}, {Nelson}, {Strait}, {Suess}, \& {Gibson}}]{gimenez2023}
{Gim{\'e}nez-Arteaga}, C., {Oesch}, P.~A., {Brammer}, G.~B., {et~al.} 2023, \apj, 948, 126

\bibitem[{Giménez-Arteaga {et~al.}(2024)Giménez-Arteaga, Fujimoto, Valentino, Brammer, Mason, Rizzo, Rusakov, Colina, Prieto-Lyon, Oesch, Espada, Heintz, Knudsen, Dessauges-Zavadsky, Laporte, Lee, Magdis, Ono, Ao, Ouchi, Kohno, \& Koekemoer}]{Gimnez-Arteaga2024}
Giménez-Arteaga, C., Fujimoto, S., Valentino, F., {et~al.} 2024

\bibitem[{{Grogin} {et~al.}(2011){Grogin}, {Kocevski}, {Faber}, {Ferguson}, {Koekemoer}, {Riess}, {Acquaviva}, {Alexander}, {Almaini}, {Ashby}, {Barden}, {Bell}, {Bournaud}, {Brown}, {Caputi}, {Casertano}, {Cassata}, {Castellano}, {Challis}, {Chary}, {Cheung}, {Cirasuolo}, {Conselice}, {Roshan Cooray}, {Croton}, {Daddi}, {Dahlen}, {Dav{\'e}}, {de Mello}, {Dekel}, {Dickinson}, {Dolch}, {Donley}, {Dunlop}, {Dutton}, {Elbaz}, {Fazio}, {Filippenko}, {Finkelstein}, {Fontana}, {Gardner}, {Garnavich}, {Gawiser}, {Giavalisco}, {Grazian}, {Guo}, {Hathi}, {H{\"a}ussler}, {Hopkins}, {Huang}, {Huang}, {Jha}, {Kartaltepe}, {Kirshner}, {Koo}, {Lai}, {Lee}, {Li}, {Lotz}, {Lucas}, {Madau}, {McCarthy}, {McGrath}, {McIntosh}, {McLure}, {Mobasher}, {Moustakas}, {Mozena}, {Nandra}, {Newman}, {Niemi}, {Noeske}, {Papovich}, {Pentericci}, {Pope}, {Primack}, {Rajan}, {Ravindranath}, {Reddy}, {Renzini}, {Rix}, {Robaina}, {Rodney}, {Rosario}, {Rosati}, {Salimbeni}, {Scarlata}, {Siana}, {Simard}, {Smidt}, {Somerville}, {Spinrad}, {Straughn}, {Strolger}, {Telford}, {Teplitz}, {Trump}, {van der Wel}, {Villforth}, {Wechsler}, {Weiner}, {Wiklind}, {Wild}, {Wilson}, {Wuyts}, {Yan}, \& {Yun}}]{grogin2011}
{Grogin}, N.~A., {Kocevski}, D.~D., {Faber}, S.~M., {et~al.} 2011, \apjs, 197, 35

\bibitem[{{Guo} {et~al.}(2012){Guo}, {Giavalisco}, {Ferguson}, {Cassata}, \& {Koekemoer}}]{guo12}
{Guo}, Y., {Giavalisco}, M., {Ferguson}, H.~C., {Cassata}, P., \& {Koekemoer}, A.~M. 2012, \apj, 757, 120

\bibitem[{{Hahn} {et~al.}(2023){Hahn}, {Kwon}, {Tojeiro}, {Siudek}, {Canning}, {Mezcua}, {Tinker}, {Brooks}, {Doel}, {Fanning}, {Gazta{\~n}aga}, {Kehoe}, {Landriau}, {Meisner}, {Moustakas}, {Poppett}, {Tarle}, {Weiner}, \& {Zou}}]{hahn2023}
{Hahn}, C., {Kwon}, K.~J., {Tojeiro}, R., {et~al.} 2023, \apj, 945, 16

\bibitem[{{Hahn} \& {Melchior}(2022)}]{Hahn2022}
{Hahn}, C. \& {Melchior}, P. 2022, \apj, 938, 11

\bibitem[{{Hainline} {et~al.}(2024){Hainline}, {Johnson}, {Robertson}, {Tacchella}, {Helton}, {Sun}, {Eisenstein}, {Simmonds}, {Topping}, {Whitler}, {Willmer}, {Rieke}, {Suess}, {Hviding}, {Cameron}, {Alberts}, {Baker}, {Baum}, {Bhatawdekar}, {Bonaventura}, {Boyett}, {Bunker}, {Carniani}, {Charlot}, {Chevallard}, {Chen}, {Curti}, {Curtis-Lake}, {D'Eugenio}, {Egami}, {Endsley}, {Hausen}, {Ji}, {Looser}, {Lyu}, {Maiolino}, {Nelson}, {Pusk{\'a}s}, {Rawle}, {Sandles}, {Saxena}, {Smit}, {Stark}, {Williams}, {Willott}, \& {Witstok}}]{Hainline2024}
{Hainline}, K.~N., {Johnson}, B.~D., {Robertson}, B., {et~al.} 2024, \apj, 964, 71

\bibitem[{{Harvey} {et~al.}(2025){Harvey}, {Conselice}, {Adams}, {Austin}, {Li}, {Rusakov}, {Westcott}, {Goolsby}, {Lovell}, {Cochrane}, {Vijayan}, \& {Trussler}}]{harvey2025}
{Harvey}, T., {Conselice}, C.~J., {Adams}, N.~J., {et~al.} 2025, \mnras, 542, 2998

\bibitem[{{Heavens} {et~al.}(2000){Heavens}, {Jimenez}, \& {Lahav}}]{heavens2000}
{Heavens}, A.~F., {Jimenez}, R., \& {Lahav}, O. 2000, \mnras, 317, 965

\bibitem[{{Iglesias-Navarro} {et~al.}(2024){Iglesias-Navarro}, {Huertas-Company}, {Mart{\'\i}n-Navarro}, {Knapen}, \& {Pernet}}]{iglesias2024}
{Iglesias-Navarro}, P., {Huertas-Company}, M., {Mart{\'\i}n-Navarro}, I., {Knapen}, J.~H., \& {Pernet}, E. 2024, \aap, 689, A58

\bibitem[{{Illingworth} {et~al.}(2013){Illingworth}, {Magee}, {Oesch}, {Bouwens}, {Labb{\'e}}, {Stiavelli}, {van Dokkum}, {Franx}, {Trenti}, {Carollo}, \& {Gonzalez}}]{illingworth2013}
{Illingworth}, G.~D., {Magee}, D., {Oesch}, P.~A., {et~al.} 2013, \apjs, 209, 6

\bibitem[{{Iyer} \& {Gawiser}(2017)}]{Iyer2017}
{Iyer}, K. \& {Gawiser}, E. 2017, \apj, 838, 127

\bibitem[{{Iyer} {et~al.}(2019){Iyer}, {Gawiser}, {Faber}, {Ferguson}, {Kartaltepe}, {Koekemoer}, {Pacifici}, \& {Somerville}}]{Iyer19}
{Iyer}, K.~G., {Gawiser}, E., {Faber}, S.~M., {et~al.} 2019, \apj, 879, 116

\bibitem[{{Iyer} {et~al.}(2025){Iyer}, {Pacifici}, {Calistro-Rivera}, \& {Lovell}}]{iyer2025spectralenergydistributionsgalaxies}
{Iyer}, K.~G., {Pacifici}, C., {Calistro-Rivera}, G., \& {Lovell}, C.~C. 2025, arXiv e-prints, arXiv:2502.17680

\bibitem[{Iyer {et~al.}(2024)Iyer, Speagle, Caplar, Forbes, Gawiser, Leja, \& Tacchella}]{Iyer_2024}
Iyer, K.~G., Speagle, J.~S., Caplar, N., {et~al.} 2024, \apj, 961, 53

\bibitem[{{Jain} {et~al.}(2024){Jain}, {Tacchella}, \& {Mosleh}}]{jain2023}
{Jain}, S., {Tacchella}, S., \& {Mosleh}, M. 2024, \mnras, 527, 3291

\bibitem[{{Jimenez Rezende} \& {Mohamed}(2015)}]{rezende2015}
{Jimenez Rezende}, D. \& {Mohamed}, S. 2015, arXiv e-prints, arXiv:1505.05770

\bibitem[{{Johnson} {et~al.}(2021{\natexlab{a}}){Johnson}, {Leja}, {Conroy}, \& {Speagle}}]{Johnson2021}
{Johnson}, B.~D., {Leja}, J., {Conroy}, C., \& {Speagle}, J.~S. 2021{\natexlab{a}}, \apjs, 254, 22

\bibitem[{{Johnson} {et~al.}(2021{\natexlab{b}}){Johnson}, {Leja}, {Conroy}, \& {Speagle}}]{prospector}
{Johnson}, B.~D., {Leja}, J., {Conroy}, C., \& {Speagle}, J.~S. 2021{\natexlab{b}}, \apjs, 254, 22

\bibitem[{{Khullar} {et~al.}(2022){Khullar}, {Nord}, {{\'C}iprijanovi{\'c}}, {Poh}, \& {Xu}}]{2022Khullar}
{Khullar}, G., {Nord}, B., {{\'C}iprijanovi{\'c}}, A., {Poh}, J., \& {Xu}, F. 2022, Machine Learning: Science and Technology, 3, 04LT04

\bibitem[{{Kingma} \& {Ba}(2014)}]{kingma2017}
{Kingma}, D.~P. \& {Ba}, J. 2014, arXiv e-prints, arXiv:1412.6980

\bibitem[{{Koekemoer} {et~al.}(2013){Koekemoer}, {Ellis}, {McLure}, {Dunlop}, {Robertson}, {Ono}, {Schenker}, {Ouchi}, {Bowler}, {Rogers}, {Curtis-Lake}, {Schneider}, {Charlot}, {Stark}, {Furlanetto}, {Cirasuolo}, {Wild}, \& {Targett}}]{koekemoer2013}
{Koekemoer}, A.~M., {Ellis}, R.~S., {McLure}, R.~J., {et~al.} 2013, \apjs, 209, 3

\bibitem[{{Koekemoer} {et~al.}(2011){Koekemoer}, {Faber}, {Ferguson}, {Grogin}, {Kocevski}, {Koo}, {Lai}, {Lotz}, {Lucas}, {McGrath}, {Ogaz}, {Rajan}, {Riess}, {Rodney}, {Strolger}, {Casertano}, {Castellano}, {Dahlen}, {Dickinson}, {Dolch}, {Fontana}, {Giavalisco}, {Grazian}, {Guo}, {Hathi}, {Huang}, {van der Wel}, {Yan}, {Acquaviva}, {Alexander}, {Almaini}, {Ashby}, {Barden}, {Bell}, {Bournaud}, {Brown}, {Caputi}, {Cassata}, {Challis}, {Chary}, {Cheung}, {Cirasuolo}, {Conselice}, {Roshan Cooray}, {Croton}, {Daddi}, {Dav{\'e}}, {de Mello}, {de Ravel}, {Dekel}, {Donley}, {Dunlop}, {Dutton}, {Elbaz}, {Fazio}, {Filippenko}, {Finkelstein}, {Frazer}, {Gardner}, {Garnavich}, {Gawiser}, {Gruetzbauch}, {Hartley}, {H{\"a}ussler}, {Herrington}, {Hopkins}, {Huang}, {Jha}, {Johnson}, {Kartaltepe}, {Khostovan}, {Kirshner}, {Lani}, {Lee}, {Li}, {Madau}, {McCarthy}, {McIntosh}, {McLure}, {McPartland}, {Mobasher}, {Moreira}, {Mortlock}, {Moustakas}, {Mozena}, {Nandra}, {Newman}, {Nielsen}, {Niemi}, {Noeske}, {Papovich}, {Pentericci}, {Pope}, {Primack}, {Ravindranath}, {Reddy}, {Renzini}, {Rix}, {Robaina}, {Rosario}, {Rosati}, {Salimbeni}, {Scarlata}, {Siana}, {Simard}, {Smidt}, {Snyder}, {Somerville}, {Spinrad}, {Straughn}, {Telford}, {Teplitz}, {Trump}, {Vargas}, {Villforth}, {Wagner}, {Wandro}, {Wechsler}, {Weiner}, {Wiklind}, {Wild}, {Wilson}, {Wuyts}, \& {Yun}}]{koekemoer2011}
{Koekemoer}, A.~M., {Faber}, S.~M., {Ferguson}, H.~C., {et~al.} 2011, \apjs, 197, 36

\bibitem[{{Kwon} \& {Hahn}(2024)}]{kwon2024}
{Kwon}, K.~J. \& {Hahn}, C. 2024, \apj, 976, 76

\bibitem[{{Lanyon-Foster} {et~al.}(2012){Lanyon-Foster}, {Conselice}, \& {Merrifield}}]{LanyonFoster2012}
{Lanyon-Foster}, M.~M., {Conselice}, C.~J., \& {Merrifield}, M.~R. 2012, \mnras, 424, 1852

\bibitem[{Leja {et~al.}(2019)Leja, Carnall, Johnson, Conroy, \& Speagle}]{Leja2019}
Leja, J., Carnall, A.~C., Johnson, B.~D., Conroy, C., \& Speagle, J.~S. 2019, \apj, 876, 3

\bibitem[{Li {et~al.}(2023)Li, Melchior, Hahn, \& Huang}]{Li2023}
Li, J., Melchior, P., Hahn, C., \& Huang, S. 2023, \aj, 167, 16

\bibitem[{{Lines} {et~al.}(2025){Lines}, {Bowler}, {Adams}, {Fisher}, {Varadaraj}, {Nakazato}, {Aravena}, {Assef}, {Birkin}, {Ceverino}, {da Cunha}, {Cullen}, {De Looze}, {Donnan}, {Dunlop}, {Ferrara}, {Grogin}, {Herrera-Camus}, {Ikeda}, {Koekemoer}, {Killi}, {Li}, {McLeod}, {McLure}, {Mitsuhashi}, {P{\'e}rez-Gonz{\'a}lez}, {Relano}, {Solimano}, {Spilker}, {Villanueva}, \& {Yoshida}}]{Lines2024}
{Lines}, N.~E.~P., {Bowler}, R.~A.~A., {Adams}, N.~J., {et~al.} 2025, \mnras, 539, 2685

\bibitem[{{Lower} {et~al.}(2020){Lower}, {Narayanan}, {Leja}, {Johnson}, {Conroy}, \& {Dav{\'e}}}]{LOWER20}
{Lower}, S., {Narayanan}, D., {Leja}, J., {et~al.} 2020, \apj, 904, 33

\bibitem[{{Lupton} {et~al.}(2004){Lupton}, {Blanton}, {Fekete}, {Hogg}, {O'Mullane}, {Szalay}, \& {Wherry}}]{lupton04}
{Lupton}, R., {Blanton}, M.~R., {Fekete}, G., {et~al.} 2004, \pasp, 116, 133

\bibitem[{{Marin} {et~al.}(2011){Marin}, {Pudlo}, {Robert}, \& {Ryder}}]{marin2011}
{Marin}, J.-M., {Pudlo}, P., {Robert}, C.~P., \& {Ryder}, R. 2011, arXiv e-prints, arXiv:1101.0955

\bibitem[{{Mosleh} {et~al.}(2025){Mosleh}, {Riahi-Zamin}, \& {Tacchella}}]{Mosleh2025}
{Mosleh}, M., {Riahi-Zamin}, M., \& {Tacchella}, S. 2025, \apj, 983, 181

\bibitem[{{Narayanan} {et~al.}(2024){Narayanan}, {Lower}, {Torrey}, {Brammer}, {Cui}, {Dav{\'e}}, {Iyer}, {Li}, {Lovell}, {Sales}, {Stark}, {Marinacci}, \& {Vogelsberger}}]{Narayanan24}
{Narayanan}, D., {Lower}, S., {Torrey}, P., {et~al.} 2024, \apj, 961, 73

\bibitem[{{Nersesian} {et~al.}(2025){Nersesian}, {van der Wel}, {Gallazzi}, {Kaushal}, {Bezanson}, {Zibetti}, {Bell}, {D'Eugenio}, {Leja}, {Martorano}, \& {Wu}}]{nersesian25}
{Nersesian}, A., {van der Wel}, A., {Gallazzi}, A.~R., {et~al.} 2025, \aap, 695, A86

\bibitem[{{Oesch} {et~al.}(2023){Oesch}, {Brammer}, {Naidu}, {Bouwens}, {Chisholm}, {Illingworth}, {Matthee}, {Nelson}, {Qin}, {Reddy}, {Shapley}, {Shivaei}, {van Dokkum}, {Weibel}, {Whitaker}, {Wuyts}, {Covelo-Paz}, {Endsley}, {Fudamoto}, {Giovinazzo}, {Herard-Demanche}, {Kerutt}, {Kramarenko}, {Labbe}, {Leonova}, {Lin}, {Magee}, {Marchesini}, {Maseda}, {Mason}, {Matharu}, {Meyer}, {Neufeld}, {Prieto Lyon}, {Schaerer}, {Sharma}, {Shuntov}, {Smit}, {Stefanon}, {Wyithe}, \& {Xiao}}]{Oesch2023}
{Oesch}, P.~A., {Brammer}, G., {Naidu}, R.~P., {et~al.} 2023, \mnras, 525, 2864

\bibitem[{{Oke} \& {Gunn}(1983)}]{oke1983}
{Oke}, J.~B. \& {Gunn}, J.~E. 1983, \apj, 266, 713

\bibitem[{{Pacifici} {et~al.}(2012){Pacifici}, {Charlot}, {Blaizot}, \& {Brinchmann}}]{pacifici12}
{Pacifici}, C., {Charlot}, S., {Blaizot}, J., \& {Brinchmann}, J. 2012, \mnras, 421, 2002

\bibitem[{{Pacifici} {et~al.}(2023){Pacifici}, {Iyer}, {Mobasher}, {da Cunha}, {Acquaviva}, {Burgarella}, {Calistro Rivera}, {Carnall}, {Chang}, {Chartab}, {Cooke}, {Fairhurst}, {Kartaltepe}, {Leja}, {Ma{\l}ek}, {Salmon}, {Torelli}, {Vidal-Garc{\'\i}a}, {Boquien}, {Brammer}, {Brown}, {Capak}, {Chevallard}, {Circosta}, {Croton}, {Davidzon}, {Dickinson}, {Duncan}, {Faber}, {Ferguson}, {Fontana}, {Guo}, {Haeussler}, {Hemmati}, {Jafariyazani}, {Kassin}, {Larson}, {Lee}, {Mantha}, {Marchi}, {Nayyeri}, {Newman}, {Pandya}, {Pforr}, {Reddy}, {Sanders}, {Shah}, {Shahidi}, {Stevans}, {Triani}, {Tyler}, {Vanderhoof}, {de la Vega}, {Wang}, \& {Weston}}]{pacifi2023}
{Pacifici}, C., {Iyer}, K.~G., {Mobasher}, B., {et~al.} 2023, \apj, 944, 141

\bibitem[{{Papamakarios} {et~al.}(2017){Papamakarios}, {Pavlakou}, \& {Murray}}]{papamakarios2017}
{Papamakarios}, G., {Pavlakou}, T., \& {Murray}, I. 2017, arXiv e-prints, arXiv:1705.07057

\bibitem[{{P{\'e}rez-Gonz{\'a}lez} {et~al.}(2003){P{\'e}rez-Gonz{\'a}lez}, {Gil de Paz}, {Zamorano}, {Gallego}, {Alonso-Herrero}, \& {Arag{\'o}n-Salamanca}}]{perez2003}
{P{\'e}rez-Gonz{\'a}lez}, P.~G., {Gil de Paz}, A., {Zamorano}, J., {et~al.} 2003, \mnras, 338, 508

\bibitem[{{P{\'e}rez-Gonz{\'a}lez} {et~al.}(2008){P{\'e}rez-Gonz{\'a}lez}, {Rieke}, {Villar}, {Barro}, {Blaylock}, {Egami}, {Gallego}, {Gil de Paz}, {Pascual}, {Zamorano}, \& {Donley}}]{perez2008}
{P{\'e}rez-Gonz{\'a}lez}, P.~G., {Rieke}, G.~H., {Villar}, V., {et~al.} 2008, \apj, 675, 234

\bibitem[{{Pirzkal} {et~al.}(2005){Pirzkal}, {Sahu}, {Burgasser}, {Moustakas}, {Xu}, {Malhotra}, {Rhoads}, {Koekemoer}, {Nelan}, {Windhorst}, {Panagia}, {Gronwall}, {Pasquali}, \& {Walsh}}]{Pirzkal2005}
{Pirzkal}, N., {Sahu}, K.~C., {Burgasser}, A., {et~al.} 2005, \apj, 622, 319

\bibitem[{Pérez-González {et~al.}(2023)Pérez-González, Barro, Annunziatella, Costantin, Ángela García-Argumánez, McGrath, Mérida, Zavala, Haro, Bagley, Backhaus, Behroozi, Bell, Bisigello, Buat, Calabrò, Casey, Cleri, Coogan, Cooper, Cooray, Dekel, Dickinson, Elbaz, Ferguson, Finkelstein, Fontana, Franco, Gardner, Giavalisco, Gómez-Guijarro, Grazian, Grogin, Guo, Huertas-Company, Jogee, Kartaltepe, Kewley, Kirkpatrick, Kocevski, Koekemoer, Long, Lotz, Lucas, Papovich, Pirzkal, Ravindranath, Somerville, Tacchella, Trump, Wang, Wilkins, Wuyts, Yang, \& Yung}]{Perez-Gonzalez2023}
Pérez-González, P.~G., Barro, G., Annunziatella, M., {et~al.} 2023, \apjl, 946, L16

\bibitem[{{Rieke} {et~al.}(2023{\natexlab{a}}){Rieke}, {Kelly}, {Misselt}, {Stansberry}, {Boyer}, {Beatty}, {Egami}, {Florian}, {Greene}, {Hainline}, {Leisenring}, {Roellig}, {Schlawin}, {Sun}, {Tinnin}, {Williams}, {Willmer}, {Wilson}, {Clark}, {Rohrbach}, {Brooks}, {Canipe}, {Correnti}, {DiFelice}, {Gennaro}, {Girard}, {Hartig}, {Hilbert}, {Koekemoer}, {Nikolov}, {Pirzkal}, {Rest}, {Robberto}, {Sunnquist}, {Telfer}, {Wu}, {Ferry}, {Lewis}, {Baum}, {Beichman}, {Doyon}, {Dressler}, {Eisenstein}, {Ferrarese}, {Hodapp}, {Horner}, {Jaffe}, {Johnstone}, {Krist}, {Martin}, {McCarthy}, {Meyer}, {Rieke}, {Trauger}, \& {Young}}]{rieke2023a}
{Rieke}, M.~J., {Kelly}, D.~M., {Misselt}, K., {et~al.} 2023{\natexlab{a}}, \pasp, 135, 028001

\bibitem[{{Rieke} {et~al.}(2023{\natexlab{b}}){Rieke}, {Robertson}, {Tacchella}, {Hainline}, {Johnson}, {Hausen}, {Ji}, {Willmer}, {Eisenstein}, {Pusk{\'a}s}, {Alberts}, {Arribas}, {Baker}, {Baum}, {Bhatawdekar}, {Bonaventura}, {Boyett}, {Bunker}, {Cameron}, {Carniani}, {Charlot}, {Chevallard}, {Chen}, {Curti}, {Curtis-Lake}, {Danhaive}, {DeCoursey}, {Dressler}, {Egami}, {Endsley}, {Helton}, {Hviding}, {Kumari}, {Looser}, {Lyu}, {Maiolino}, {Maseda}, {Nelson}, {Rieke}, {Rix}, {Sandles}, {Saxena}, {Sharpe}, {Shivaei}, {Skarbinski}, {Smit}, {Stark}, {Stone}, {Suess}, {Sun}, {Topping}, {{\"U}bler}, {Villanueva}, {Wallace}, {Williams}, {Willott}, {Whitler}, {Witstok}, \& {Woodrum}}]{Rieke2023}
{Rieke}, M.~J., {Robertson}, B., {Tacchella}, S., {et~al.} 2023{\natexlab{b}}, \apjs, 269, 16

\bibitem[{{Ryon}(2023)}]{ryon2023}
{Ryon}, J.~E. 2023, in ACS Instrument Handbook for Cycle 31 v. 22.0, Vol.~22 (STScI), 22

\bibitem[{{Sawicki} \& {Yee}(1998)}]{sawicki98}
{Sawicki}, M. \& {Yee}, H.~K.~C. 1998, \aj, 115, 1329

\bibitem[{{Smith} \& {Hayward}(2018)}]{Smith2018}
{Smith}, D. J.~B. \& {Hayward}, C.~C. 2018, \mnras, 476, 1705

\bibitem[{Song {et~al.}(2023)Song, Fang, Lin, Gu, \& Kong}]{Song2023}
Song, J., Fang, G., Lin, Z., Gu, Y., \& Kong, X. 2023, \apj, 958, 82

\bibitem[{{Sorba} \& {Sawicki}(2015)}]{Sorba2015}
{Sorba}, R. \& {Sawicki}, M. 2015, \mnras, 452, 235

\bibitem[{{Sorba} \& {Sawicki}(2018)}]{Sorba2018}
{Sorba}, R. \& {Sawicki}, M. 2018, \mnras, 476, 1532

\bibitem[{{Speagle}(2020)}]{dynesty}
{Speagle}, J.~S. 2020, \mnras, 493, 3132

\bibitem[{{Tacchella} {et~al.}(2022){Tacchella}, {Conroy}, {Faber}, {Johnson}, {Leja}, {Barro}, {Cunningham}, {Deason}, {Guhathakurta}, {Guo}, {Hernquist}, {Koo}, {McKinnon}, {Rockosi}, {Speagle}, {van Dokkum}, \& {Yesuf}}]{Tacchella2021}
{Tacchella}, S., {Conroy}, C., {Faber}, S.~M., {et~al.} 2022, \apj, 926, 134

\bibitem[{Tacchella {et~al.}(2020)Tacchella, Forbes, \& Caplar}]{Tacchella_2020}
Tacchella, S., Forbes, J.~C., \& Caplar, N. 2020, \mnras, 497, 698–725

\bibitem[{{Talts} {et~al.}(2018){Talts}, {Betancourt}, {Simpson}, {Vehtari}, \& {Gelman}}]{TALTS2018}
{Talts}, S., {Betancourt}, M., {Simpson}, D., {Vehtari}, A., \& {Gelman}, A. 2018, arXiv e-prints, arXiv:1804.06788

\bibitem[{{Tejero-Cantero} {et~al.}(2020){Tejero-Cantero}, {Boelts}, {Deistler}, {Lueckmann}, {Durkan}, {Gon{\c{c}}alves}, {Greenberg}, \& {Macke}}]{tejero2020}
{Tejero-Cantero}, A., {Boelts}, J., {Deistler}, M., {et~al.} 2020, {JOSS}, 5, 2505

\bibitem[{{Tojeiro} {et~al.}(2007){Tojeiro}, {Heavens}, {Jimenez}, \& {Panter}}]{tojeiro07}
{Tojeiro}, R., {Heavens}, A.~F., {Jimenez}, R., \& {Panter}, B. 2007, \mnras, 381, 1252

\bibitem[{{Vazdekis} {et~al.}(2010){Vazdekis}, {S{\'a}nchez-Bl{\'a}zquez}, {Falc{\'o}n-Barroso}, {Cenarro}, {Beasley}, {Cardiel}, {Gorgas}, \& {Peletier}}]{vazdekis2010}
{Vazdekis}, A., {S{\'a}nchez-Bl{\'a}zquez}, P., {Falc{\'o}n-Barroso}, J., {et~al.} 2010, \mnras, 404, 1639

\bibitem[{{Villaume} {et~al.}(2015){Villaume}, {Conroy}, \& {Johnson}}]{Villaume2015}
{Villaume}, A., {Conroy}, C., \& {Johnson}, B.~D. 2015, \apj, 806, 82

\bibitem[{{Wan} {et~al.}(2024){Wan}, {Tacchella}, {Johnson}, {Iyer}, {Speagle}, \& {Maiolino}}]{wan2024}
{Wan}, J.~T., {Tacchella}, S., {Johnson}, B.~D., {et~al.} 2024, \mnras, 532, 4002

\bibitem[{{Wang} {et~al.}(2025){Wang}, {Leja}, {Atek}, {Bezanson}, {Burnham}, {Dayal}, {Feldmann}, {Greene}, {Johnson}, {Labb{\'e}}, {Maseda}, {Nanayakkara}, {Price}, {Suess}, {Weaver}, \& {Whitaker}}]{wang2025}
{Wang}, B., {Leja}, J., {Atek}, H., {et~al.} 2025, \apj, 987, 184

\bibitem[{{Watson} {et~al.}(2025){Watson}, {Vulcani}, {Werle}, {Poggianti}, {Gullieuszik}, {Trenti}, {Wang}, \& {Roy}}]{Watson2024}
{Watson}, P.~J., {Vulcani}, B., {Werle}, A., {et~al.} 2025, \aap, 699, A365

\bibitem[{{Williams} {et~al.}(2023){Williams}, {Tacchella}, {Maseda}, {Robertson}, {Johnson}, {Willott}, {Eisenstein}, {Willmer}, {Ji}, {Hainline}, {Helton}, {Alberts}, {Baum}, {Bhatawdekar}, {Boyett}, {Bunker}, {Carniani}, {Charlot}, {Chevallard}, {Curtis-Lake}, {de Graaff}, {Egami}, {Franx}, {Kumari}, {Maiolino}, {Nelson}, {Rieke}, {Sandles}, {Shivaei}, {Simmonds}, {Smit}, {Suess}, {Sun}, {{\"U}bler}, \& {Witstok}}]{williams2023}
{Williams}, C.~C., {Tacchella}, S., {Maseda}, M.~V., {et~al.} 2023, \apjs, 268, 64

\bibitem[{{Wuyts} {et~al.}(2012){Wuyts}, {F{\"o}rster Schreiber}, {Genzel}, {Guo}, {Barro}, {Bell}, {Dekel}, {Faber}, {Ferguson}, {Giavalisco}, {Grogin}, {Hathi}, {Huang}, {Kocevski}, {Koekemoer}, {Koo}, {Lotz}, {Lutz}, {McGrath}, {Newman}, {Rosario}, {Saintonge}, {Tacconi}, {Weiner}, \& {van der Wel}}]{Wuyts2012}
{Wuyts}, S., {F{\"o}rster Schreiber}, N.~M., {Genzel}, R., {et~al.} 2012, \apj, 753, 114

\bibitem[{{Zibetti} {et~al.}(2009){Zibetti}, {Charlot}, \& {Rix}}]{zibetti2009}
{Zibetti}, S., {Charlot}, S., \& {Rix}, H.-W. 2009, \mnras, 400, 1181

\end{thebibliography}

\begin{appendix}

\onecolumn
\section{Posterior distributions using different SFH priors for a simulated galaxy}

\label{app_corner}

\begin{figure}[h!]
    \centering
    \includegraphics[width=\linewidth]{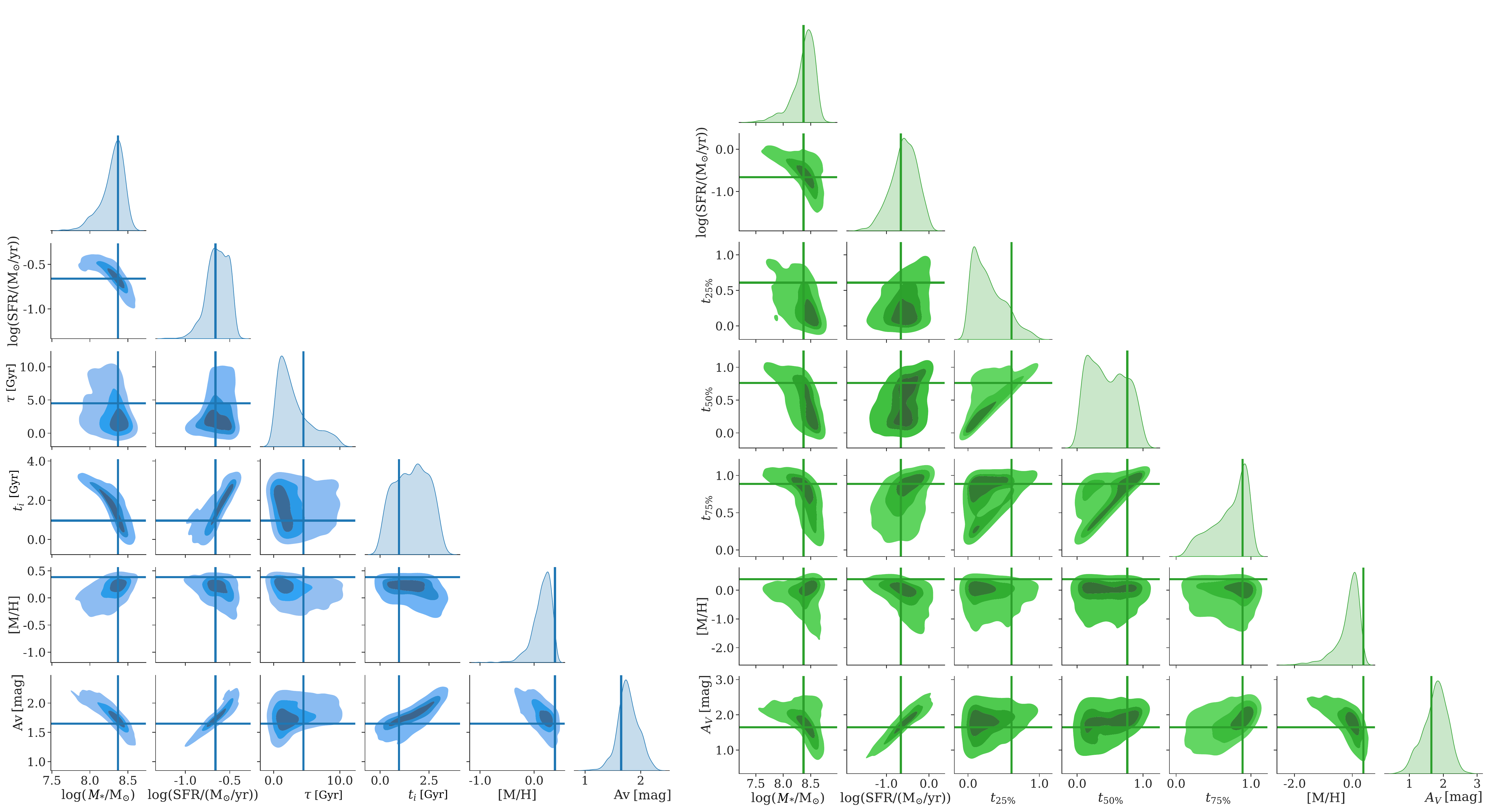}
    \caption{Corner plots showing the posterior distributions for a simulated galaxy (using a $\tau$-delayed SFH) at redshift $1.89$, obtained with the $\tau$-delayed model (left) and Dirichlet model (right). Each posterior uses 500 samples. Kernel density estimation contours are shown with different shades at iso-proportions of the density of samples. The solid lines correspond to the true values.}
    \label{corner}
\end{figure}

We include in Fig.~\ref{corner} two corner plots showing the posterior distributions obtained for a single simulated galaxy (using a $\tau$-delayed simulation prior) when inference is performed with the model trained on the $\tau$-delayed simulation (left) and the model trained on the Dirichlet simulation (right). The posterior distributions are almost identical for the two models, although the distributions are slightly wider for the model trained on the Dirichlet prior. We can also see the correlations and anticorrelations that exist between the different properties. On the left we find positive correlations between $\log_{10}(M_{*}/\rm{M}_{\odot})$ and $\log_{10}$(SFR/($\rm{M}_{\odot}$ yr$^{-1}$)), $\log_{10}$(SFR/($\rm{M}_{\odot}$ yr$^{-1}$)) and $A_{V}$, and $t_i$ (cosmic time) and [$M$/H], as well as negative correlations between $\log_{10}(M_{*}/\rm{M}_{\odot})$ and [$M$/H], and [$M$/H] and $A_{V}$. On the right, these are less obvious because the distributions are wider, but we observe positive correlations between different $t_x$ (cosmic time normalised to the age of the Universe at the redshift of the source), and negative correlations between [$M$/H] and $A_{V}$. \\

\FloatBarrier   

\section{Simulation-based calibration}
\label{sbc}

\begin{figure}[h!]
    \centering
    \includegraphics[width=0.2\linewidth]{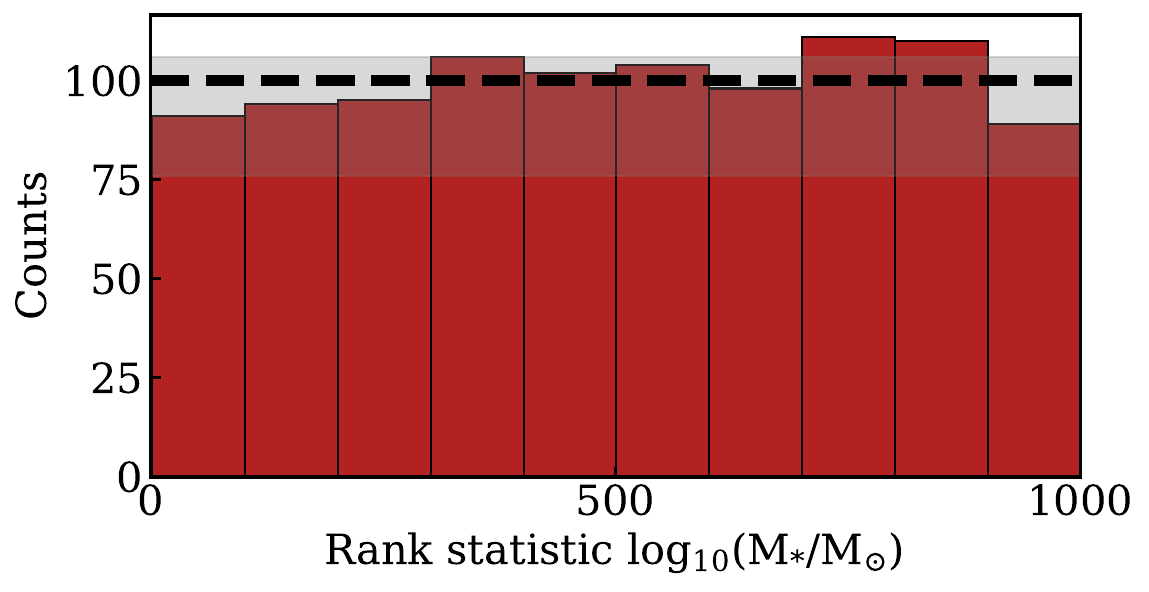}
    \includegraphics[width=0.2\linewidth]{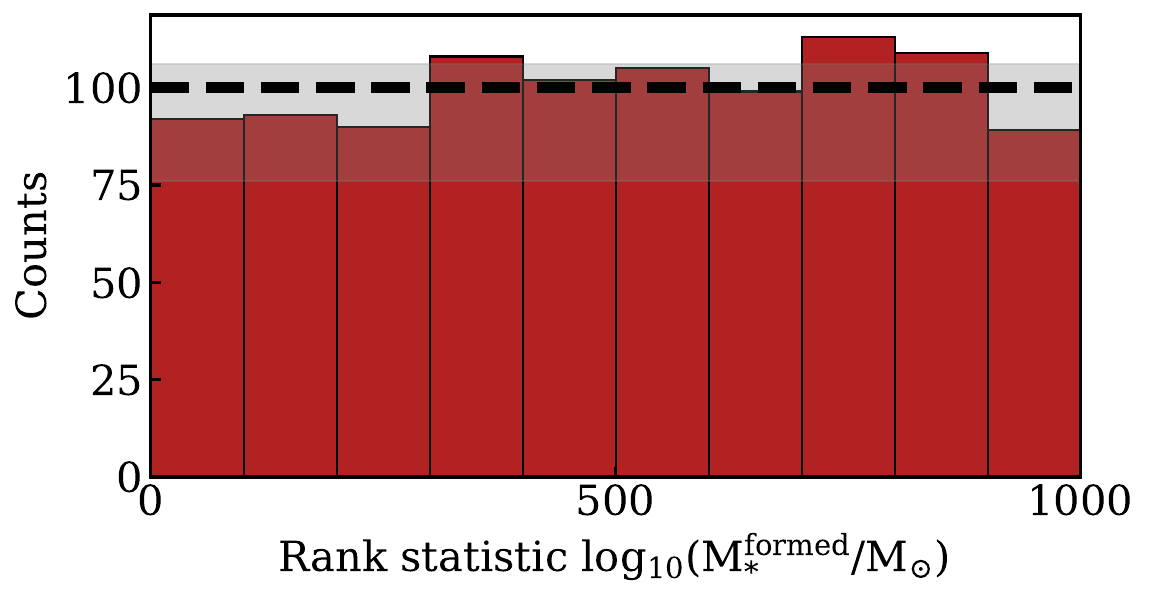}
    \includegraphics[width=0.2\linewidth]{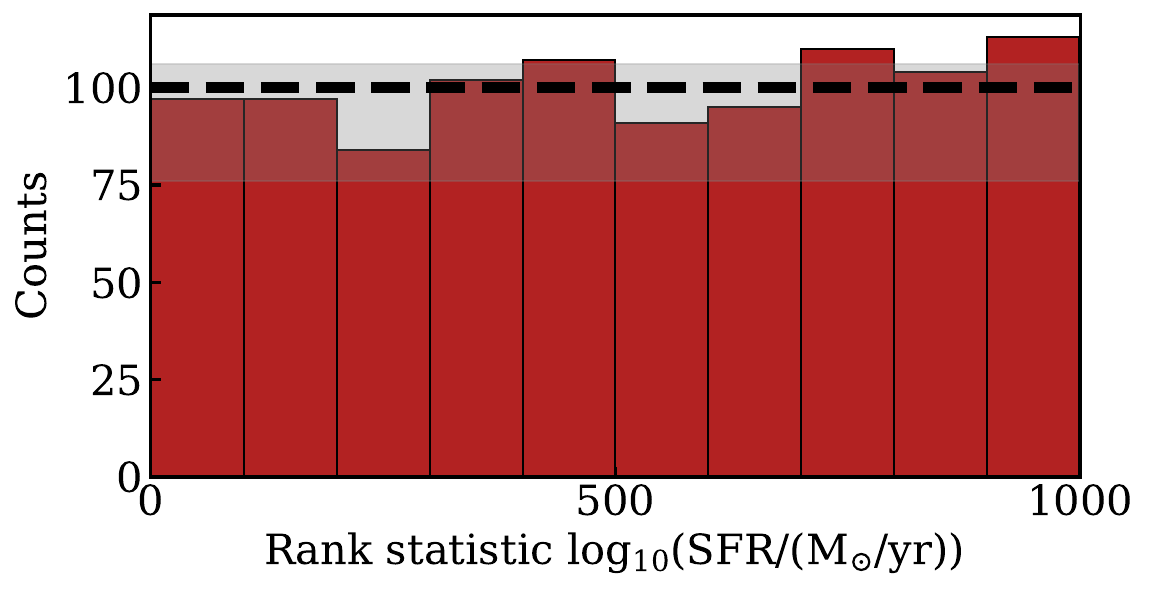}
    \includegraphics[width=0.2\linewidth]{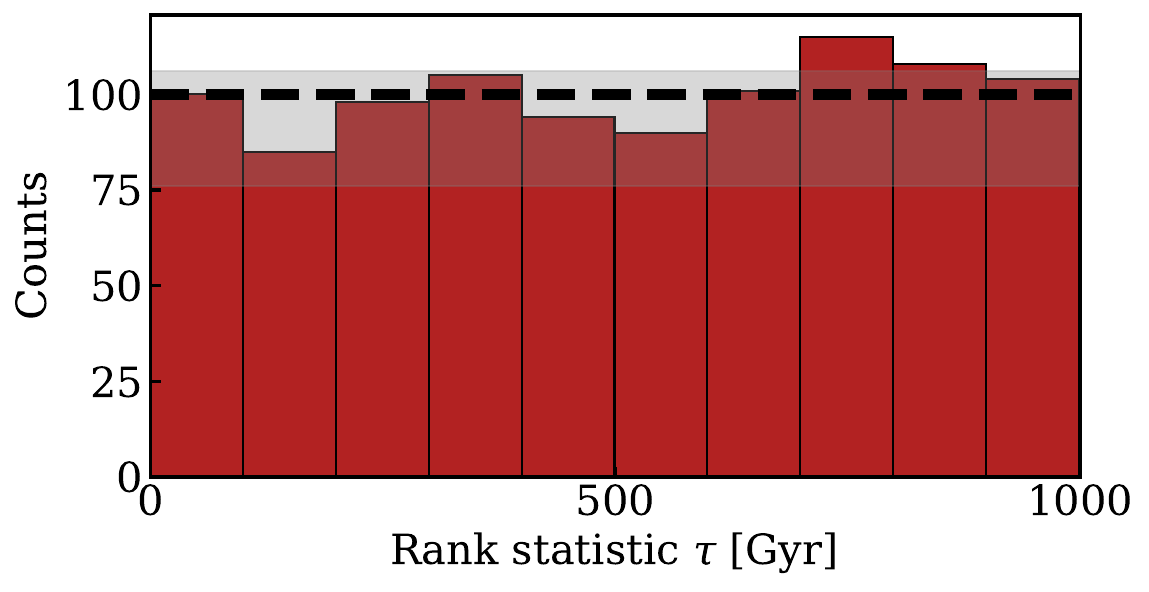}
    \includegraphics[width=0.2\linewidth]{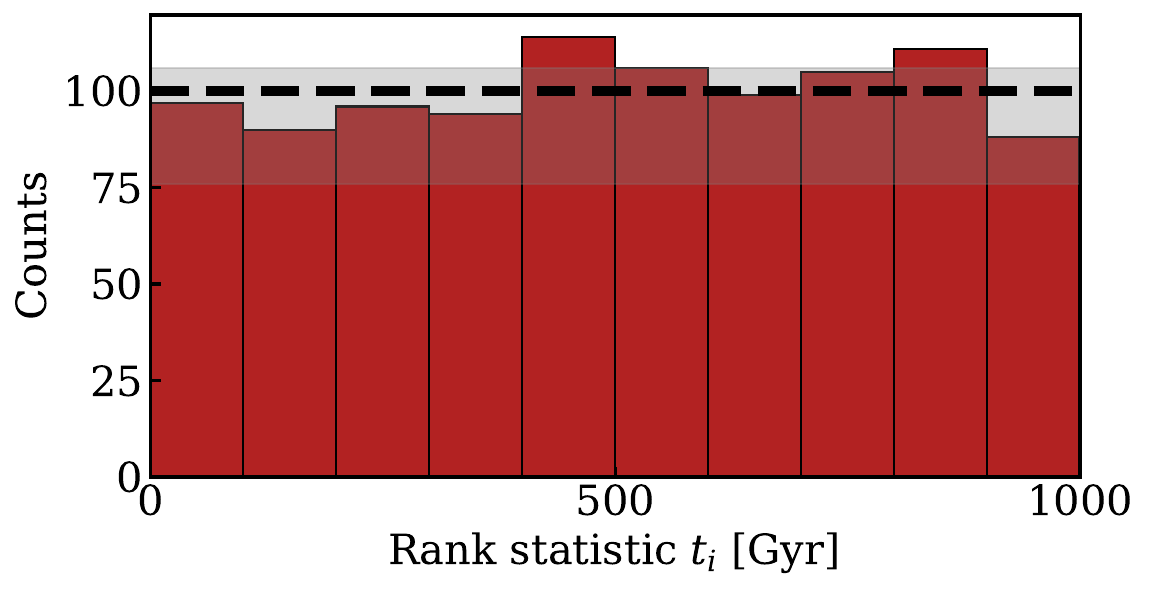}\includegraphics[width=0.2\linewidth]{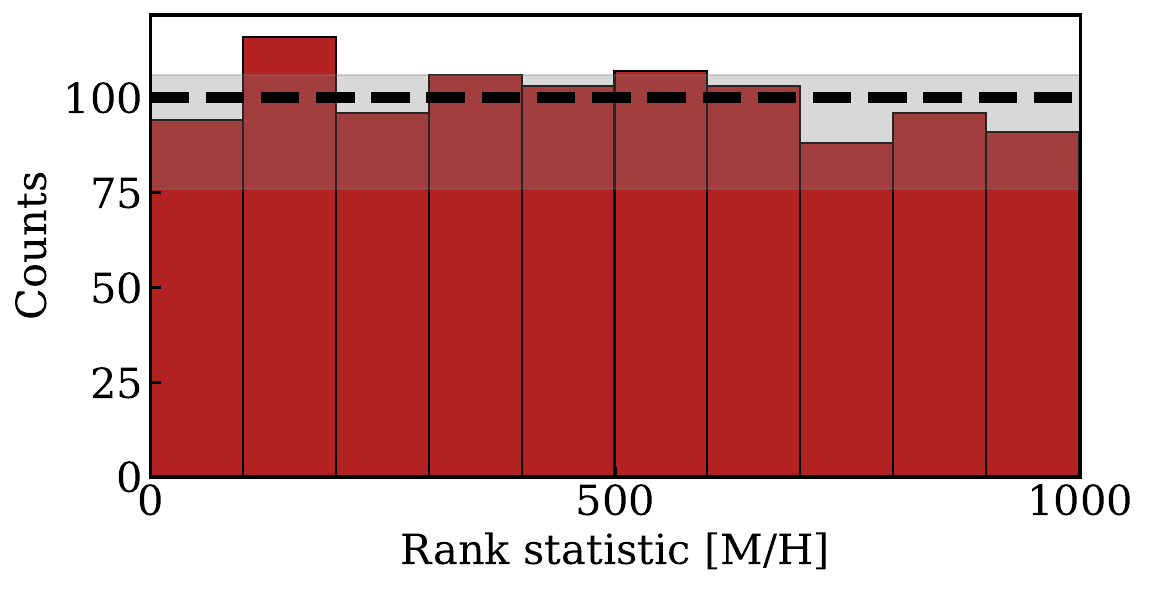}
    \includegraphics[width=0.2\linewidth]{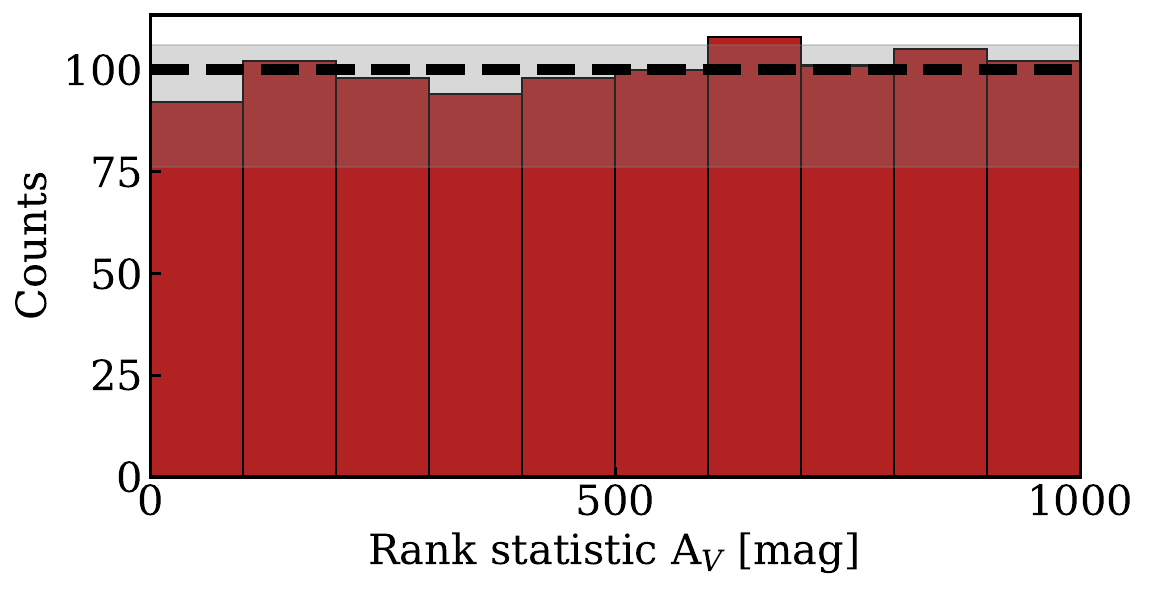} 
    \caption{SBC test of the posteriors for $1.000$ simulations using a $\tau$-delayed prior for the SFHs. The histograms show the distribution of the rank statistic of the true value within the marginalised posterior for all the  properties. For a well-calibrated posterior, the rank statistics should have a uniform distribution (dashed black line). The grey band indicates $99\%$ of the variation expected from a uniform histogram. The rank statistic distributions of our posteriors are nearly uniform for all the parameters, and therefore, the model provides unbiased and accurate posteriors.}
    \label{sbc_ranks_tau}
\end{figure}

\begin{figure}[h!]
    \centering
    \includegraphics[width=0.2\linewidth]{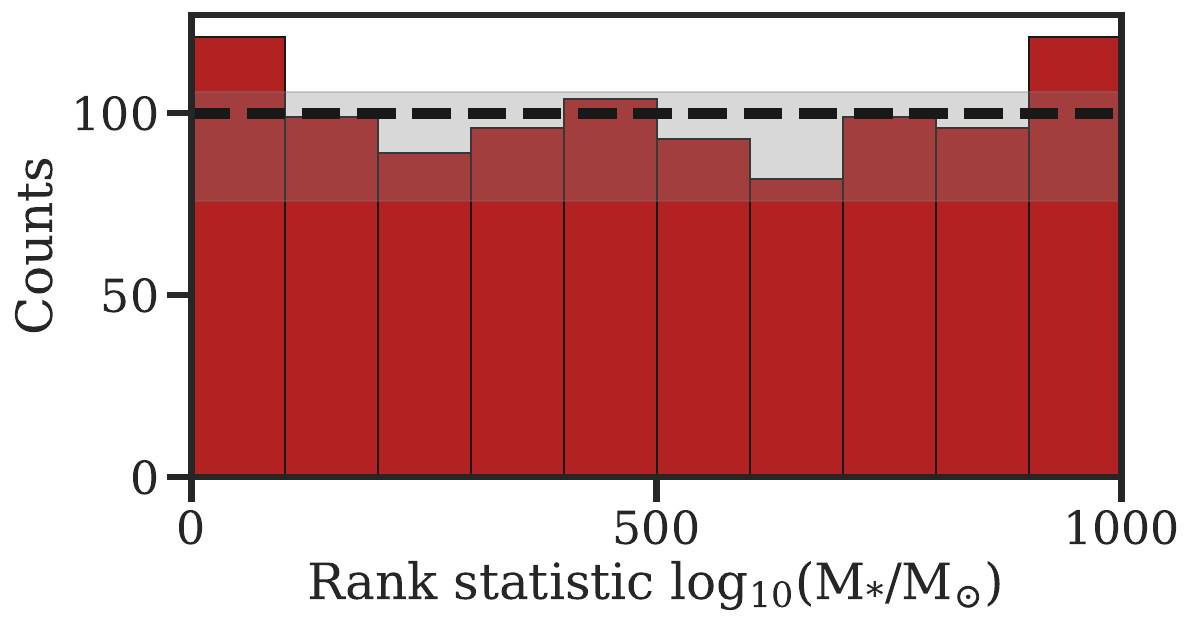} \includegraphics[width=0.2\linewidth]{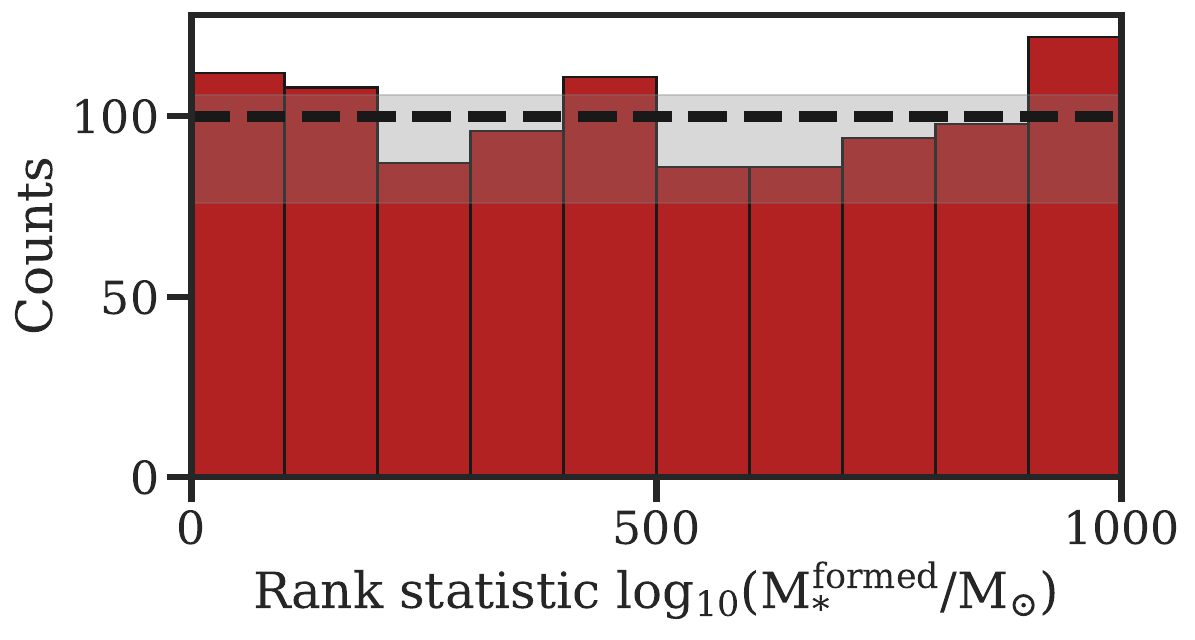}
    \includegraphics[width=0.2\linewidth]{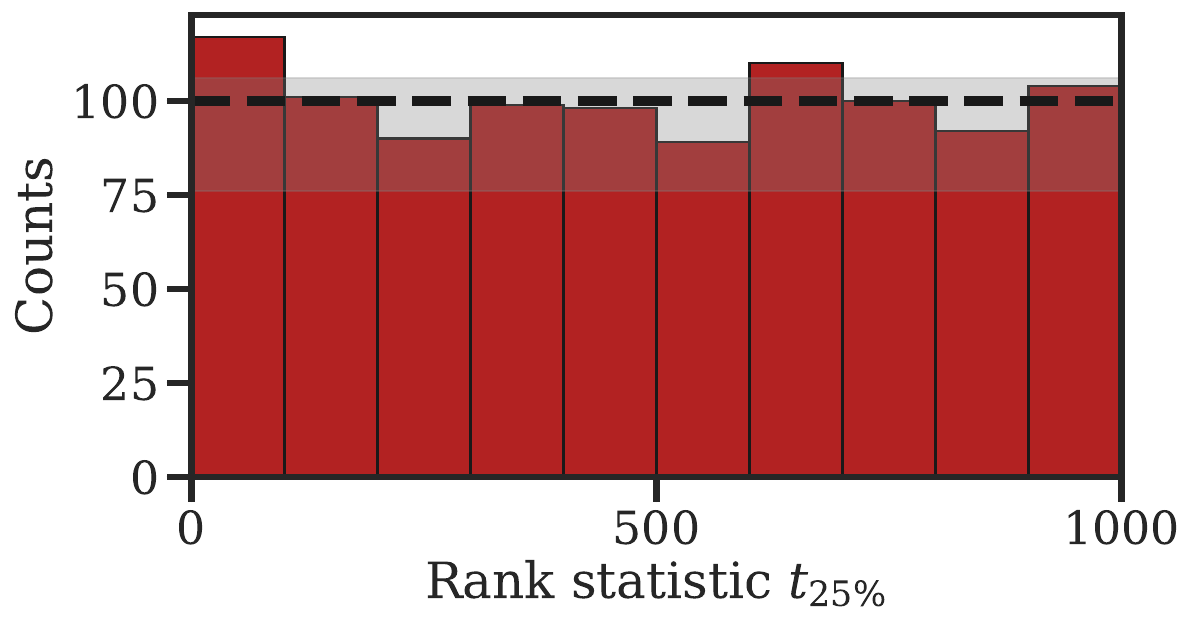}
    \includegraphics[width=0.2\linewidth]{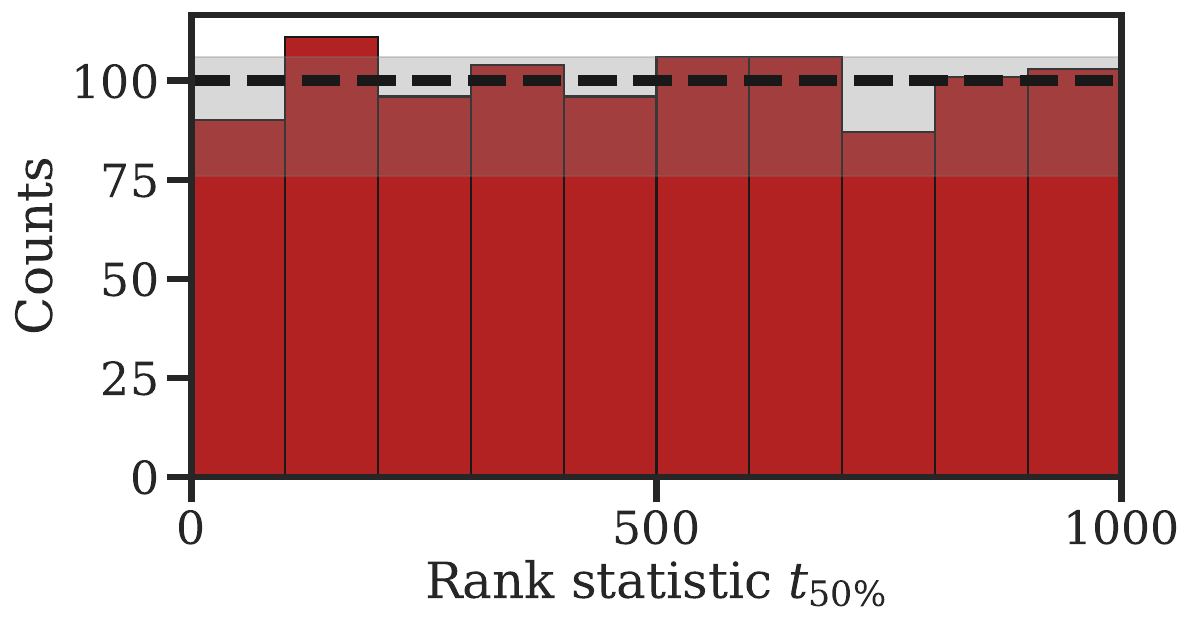}
    \includegraphics[width=0.2\linewidth]{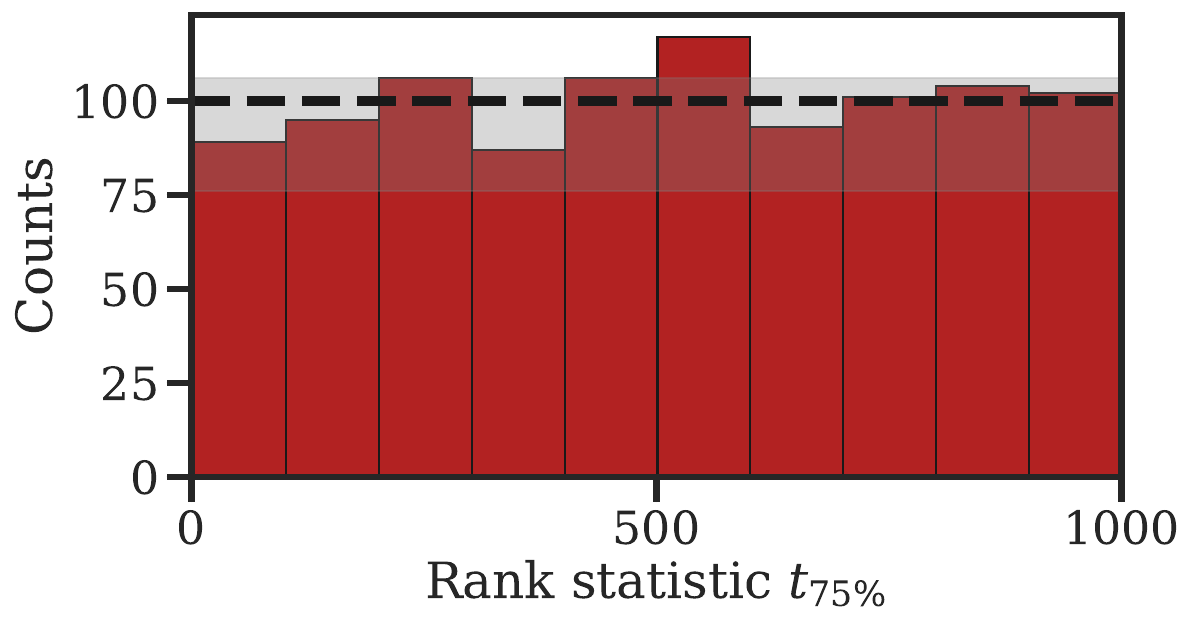}\includegraphics[width=0.2\linewidth]{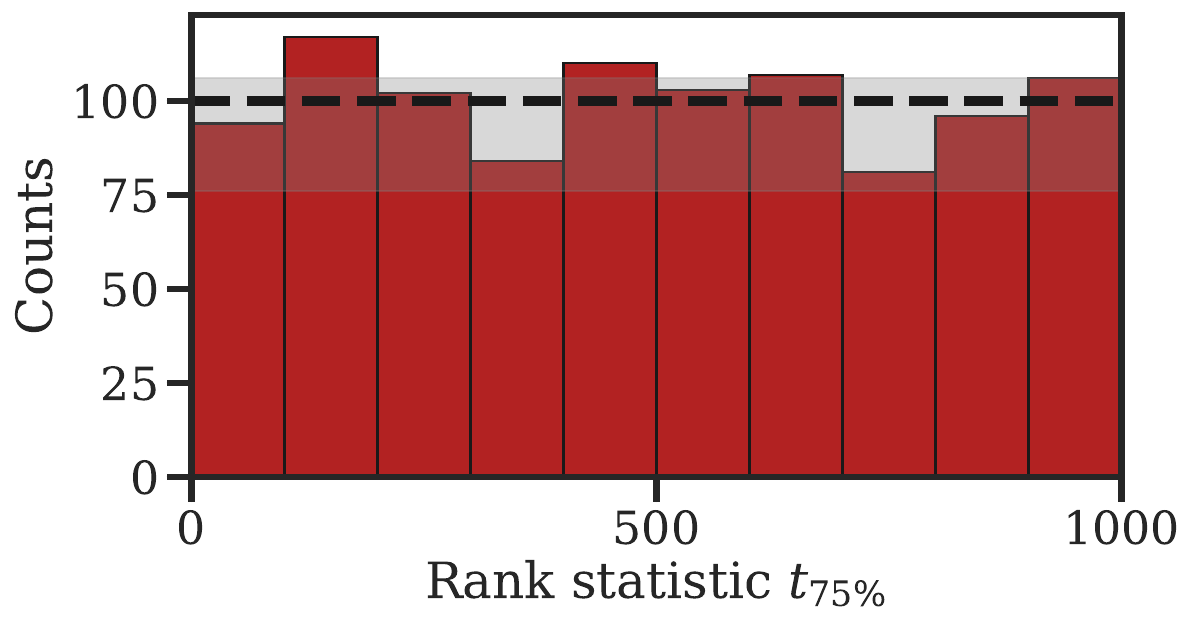}
    \includegraphics[width=0.2\linewidth]{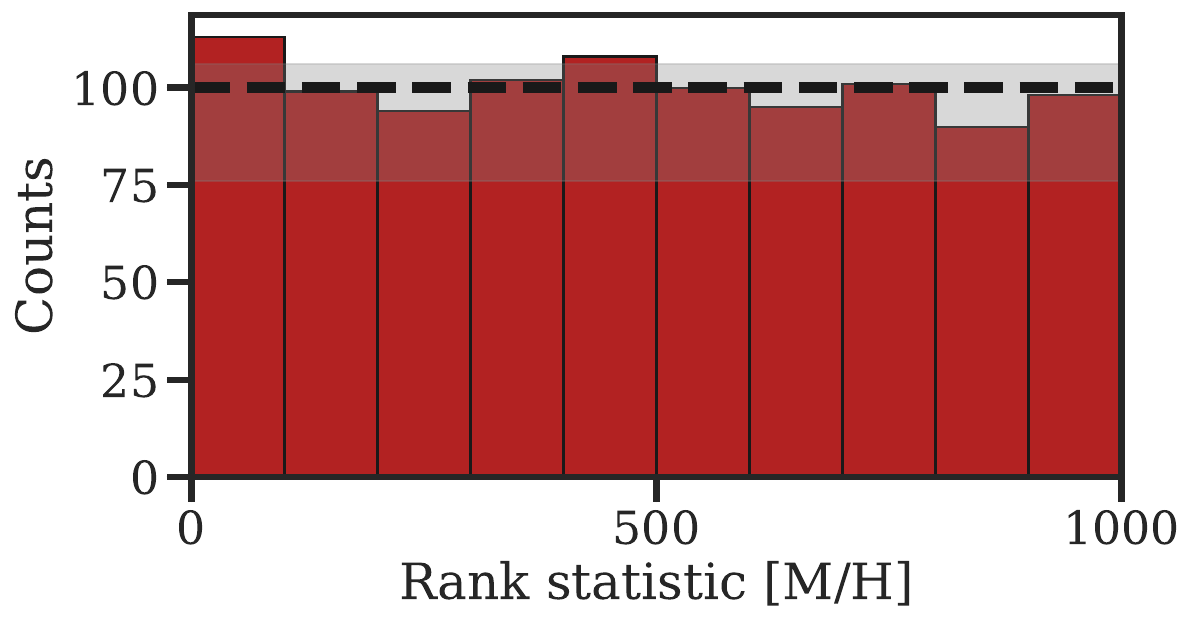} 
    \includegraphics[width=0.2\linewidth]{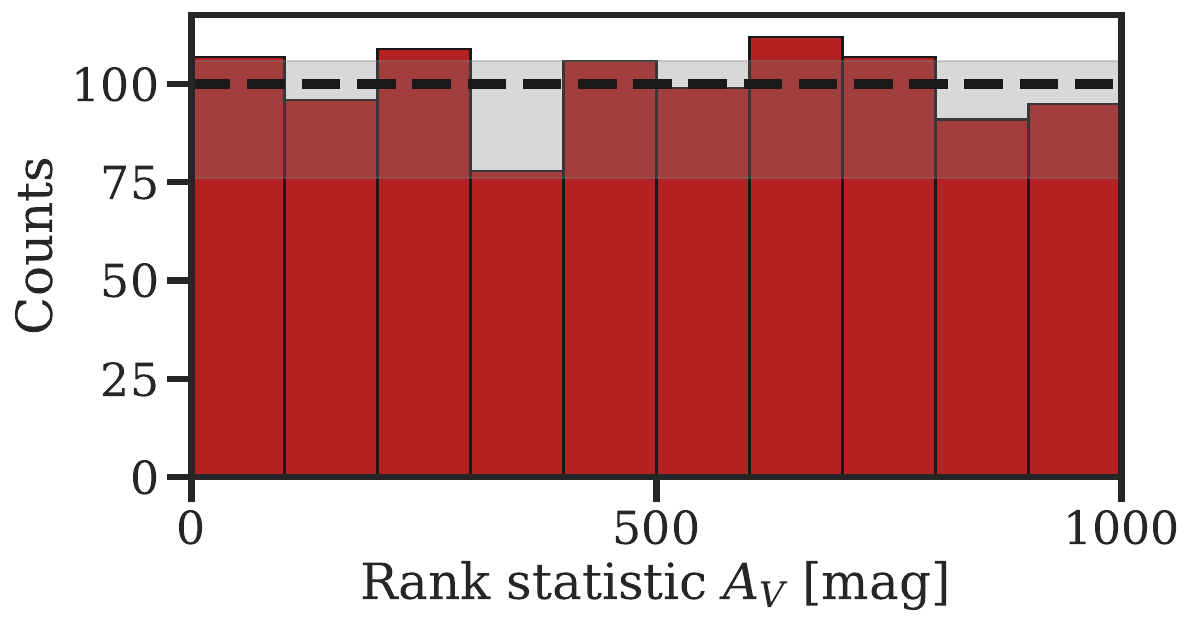} 
    \caption{Same than Fig.~\ref{sbc_ranks_tau}, but using a Dirichlet prior for the SFHs.}
    \label{sbc_ranks_dir}
\end{figure}

\begin{figure}[h!]
    \centering
    \includegraphics[width=0.4\linewidth]{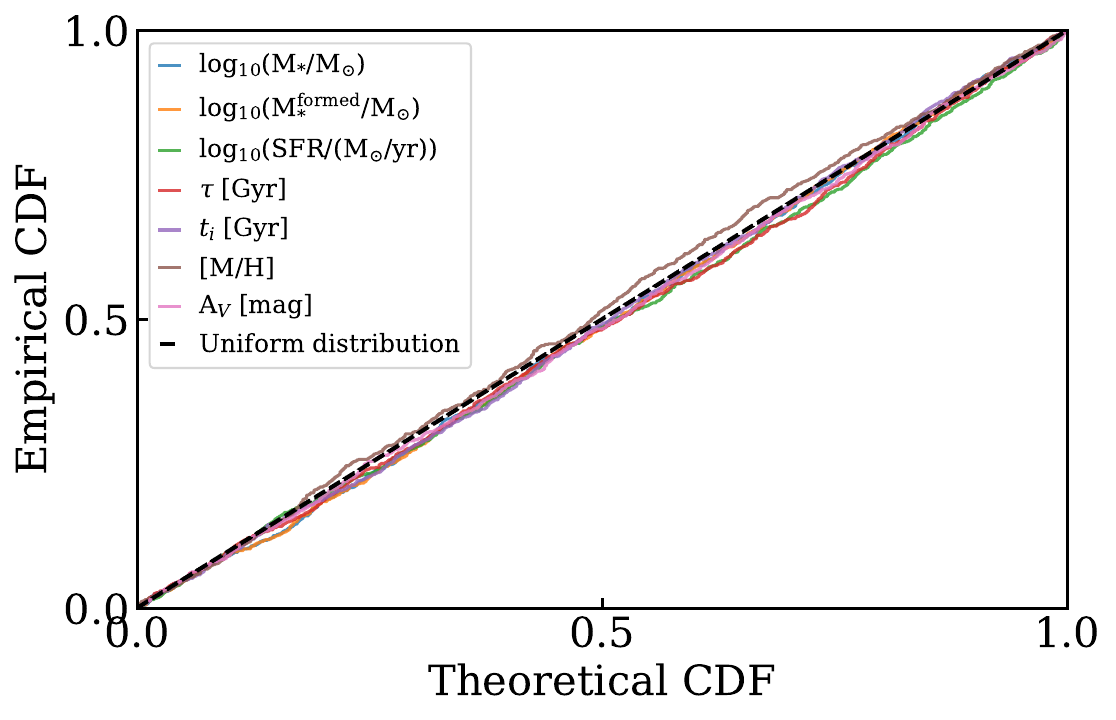}  
    \includegraphics[width=0.4\linewidth]{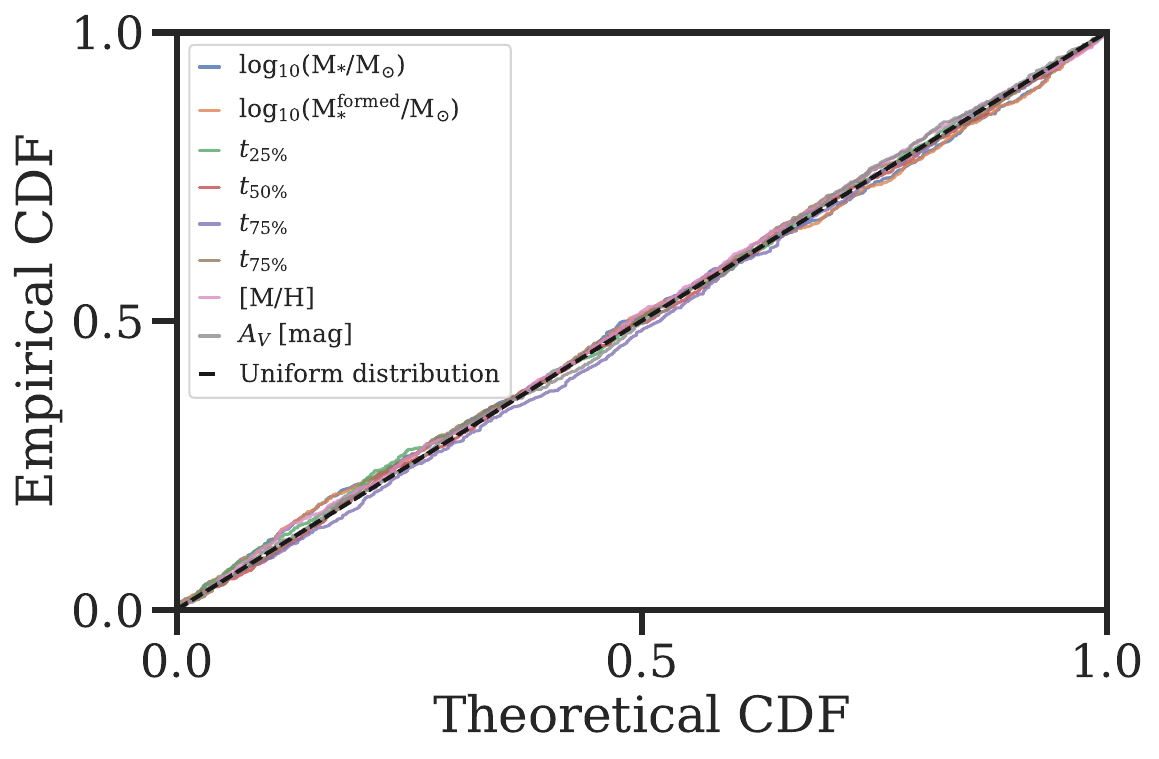}

    \caption{Empirical CDFs for all the parameters obtained with $1000$ simulations using a $\tau$-delayed prior (left) and a Dirichlet prior (right) for the SFHs.  If the posteriors are well calibrated, the nominal coverage probability (the fraction of the probability volume), on the $x$-axis, will be equal to the coverage probability (the fraction of actual values in such a volume), on the $y$-axis, leading to a diagonal CDF (dashed black line). Our CDFs closely follow the diagonal, indicating that the posteriors are properly computed.}
    \label{sbc_cdf_tau}
\end{figure}

We perform a simulation-based calibration \citep[SBC;][]{TALTS2018} test to ensure that the NFs are well trained and thus the posteriors are properly calibrated. In Fig.~\ref{sbc_ranks_tau} we show the distribution of the rank statistics of the true parameter values within the marginalised posteriors for the $\tau$-delayed model, which must be uniform if the posterior samples are consistent with the prior used in the simulations. We show in Fig.~\ref{sbc_ranks_dir} the analogue for the Dirichlet prior. Figure ~\ref{sbc_cdf_tau} shows the empirical cumulative distribution function for both the $\tau$-delayed model and the Dirichlet model, which confirms that they produce posteriors that are fully consistent with the simulations, without underestimating or overestimating the uncertainties. However, we point out that the performance of any model when obtaining posteriors distributions for real galaxies is unknown, and the ability to get accurate distributions, as in any inference pipeline, ultimately depends only on how well the forward model (or likelihood in traditional approaches) matches the observations.\\

\FloatBarrier
\section{Posterior distributions as a function of the S/N}
\label{sigma_sn}

\begin{figure*}[h!]
    \centering
    \includegraphics[width=0.99\linewidth]{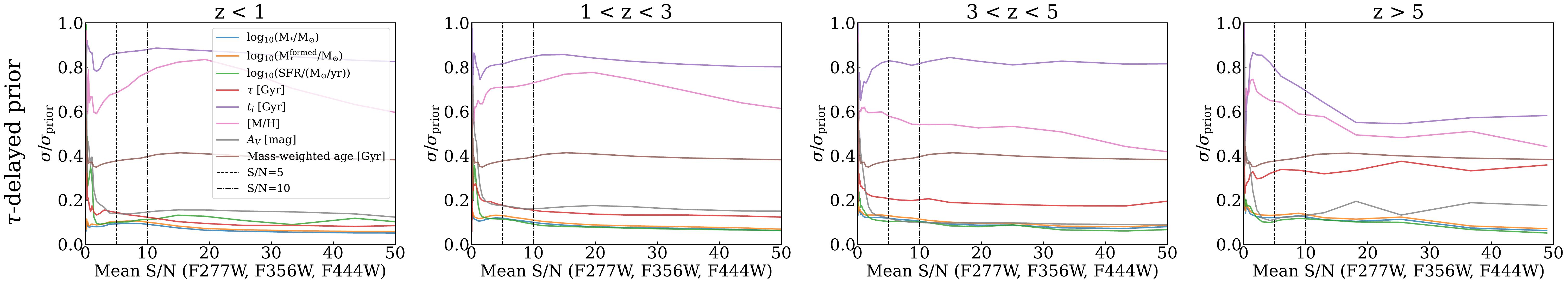}
    \includegraphics[width=0.99\linewidth]{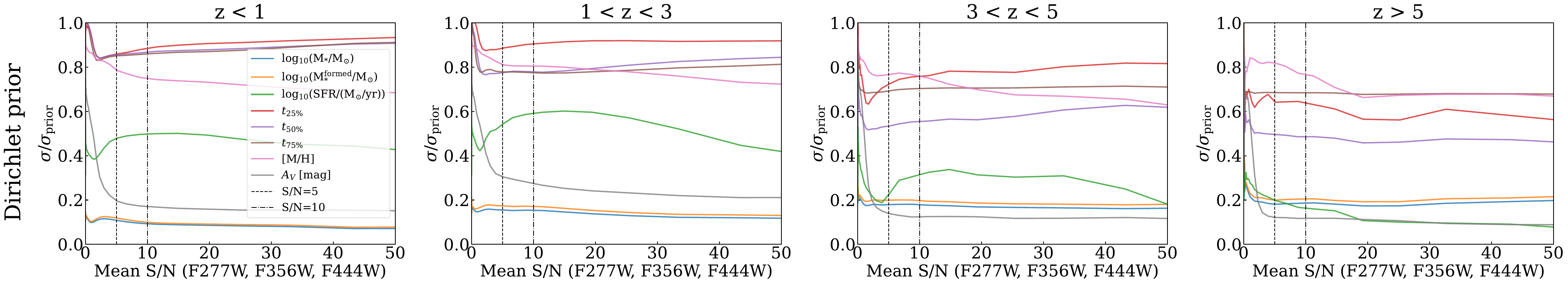}
    \caption{Posterior widths for each property. The average ratio of the posterior standard deviation (obtained with 500 samples) and the prior standard deviation for parameters using a $\tau$-delayed (top row) and Dirichlet (bottom row) prior for the SFHs. This is calculated across all the pixels for each of the $1.083$ galaxies in the dataset, as a function of the average S/N in filters F277W, F356W, and F444W, at different redshift bins.}
    \label{sigma_sn_fig}
\end{figure*}

In a Bayesian analysis, the degree of uncertainty is reflected in the posterior widths, as prescribed by Bayes' theorem. When little to no information can be extracted from the observations, the posteriors tend to resemble the priors. In such cases, where the data have limited constraining power, the posterior is said to be prior-dominated. We fit all the pixels of the $1083$ galaxies with the trained model and obtain posterior distributions with $500$ samples for each of them. We average in Fig.~\ref{sigma_sn_fig} the ratios between the standard deviation of the posteriors and that of the prior for all the parameters for the $\tau$-delayed and Dirichlet priors for the SFHs, in different panels binning them by redshift. As for $\tau$, $t_{25\%}$ and $t_{75\%}$ the prior distributions are highly skewed,  we use another metric for the widths of both prior and posterior distributions, namely the ratio of the percentiles $75\%$ and $25\%$ rather than $\sigma$. For consistence we also use this ratio for $t_{50\%}$.\\

Both the surviving and formed stellar masses can be well-constrained up to  15\% of the prior distribution width, without important changes with the S/N, as low S/N are also informative of the mass content of each of pixels up to S/N~$\sim 1$. There are no strong differences with redshift either, with only a slight increment across all S/N. The dust attenuation index $A_V$ can also be recovered with narrow posterior distributions ($\sigma \sim 0.2 \sigma_{\rm{prior}}$), but experiment a sudden broadening at  S/N~$< 5$, meaning that we start to be prior-dominated. Interestingly, $A_V$ can be better constrained at higher redshifts, probably because most galaxies are  dust free, but also because we have access to the restframe UV slope and Balmer break for these galaxies.\\

The behaviour of the rest of the properties depends on the prior used for the SFHs. The metallicity cannot be retrieved with precision, as already known from photometry only \citep[e.g.][]{nersesian25}, we only reach $\sigma \sim 0.6 \sigma_{\rm{prior}}$ for the $\tau$-delayed prior, and $\sigma \sim 0.7 \sigma_{\rm{prior}}$  for the Dirichlet prior, experimenting also a significant broadening for the posteriors at S/N $< 5$. While for the Dirichlet prior we do not find  important changes across redshifts, for the $\tau$-delayed prior we find slightly narrower posteriors at high redshift, maybe highlighting a strong degeneracy with the age that is only significant at low redshift due to the larger possible ages.\\

The SFR behaviour is very different from one prior to another, with  $\sigma \sim 0.15 \sigma_{\rm{prior}}$ and $\sigma \sim 0.4 \sigma_{\rm{prior}}$ at low redshifts respectively, as expected by restricting the shapes of the SFHs by using a functional form. We obtain overall narrower distributions at high-redshift, especially for the Dirichlet prior. This can again be due to lower dust attenuation but also due to selection effects at higher redshifts in the ranges of masses analysed. The parameter $t_i$ can be poorly constrained, with values ranging from $\sigma \sim 0.6-0.8 \sigma_{\rm{prior}}$,  slightly better for $z>5$. For $\tau$, $t_{25\%}$, $t_{50\%}$ and $t_{75\%}$, we should not compare the values with the other properties because the metric is different. However, we find strong broadening of the posterior distributions for S/N~$< 5$. The width of the $\tau$ posterior distributions is higher with redshift, but slightly lower for $t_{25\%}$ and $t_{75\%}$. As expected, in most of the redshift bins  $t_{25\%}$ is wider than $t_{75\%}$, but this result is inverted for the last bin of redshift. We expect this to be a consequence of the prior for $t_{25\%}$ that highly favours short cosmic times.  For the parametric prior, we also combine the $\tau$ and $t_i$ samples of the posteriors to derive samples for the mass-weighted age, which is known to be a more robust and less degenerated quantity, and find that we can indeed constrain better this property with  $\sigma \sim 0.4 \sigma_{\rm{prior}}$ for all redshift bins.\\

\FloatBarrier

\section{Uncertainty maps}
\label{uncertainty_maps}
\begin{figure*}[h!]
    \centering
    \includegraphics[width=0.6\textwidth]{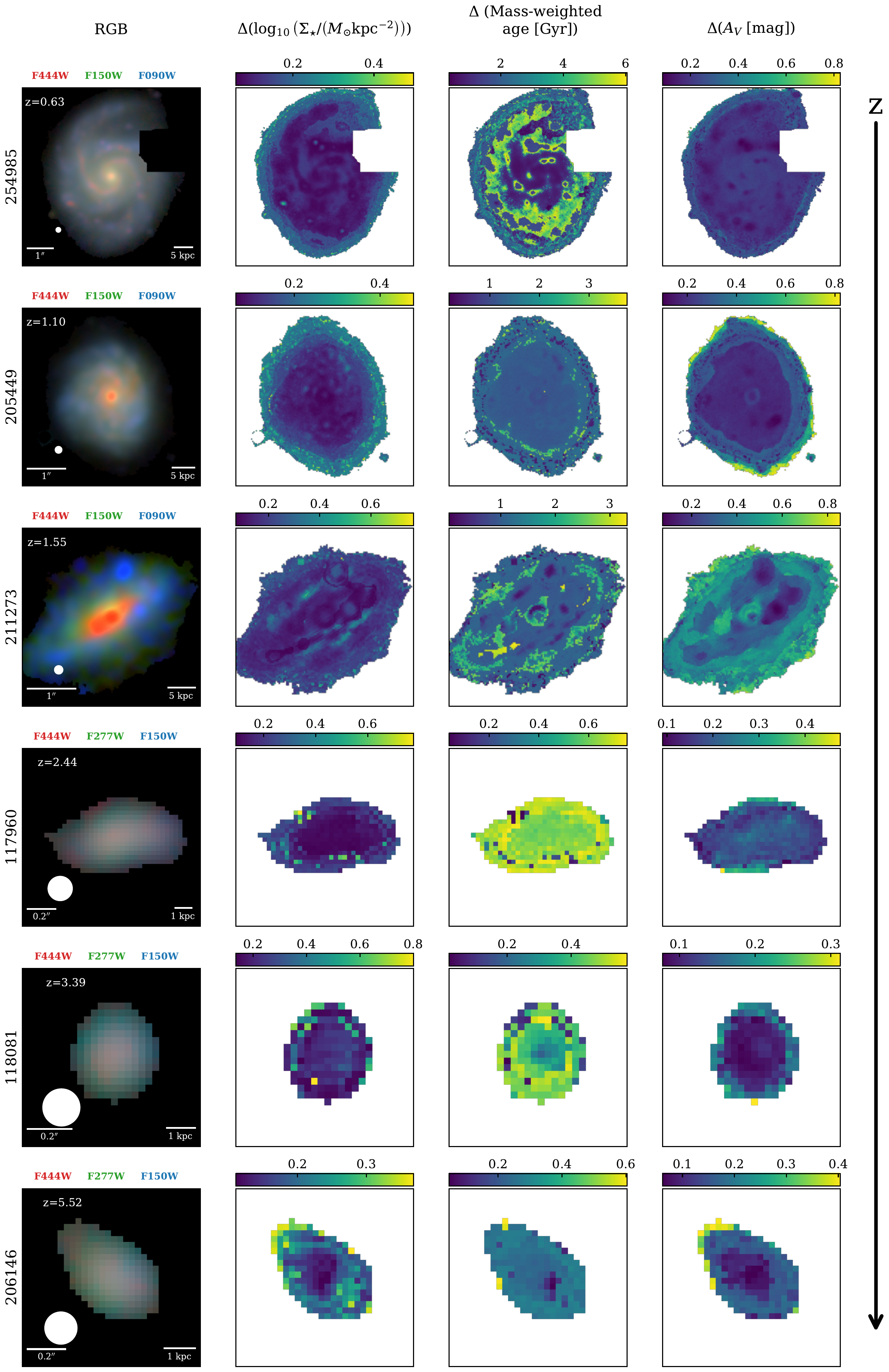}
    \caption{Same as Fig.~\ref{maps_tau} but with the uncertainties from the standard deviations of the posterior distributions.}
    \label{uncertainty_maps_fig}
\end{figure*}

We repeat Fig.~\ref{maps_tau} with the propagated uncertainties from the standard deviations of the posterior samples per pixel in Fig.~\ref{uncertainty_maps_fig}. The maps for the properties of the six sample galaxies show higher uncertainties in the outskirts of the galaxies, which also correspond with the pixels with lower S/N, for $\log _{10}\left(\Sigma_{\star} /\left(M_{\odot} \mathrm{kpc}^{-2}\right)\right)$ and $A_V$, while the higher uncertainties in the mass-weighted age maps correspond to the oldest regions of the galaxy, whose timescales are more difficult to determine, especially at lower redshifts.

\FloatBarrier

\section{Recovered SEDs: posterior predictive checks}
\label{ppc}
\begin{figure*}[h!]
    \centering
    \includegraphics[width=0.75\textwidth]{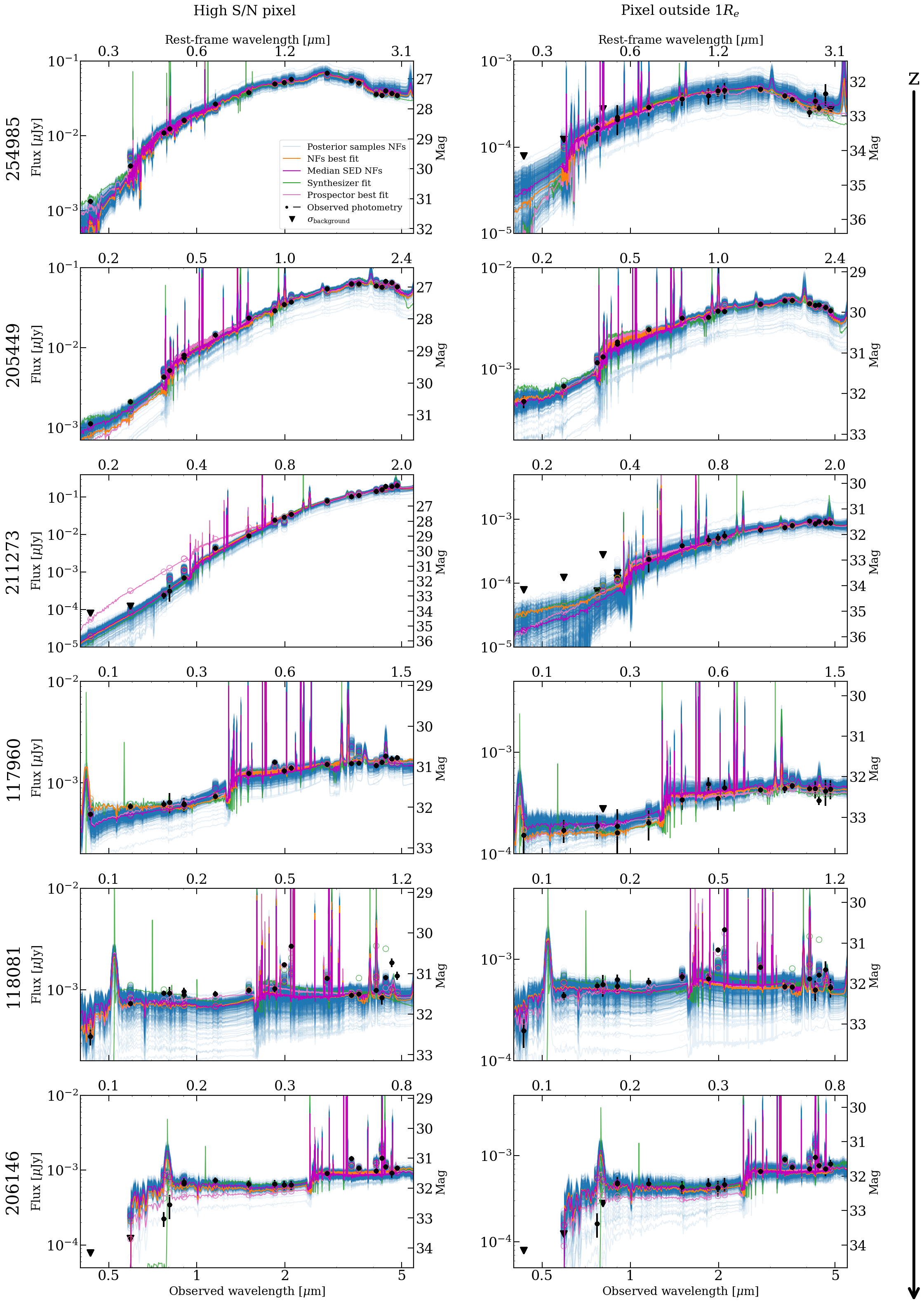}
    \caption{Fit for the two pixels of the six galaxies, repeating the simulation with $500$ samples of the posterior distributions for the pixels with maximum average S/N of the galaxy. In black, we include the photometry with its error, in blue the templates fitted for each posterior sample, with blue circles for the corresponding photometry, in purple the median of these templates,  and in orange the best-fit sample. We also include black triangles with the $1\sigma$ upper limits. The best fit of \texttt{Prospector} is shown in pink and the fit done with \texttt{Synthesizer} is shown in green. }
    \label{templates}
\end{figure*}

We perform a PPC \citep[][]{TALTS2018}, repeating the simulation for every sample of the posteriors for the same two pixels of each of the six galaxies using the parametric model.  We compare the results of the PPC with the actual observed photometry in Fig.~\ref{templates}. We find a general agreement across all galaxies. However, certain sets of SEDs deviate from the true photometry in some filters, typically the bluest ones, for instance F775W and F814W for the pixel with highest S/N of 206146. This discrepancy arises from model misspecification when none of the simulation templates align with the observed photometry for any noise realisation, making the posterior distributions unreliable. This misalignment could stem from problems in the photometry of that pixel, such as contamination from other objects or improper background subtraction. However, it is more likely due to assumptions in the SPS models, including the prior ranges. On the other hand, we observe for some of the pixels there multiple possible SED templates and the model correctly interprets this, so that the posterior distributions are wider, especially at low S/N and at high redshifts where several filters are only fitted with upper limits. The best example for this is the pixel outside $1R_{\rm{eff}}$ of 211273. Finally, a few posterior samples can produce SEDs that appear inconsistent with the photometry, as seen for example in galaxy 188081. This object is very compact and dominated by emission lines, making it particularly difficult to model within our simulation framework, which uses Cloudy from FSPS without emission line fine-tuning. The apparent deviations mainly stem from imperfect modelling of emission lines in the redder part of the spectrum. These cases highlight the non-linear relationship between physical parameters and flux space, where modest parameter shifts can lead to significant SED variation.\\

We include in the same figure the best fit of \texttt{Prospector} in pink, as well as the \texttt{Synthesizer} fit in green. While the fit of \texttt{Prospector} uses the same SPS modelling assumptions, and the deviations we see with respect of our model are mainly random, the differences with \texttt{Synthesizer} are systematic, as we can clearly see different spectral templates. This is more evident for high-redshift galaxies and show the importance of the choice of libraries and parameters for stellar population properties estimation and of marginalising over all the different templates to properly capture the true uncertainties.

\FloatBarrier

\section{Profiles of stellar population properties}
\label{profiles}
\begin{figure*}[h!]
    \centering
    \includegraphics[width=0.8\linewidth]{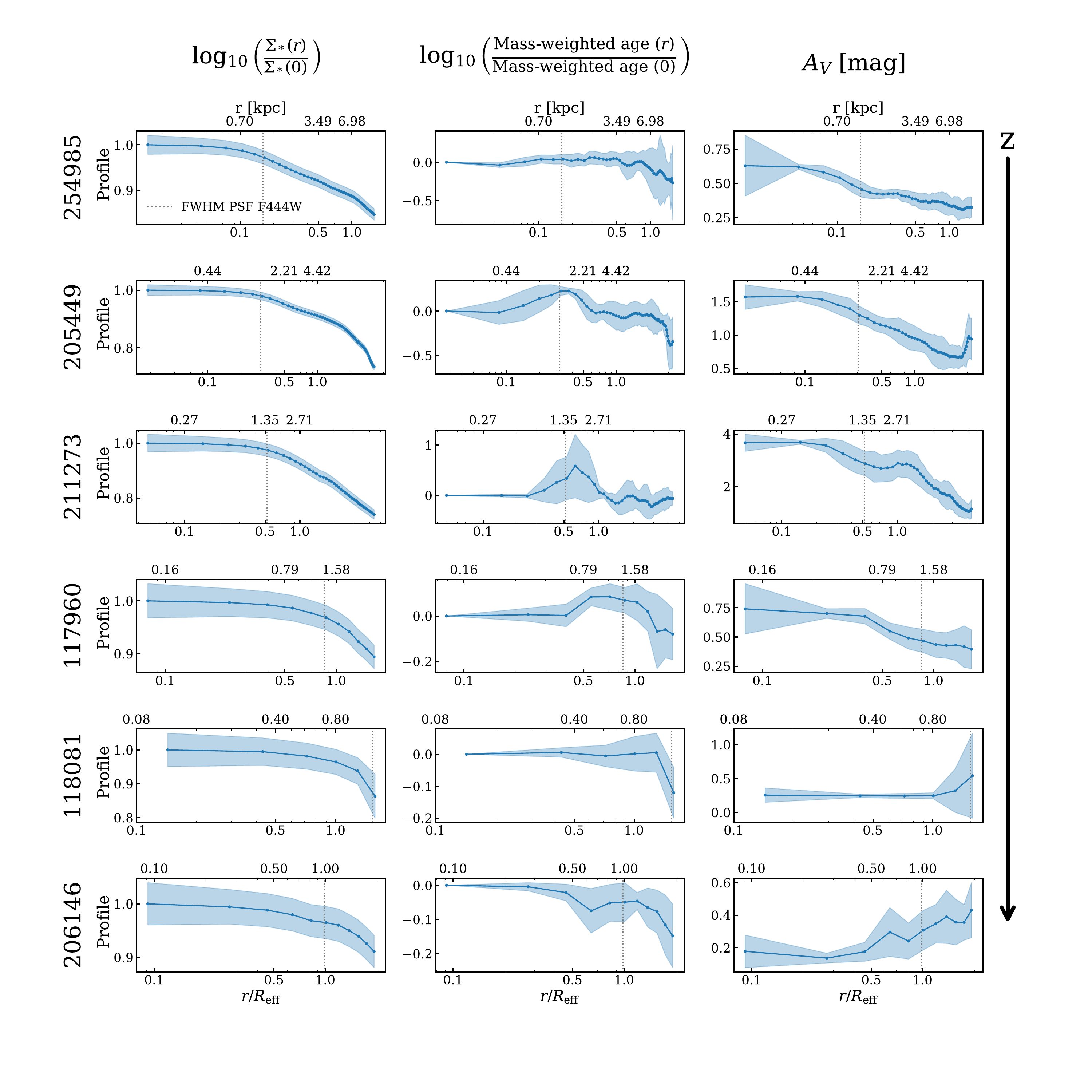}
    \caption{Profiles of stellar mass density, mass-weighted age and dust attenuation, obtained with the model trained on $\tau$-delayed SFHs. The first two were normalised to the value in the centre, for the median of the posterior distributions obtained with the NFs. The errors correspond to the $1\sigma$ of the distributions.}
    \label{prof}
\end{figure*}

We compute profiles of stellar mass density, mass-weighted age and dust attenuation, from posterior distributions obtained with the parametric model. In Fig.~\ref{prof}, we show in blue the profiles of the properties computed from elliptical apertures with  the semimajor axis,  the ellipticity and  the position angle obtained from the JADES photometry catalogue \citep{eisenstein2023b}, and the centres fixed as the pixel with higher stellar masses (medians of the posteriors). The effective radius we use to normalise the distances to the centre was also obtained from the JADES catalogue. The profiles correspond to the medians of the posterior distributions and the errors to the $1\sigma$ level, normalised to the values in the centre of the galaxy. It is important to note again that the profiles below the PSF FWHM of F444W are mainly flat, and after this radial scale vary significantly.\\

We recover what we see in the previous maps, the stellar mass density profiles are smooth and do not suffer strong variations, while the other properties, especifically for the first three galaxies where we reach larger radius, show important variations. While the uncertainties ($1\sigma$) are considerable, we find similar profiles for both three low redshift galaxies (first row), with medium-age centres, old regions between $0.2-0.5$ $R_{\rm{eff}}$ and younger outskirts. For the high-redshift galaxies, we obtain flatter profiles, specially for 118081 where we are mostly inside the FHWM of F444W due to the compactness of the galaxy. For the first four galaxies the centres are more dust attenuated than the outskirts, but both 118081 and 206146 have very little dust contribution in the centre, compatible with the high redshift of these sources.

\FloatBarrier

\section{Offset in stellar mass estimations with different priors for the SFHs}

\begin{figure*}[h!]
\centering
    \label{offset_simulations}
    \includegraphics[width=0.45\linewidth]{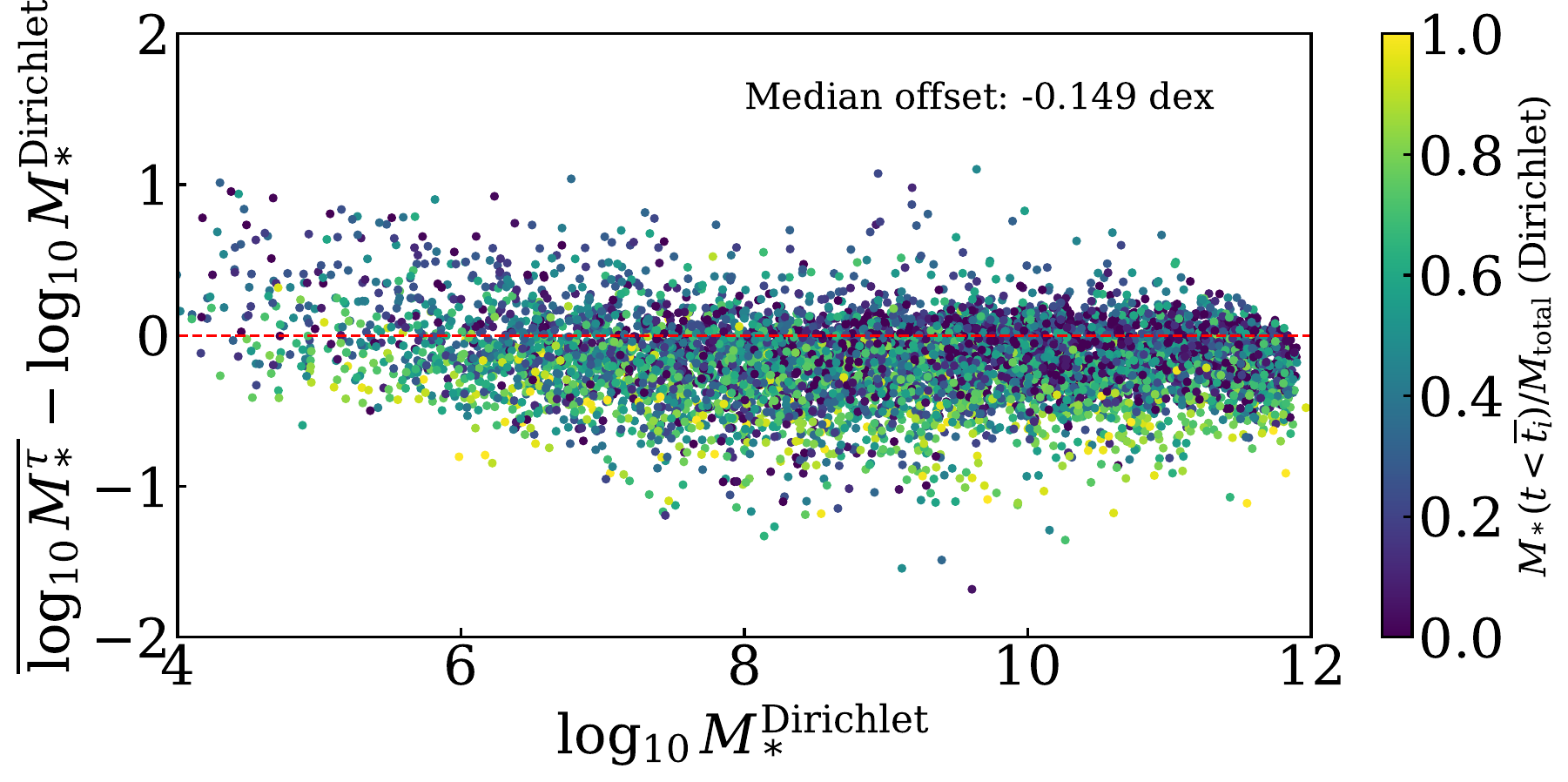}
    \includegraphics[width=0.45\linewidth]{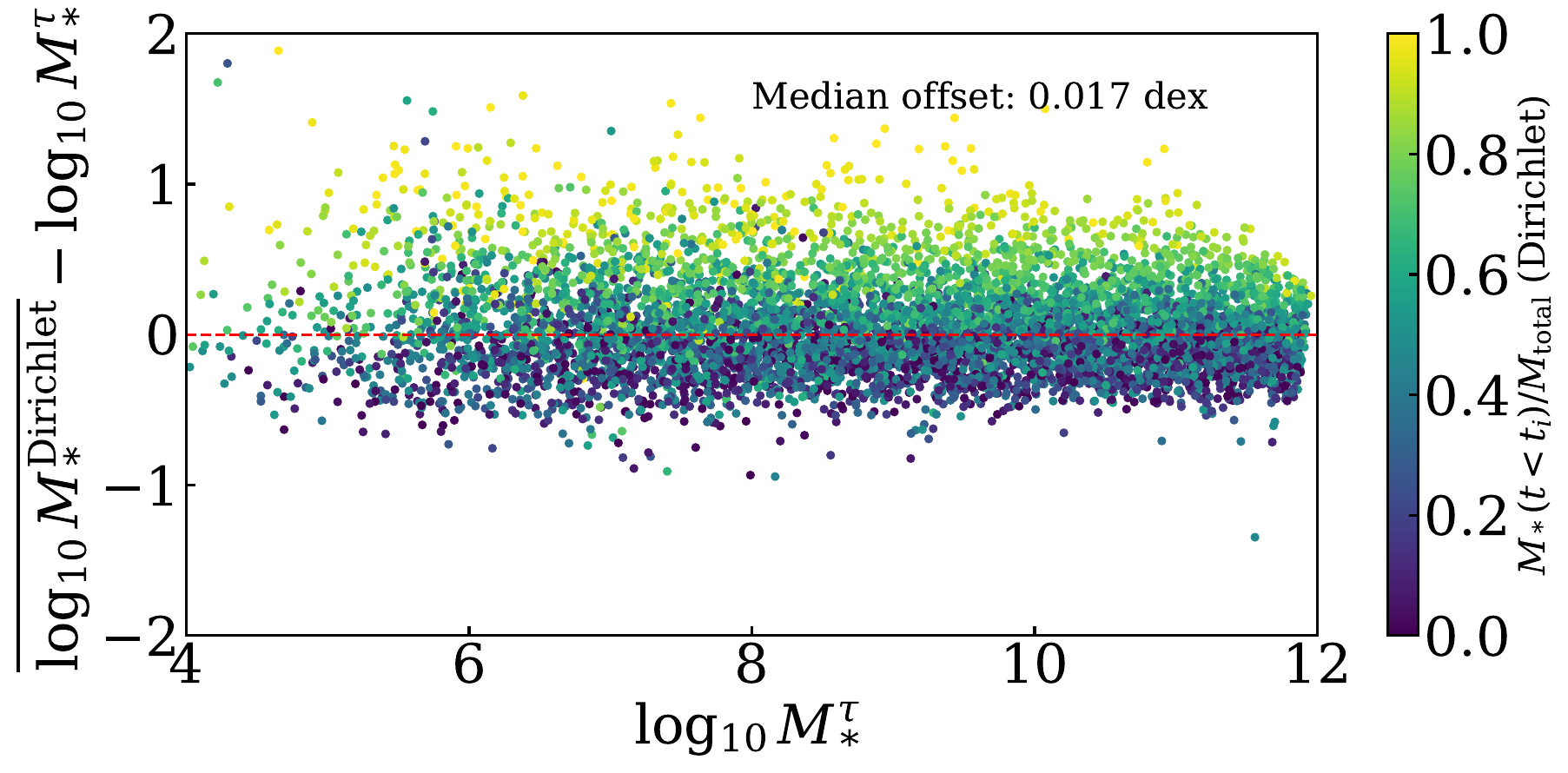}
    \caption{Residuals for the stellar mass when a model trained on one simulation is applied to the other. On the left, we apply the model trained on $\tau$-delayed SFHs to the test set of the simulations performed with the Dirichlet prior. On the right, the model is trained on Dirichlet SFHs and applied to $\tau$-delayed simulations. We colour code each simulation with the fraction of the total stellar mass, computed from the true (inferred) Dirichlet SFH, that is formed before the inferred (true) value for $t_i$ from the $\tau$-delayed SFH. We find a significant average underestimation of the stellar mass for the parametric model applied to the Dirichlet simulations, which correlates with the fraction of mass formed before the inferred starting time for the star formation.}
\end{figure*}

In Fig.~\ref{resolved_vs_integrated} we found a systematic difference between both the pixel-by-pixel and the integrated stellar mass estimates for the observations depending on the two different priors used throughout the paper for the SFHs: the parametric $\tau$-delayed prior and the non-parametric Dirichlet prior. Here, we perform a cross-validation study on the simulated test set, using the model trained with one prior against the simulation with the other prior. In Fig.~\ref{offset_simulations} we show the residuals on the stellar mass for this cross-validation test.\\

For the $\tau$-delayed model applied to the Dirichlet simulation we observe a significant average underestimation of $0.15$~dex. This is expected because the fixed form of the $\tau$-delayed prior is not able to capture early episodes of star formation \citep[e.g.][]{carnall2019,LOWER20}. We also compute the fraction of the true total stellar mass formed before the median $t_i$ obtained with the $\tau$-delayed model, according to the simulated Dirichlet SFH. We find a correlation between the residuals and this fraction, showing that the largest underestimations of total stellar mass for the $\tau$-delayed model actually occur when there are episodes of star formation that contribute significantly to the mass, but are obscured by more moderate and recent bursts of star formation.\\

If instead we apply the model trained on complex SFHs (Dirichlet prior) to the simple $\tau$-delayed simulation, the averaged offset is almost negligible (overestimation of $0.016$~dex). However, there is significant scatter, and when we calculate the fraction of the total stellar mass (according to the Dirichlet fit of the SFHs) formed before the true $t_i$ that is fixed by the $\tau$-delayed simulation, we find again a correlation with the offset, showing that in the cases of overestimation of the stellar mass the model predicts early episodes of star formation that are not present in the simulation.\\

\begin{figure*}[h!]
\centering
    \label{sfh_example}
    \includegraphics[width=0.49\linewidth]{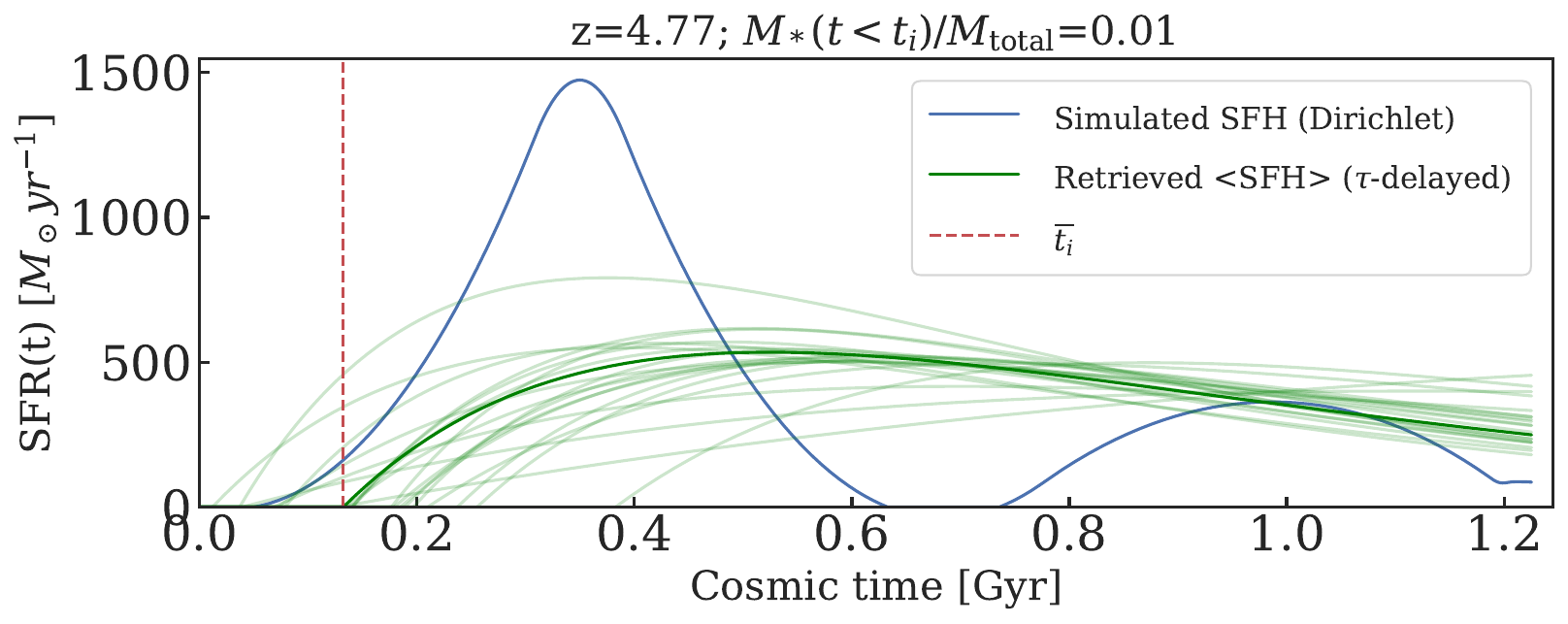}
    \includegraphics[width=0.49\linewidth]{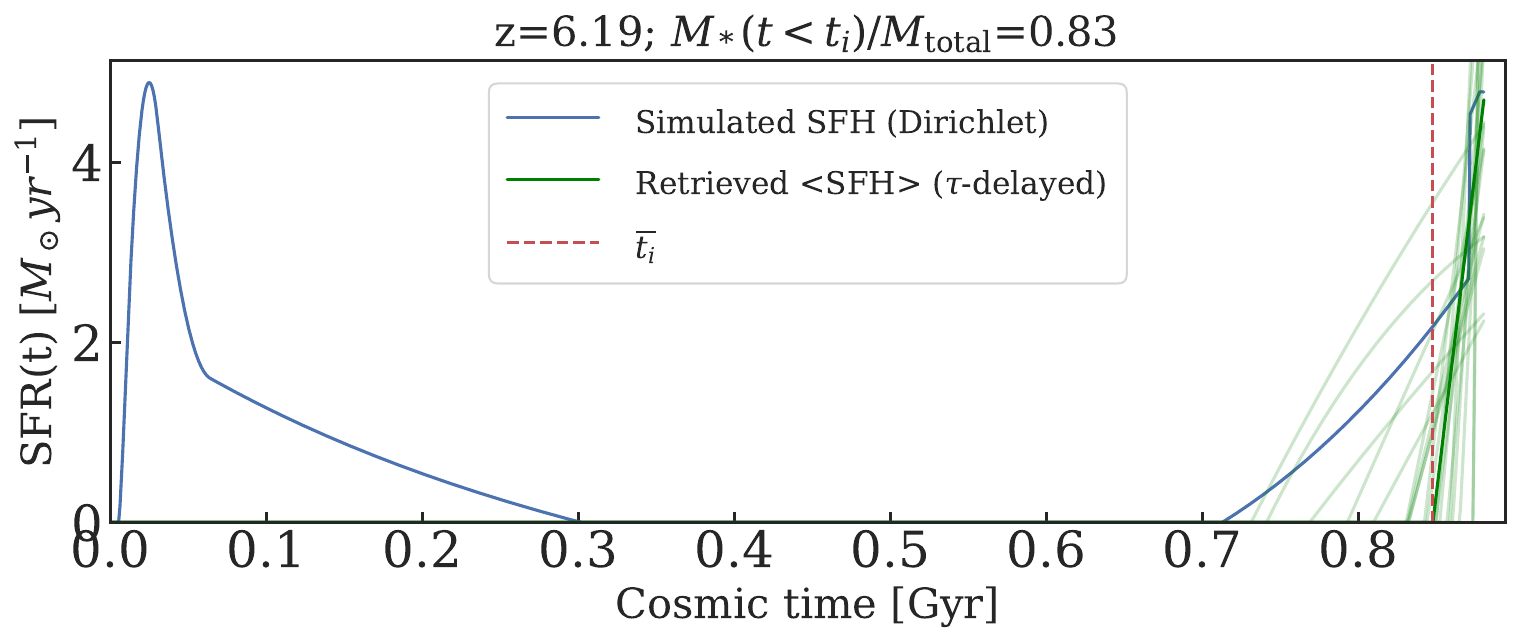}
    \includegraphics[width=0.49\linewidth]{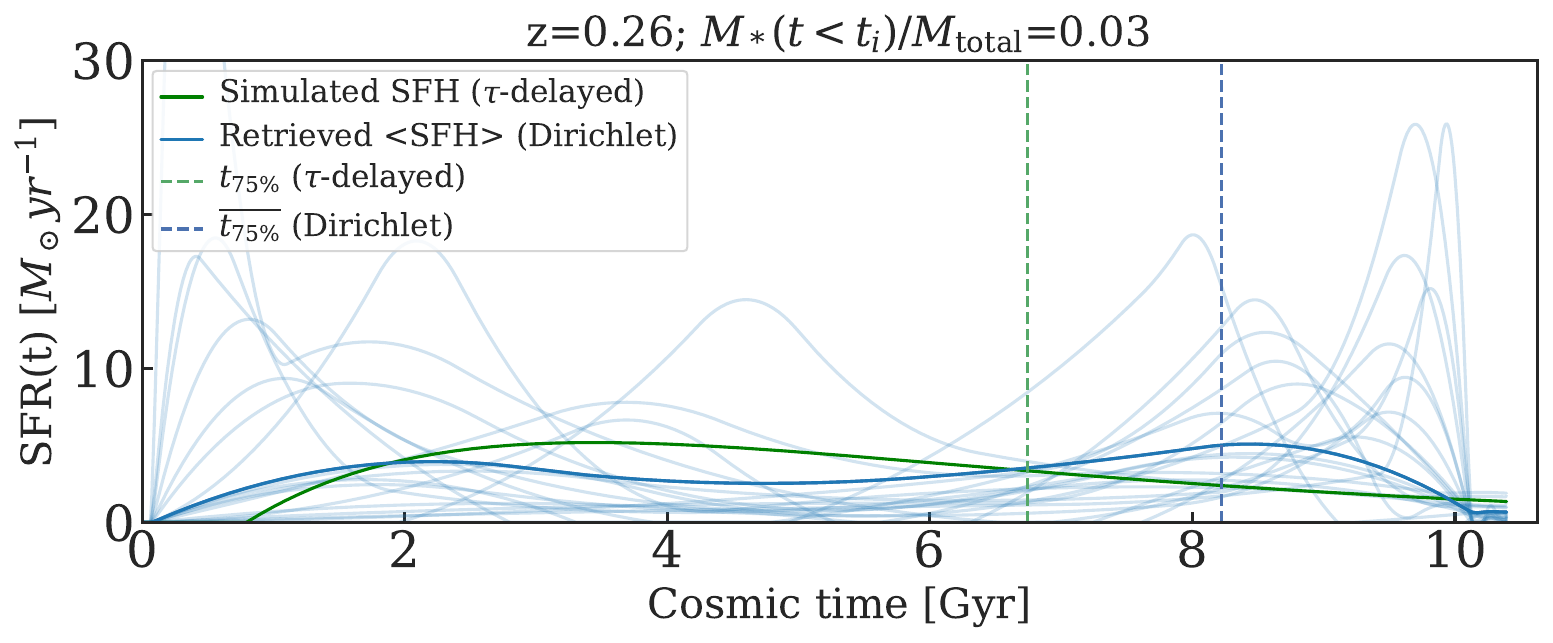}
    \includegraphics[width=0.49\linewidth]{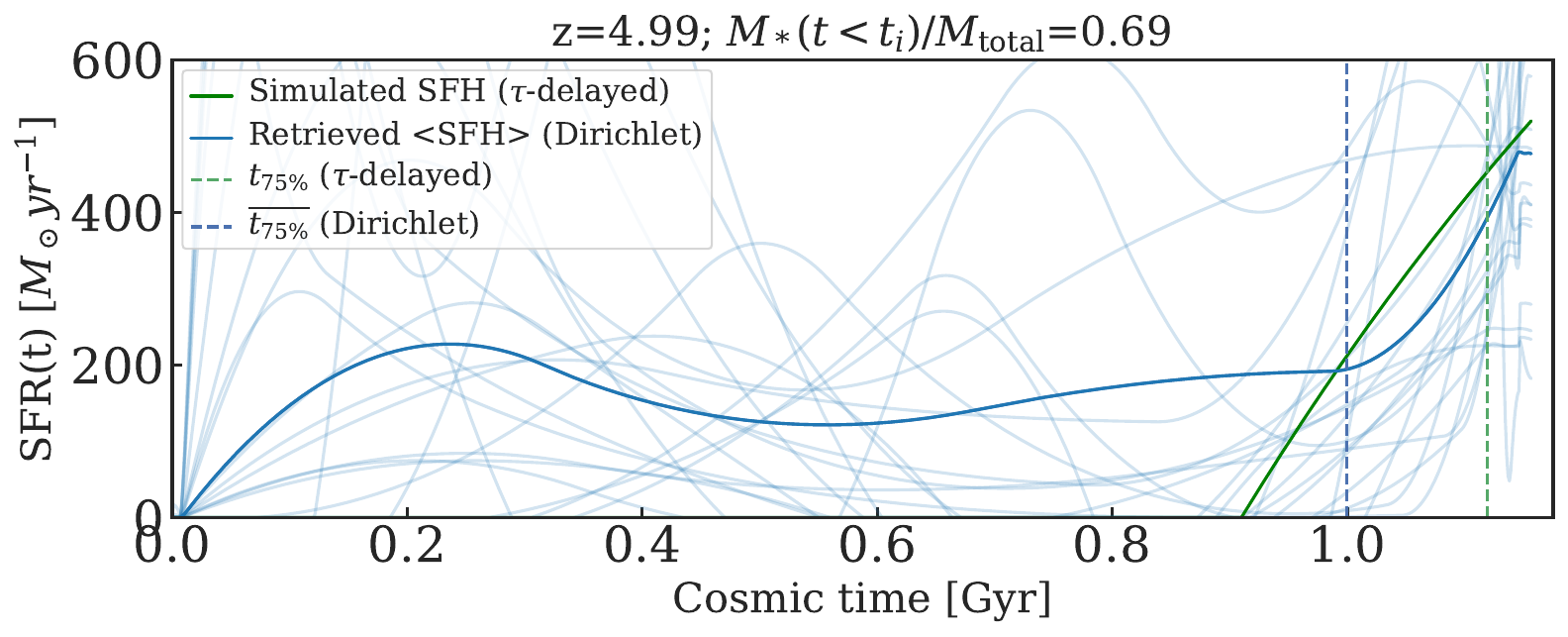}
    \caption{
    Examples of SFHs simulated and inferred with the $\tau$-delayed and Dirichlet priors. The top row shows SFHs simulated with the Dirichlet prior (in blue), and the SFHs inferred from the posterior samples of the parameters obtained with the model trained on $\tau$-delayed SFHs (in low-opacity green), as well as the SFH computed from the median of these posterior samples (in high-opacity green). The bottom row shows simulated $\tau$-delayed SFHs and their corresponding Dirichlet SFHs retrieved from samples of the posterior distribution (in low-opacity blue) and from the median (in high-opacity blue). The left-hand column illustrates examples where the stellar mass estimates are accurate, and the fraction of the total stellar mass, computed from the true (inferred) Dirichlet SFH, that was formed before the inferred (true) value for $t_i$, is close to zero. In the right-hand column, we show examples where the mass estimates are inaccurate, and a significant fraction of the mass was formed (incorrectly assumed to be formed) before the inferred (true) value of $t_i$.}

\end{figure*}

In Fig.~\ref{sfh_example} we randomly select four CSPs from the test sample, simulated in the first row with a Dirichlet prior on the SFHs and in the second with a $\tau$-delayed prior.
The first column corresponds to simulations with small fractions of the mass formed before (the derived or true) $t_i$, while the second corresponds to large values of this fraction.
We show the true SFH, the SFH derived with the median of the posterior samples for the parameters that determine the SFH, and with lower opacity SFHs derived from $20$ samples of the posteriors. For the simulations with the Dirichlet prior, on the left, the $\tau$-delayed model predicts a large episode of star formation to mimic the two different simulated bursts, leading to correct estimates of $t_{i}$, but also correctly estimating the recent SFR and the total stellar mass formed. On the right, due to the intensity of the last simulated burst, the $\tau$-delayed model completely ignores the first episode of star formation, leading to a loss of $83\%$ of the total stellar mass. We also see from the SFHs derived from the posterior samples that the predicted SFHs are not flexible enough to account for the complex episodes, and the uncertainties are certainly underestimated. \\

In the second row, the simulation is performed with a $\tau$-delayed prior. While in the left case, where $\tau$ is large and thus the SFH is extended, the Dirichlet model performs well and also obtains a recent SFR that is close to the true value, in the right case the Dirichlet model predicts previous episodes of star formation that are not realistic, leading to $69\%$ of the mass predicted to form before the star formation actually starts. Nevertheless, in this case we observe that some of the SFHs drawn from the posterior samples are fully compatible with the simulation. This flexibility of the Dirichlet prior, perhaps excessive, allows us to marginalise through a complete set of behaviours, and in this case the large uncertainties in constraining the SFHs are present in the posterior distributions of all the parameters. This leads to broader posterior distributions, as already shown in Fig.~\ref{sigma_sn_fig}, which are more representative of the ignorance of the past SFHs of true galaxies. This effect is also found in the pixel-by-pixel analysis of true JWST galaxies. While for most galaxies the flexibility of the pixel-by-pixel analysis probably does not require complex SFHs at the pixel level, one pixel may represent several SSPs with non-trivial SFHs, especially at high redshift. Ultimately, we should look for physically motivated priors that match the burstiness and star formation scales of the real Universe. This approach, already explored in the works of \cite{Caplar_2019,Tacchella_2020,Iyer_2024,wan2024}, and \cite{wang2025}, for instance, is crucial to properly assess the estimation of stellar population parameters with well-calibrated uncertainties from photometry alone.

\end{appendix}

\end{document}